\documentclass[acmlarge]{acmart}
\AtBeginDocument{%
  }

\setcopyright{acmlicensed}
\copyrightyear{2025}
\acmYear{2026}
\acmDOI{XXXXXXX.XXXXXXX}

\acmJournal{POMACS}
\acmVolume{37}
\acmNumber{4}
\acmArticle{111}
\acmMonth{07}

\usepackage{makecell}
\usepackage{tcolorbox}
\usepackage{tikz}
\usepackage{tikz-uml}
\usepackage{multirow}
\usepackage{booktabs}
\usepackage{multirow}
\usepackage{subcaption}

\usepackage{booktabs}
\usepackage{tabularx}
\begin{document}

\title{How Do AI Coding Agents Contribute to Software Development? an Empirical Study of Agentic Pull Requests}

\author{Iren Mazloomzadeh}
\authornote{Both authors contributed equally to this research.}
\email{iren.mazloomzadeh@polymtl.ca}
% \orcid{0000-0003-2263-2984}

\author{Mohammad Mehdi Morovati}
\authornotemark[1]
\email{mehdi.morovati@polymtl.ca}

\author{Foutse Khomh}
% \authornote{Corresponding author.}
\email{foutse.khomh@polymtl.ca}

\affiliation{%
  \institution{Polytechnique Montréal}
  \city{Montréal}
  \state{Quebec}
  \country{Canada}
}

% \affiliation{%

\renewcommand{\shortauthors}{Mazloomzadeh et al.}

\begin{abstract}

Recent advances in large language models and their rapid adoption across software engineering tasks have made Artificial Intelligence (AI) coding agents an integral component of modern software development workflows. 
While developers increasingly benefit from these coding agents, their impact on software quality remains insufficiently understood. In particular, how agentic contributions evolve across the software development lifecycle has not been thoroughly investigated.
This study aims to characterize agentic pull requests (PR) in comparison to human generated PRs and to examine how their properties change across different stages of the development lifecycle. Using the AIDev dataset, we first analyze how differences in merge rates between agentic and human generated PRs vary over time. We then identify the types of development tasks where AI coding agents are predominantly applied and investigate how these task distributions evolve across development quarters. Finally, we compare a set of key characteristics of agentic and human generated PRs, focusing on their implications for software quality and their temporal dynamics.
Overall, our findings provide an empirical and longitudinal perspective on the role of AI coding agents in software development, offering a more nuanced understanding of their benefits and limitations in real-world practices.

\end{abstract}





\maketitle

% ==========================================
% --------  Introduction Section -------------
% ==========================================
\section{Introduction}\label{sec:introduction}

% ...................................................................................................................................................................................

The rapid emergence of AI-powered coding agents is reshaping the landscape of collaborative software development ~\cite{hassan2025agentic, horikawa2025agentic,watanabe2025use,wang2025ai}.
Agentic coding is based on Artificial Intelligence (AI) agents that can interpret high-level objectives, decompose them into subtasks, plan and execute development actions, and iteratively refine their behavior using real-time feedback and outcomes with minimal human intervention ~\cite{horikawa2025agentic,sapkota2025vibe,watanabe2025use}.
For instance, an AI agent can translate a natural language feature description into a sequence of development activities, including code generation, test creation, test execution, bug analysis and fixing, and ultimately the preparation of a pull request (PR) ~\cite{wang2025ai}.

Despite growing adoption, there is still limited empirical understanding of how coding agents influence long-term project outcomes ~\cite{he2025speed}.
It is important to examine agentic coding tools implications for software quality, team dynamics, and software engineering processes, because they move beyond the capabilities of Large Language Model (LLM)-based chatbots such as ChatGPT by autonomously handling more complex and integrated development tasks ~\cite{watanabe2025use}.
For instance, it remains unclear whether agent-generated contributions improve over time as repositories evolve, or how human developers respond to and integrate them.
Another important gap concerns the functional usage patterns of coding agents. While prior studies suggest that agents are often used for documentation, refactoring, or test improvement ~\cite{watanabe2025use}, there is no comprehensive characterization of the types of tasks where agents are most frequently employed in practice. Understanding these usage patterns is essential for identifying both strengths and limitations of current agent-based development workflows. Finally, a critical concern is the comparative quality of agentic contributions relative to human-authored ones ~\cite{yoshimoto2026testing, ogenrwot2026ai}.

To address these gaps, we conduct a large-scale empirical study of how AI coding agents contribute to collaborative software development and how their contributions vary across the agentic repository lifecycle. Using the AIDev dataset~\cite{li2025rise}, supplemented with PR- and commit-level data collected from GitHub, we analyze 220,612 closed PRs from 489 popular Python repositories. This dataset includes 9,428 PRs generated by five widely used coding agents: OpenAI Codex~\cite{openai_codex}, GitHub Copilot~\cite{github_copilot}, Claude Code~\cite{claude_code}, Cursor~\cite{cursor}, and Devin~\cite{devin}. We examine agentic contributions from three complementary perspectives. First, we investigate how the likelihood of an agentic PR being merged varies across coding agents and software development lifecycle. Second, we characterize the development tasks addressed by agentic PRs and examine how their prevalence and mergeability vary across software development lifecycle quarters. Third, using balanced samples of 2,275 merged agentic and 2,275 merged human-generated PRs, we compare their change, review, and integration characteristics and assess their respective associations with bug-inducing commits. By integrating longitudinal, task-oriented, and defect-related analyses, our study provides a multifaceted understanding of the roles and outcomes of agentic contributions in collaborative software development.

Our findings show that the mergeability of agentic PRs varies across coding agents but remains relatively stable throughout the software development lifecycle. Our descriptive task-level analysis indicates that coding agents are used predominantly for repetitive, well-structured tasks. Agentic PRs addressing narrowly scoped and semantically well-defined tasks generally exhibit higher and more stable merge rates than those requiring extensive contextual understanding, although these differences become less pronounced
in the final development quarter. Although agentic and human-generated PRs differ in several contribution characteristics, most differences are limited in practical magnitude. 
It is also worth noting that agentic PRs show comparable or lower defect proneness than human PRs, with mostly non-significant differences, 
% while their defect proneness declines significantly from the second to the third lifecycle quarter. 
Taken together, these findings provide a nuanced understanding of the roles, outcomes, and potential risks of AI coding agent contributions in collaborative software development.

This paper makes three principal contributions. First, we operationalize the software development lifecycle and use it to conduct a longitudinal analysis of PR acceptance across five coding agents, enabling standardized comparisons among repositories with different periods of agent activity. Second, we provide a task-level characterization of agentic PRs and examine how the prevalence and mergeability of PRs related to development tasks change across various development quarters. Third, through balanced comparisons with human-generated PRs, we quantify differences in change scope, collaboration, and integration and assess whether the two groups differ in their propensity to introduce bug-inducing commits. To facilitate replication and further research, we publicly release the data and supplementary materials used in this study~\cite{replication_package}.

The remainder of this paper is structured as follows. Section~\ref{sec:background} provides background of coding agents, PRs analysis, and repository evolution. Section~\ref{sec:methodology} 
describes the study goal, perspective, context, and research questions, followed by the data-collection process, measures, and analysis procedures.
Sections \ref{sec:result} 
present our findings for each research question. 
The closest related works to this study are reviewed in Section~\ref{sec:related_work}. 
Finlay, Section~\ref{sec:threats} explains threats to the validity of our findings, and Section~\ref{sec:conclusion} concludes the research.

\section{Background}
\label{sec:background}

\subsection{GitHub and PR}
GitHub serves as a comprehensive platform for collaborative software development, where project artifacts are organized in a structured repository system that supports version-controlled history, enabling teams to track changes, manage project components efficiently, and restore earlier states when necessary, thereby ensuring both data integrity and reliable project evolution ~\cite{wunschiers2025github,zhang2025small}.
GitHub has become a central ecosystem for open-source software development by combining large-scale repository hosting with social collaboration features, such as issues, PRs, stars, and forks, which facilitate developer interaction, project visibility, and community-driven software evolution across academia and industry ~\cite{alrashedy2024software}.

The PR mechanism ~\cite{gousios2014exploratory,vasilescu2015quality} integrates code repositories, issue tracking systems, code review processes, and automated DevOps tools into a unified collaboration environment for distributed Open-source Software  (OSS) development, thereby significantly lowering the entry barrier for contributors and simplifying collaboration compared to traditional patch-based approaches ~\cite{li2021you}. 
As a widely adopted mechanism in both open-source and commercial projects, PRs provide developers with a convenient way to contribute to software development while enabling systematic review of contributions from different developers ~\cite{lenarduzzi2021does}.
In open-source projects, global collaboration is enabled through the pull-based development model, in which contributors propose changes by submitting PRs from their forked repositories for review ~\cite{zhang2022pull}.

\begin{figure}
    \centering
    \includegraphics[width=0.9\linewidth]{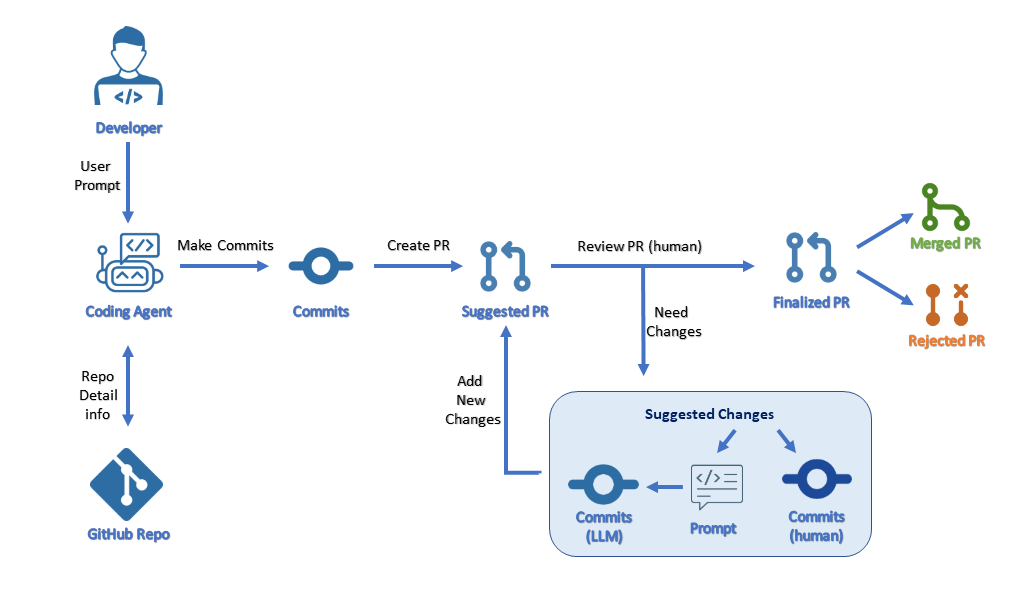}
    \caption{Workflow of PR generated by AI coding agents}
    \label{fig:placeholder}
\end{figure}

\subsection{AI Coding Agent}

The landscape of software engineering is being transformed by autonomous AI agents capable of handling complex tasks, ranging from feature development to debugging and code review, often requiring little human intervention ~\cite{li2025rise}.
LLM-based agents or agentic AI in the LLM era combine language model reasoning with autonomous capabilities—such as memory, planning, tool access, and environmental interaction—to achieve more sophisticated reasoning, deeper contextual understanding, and superior tool utilization when tackling complex, multi-step objectives~\cite{krishnan2025ai,sapkota2025vibe}.
Autonomous coding agents like Google’s Jules, OpenAI’s Codex, Anthropic’s Claude Code, and Congition’s Devin are flooding repositories with PRs~\cite{li2025rise}, yet this hyper-productivity is widening the gap between rapid delivery and verified trust~\cite{hassan2025agentic}.
Agentic AI systems face compounded challenges—including weak causal reasoning, fragile coordination and communication, unpredictable emergent behavior, poor scalability and debuggability, limited explainability and verification, heightened security risks, unresolved ethical and governance issues, and immature theoretical foundations—that together undermine their reliability, safety, and trustworthiness at scale~\cite{sapkota2025ai}.

\section{Methodology}
\label{sec:methodology}

\subsection{Study Goal, Perspective, and Context}

The primary \textit{goal} of this study is to understand how AI coding agents contribute to collaborative software development and how the characteristics of their contributions change across different stages of the software development lifecycle.
To this end, 
we examine agentic contributions from three complementary perspectives: (i) the likelihood of acceptance of agentic PRs across different agents and stages of the software development lifecycle, (ii) the development tasks addressed by agentic PRs and the agentic behavior changes related to each development task over time, 
and (iii) the characteristics and defect proneness of agentic PRs compared with human-generated PRs. Overall, these analyses provide empirical evidence about the acceptance, development roles, characteristics, and potential risks of coding agent contributions in real-world software development.

The \textit{perspective} is primarily that of software engineering researchers interested in understanding the evolution and impact of AI coding agents in real-world software development. 
The findings may also be of interest to repository maintainers and developers, who review and integrate agentic contributions and need to understand the development activities for which coding agents are more frequently used and the quality implications of their contributions. 
Furthermore, engineering managers and quality assurance personnel may use these findings when assessing the adoption of coding agents and their potential long-term impact on software systems.

\begin{figure}
    \centering
    \includegraphics[width=0.95\linewidth]{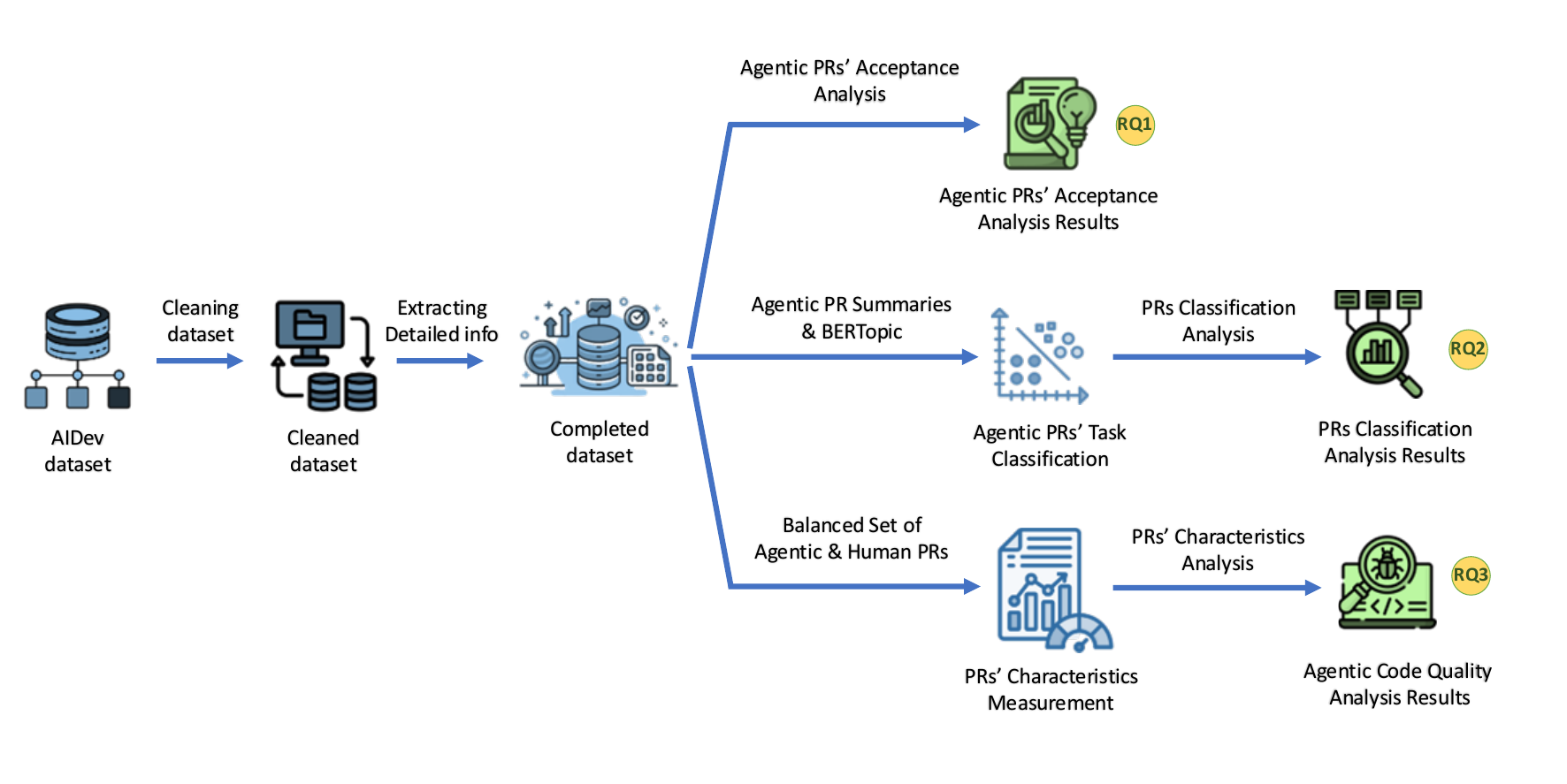}
    \caption{High-level view of the used methodology for this paper}
    \label{fig:method}
\end{figure}

The \textit{context} of this study consists of agentic and human-generated PRs from open-source software repositories. The analyzed repositories include contributions from multiple coding agents and their development histories, allowing us to study agentic PRs across software development lifecycle stages and compare agentic contributions with human-generated contributions.
Fig.~\ref{fig:method} provides a high-level overview of the overall research process. 
Accordingly, we formulate the following research questions.

\begin{itemize}
    \item[\textbf{RQ1.}] \textbf{How does the mergeability of agentic PRs evolve across different stages of the software development lifecycle?} The increasing use of coding agents does not necessarily imply that their contributions are successfully integrated into software projects. The acceptance of agentic PRs may vary across coding agents and over time as 
    software systems evolve. Understanding these variations is important for assessing whether coding agents contributors become more effective throughout repository evolution. Therefore, we investigate the mergeability of agentic PRs across coding agents and software development stages.
    \item[\textbf{RQ2.}] \textbf{What development tasks are most frequently addressed by agentic PRs?} Although coding agents can potentially contribute to a broad range of software development activities, their actual roles in real-world software systems remain insufficiently understood. Identifying the development tasks addressed by agentic PRs can reveal the activities for which coding agents are most frequently employed and whether their contributions concentrate on particular types of development work. Moreover, examining these tasks throughout software development evolution allows us to investigate whether the functional roles of coding agents change over time.
    \item[\textbf{RQ3.}] \textbf{How do agentic PRs differ from human-generated PRs in terms of code quality and defect proneness?} The acceptance of an agentic PR does not necessarily indicate that its contribution has the same quality characteristics as a human-generated contribution. Agentic and human-generated PRs may differ in terms of complexity, collaboration and review effort, and integration time. More importantly, agentic and human-generated PRs may have different propensities to introduce defects into the codebase.
    Comparing the characteristics and defect proneness of agentic and human-generated PRs helps us evaluate how agentic contributions differ from human contributions and assess their potential long-term implications for software quality.
\end{itemize}

\subsection{Data Collection}
\label{subsec:data_collection}

The primary data source for this study is the AIDev dataset provided by Li et al.~\cite{li2025rise}.
AIDev provides a dataset of detailed information regarding GitHub PRs generated by five widely used AI coding agents including
OpenAI Codex~\cite{openai_codex}, GitHub Copilot~\cite{github_copilot}, Cursor~\cite{cursor}, Devin~\cite{devin}, and Claude Code~\cite{claude_code}.
The dataset also includes several derived sub-datasets created by applying additional filtering criteria. We utilize the updated repository dataset from Li et al. \cite{li2025rise}, available on Hugging Face\footnote{https://huggingface.co/datasets/hao-li/AIDev} as of the August 10, 2025 update.
We focus our analysis on \textit{Python} repositories associated with the selected software engineering agents. We chose \textit{Python} because it is one of the most widely used programming languages in software engineering research and practice~\cite{jimenez2023swe,morovati2024bug}, and it is also well represented in the AIDev dataset~\cite{li2025rise}. Following the filtering strategy of the AIDev dataset, we further limited our analysis to repositories with more than 100 GitHub stars. This threshold has been widely adopted in previous software engineering studies to select repositories with sufficient popularity and visibility~\cite{moriconi2025ghalogs,tang2022towards,romeo2025uml}. It also serves as an indicator of repository activity and community engagement~\cite{sakib2025understanding}. Applying this criterion helps focus the analysis on repositories with sufficient development history while reducing noise from inactive or experimental projects.
Using this filtration method, we identified 539 unique repositories. 
The number of \texttt{Python} repositories with at least 100 stars 
associated with each AI agent is as follows: OpenAI Codex has 304 repositories, GitHub Copilot has 136 repositories, Claude Code has 58 repositories, Cursor has 57 repositories, and Devin has 51 repositories.

Since our analysis requires examining both merged and rejected PRs generated by coding agents, we collect all closed PRs (rejected and merged) from these 539 unique repositories.
To extract this information, we use the GitHub REST API v3~\cite{github_api_v3}.
To ensure that all collected repositories for each agent contained at least one PR generated by that agent, we validated every repository following the methodology described by Li et al. \cite{li2025rise}. 
By agentic PRs associated with a specific agent, we refer exclusively to PRs generated by that targeted agent, excluding PRs produced by any other AI coding agents.
For the Open-AI Codex, GitHub Copilot and Cursor agents, we included repositories that contained at least one PR whose branch prefix began with \texttt{"head:codex/"}, \texttt{"head:copilot/"}, and \texttt{"head:cursor/"}, respectively. For the Devin agent, we selected repositories that contained at least one PR authored by \texttt{"devin-ai-integration[bot]"}. For Claude Code, we included repositories with at least one PR whose body explicitly contained the message \texttt{"Co-Authored-By: Claude"}.
Based on the release dates of the agents—Open-AI Codex \footnote{https://openai.com/index/introducing-codex/} (May 2025), GitHub Copilot\footnote{https://github.blog/news-insights/product-news/introducing-github-copilot-ai-pair-programmer/} (January 2025; no official release date availables), Claude Code\footnote{https://www.anthropic.com/news/claude-3-7-sonnet} (February 2025), Devin\footnote{https://cognition.ai/blog/introducing-devin} (November 2024), and Cursor\footnote{https://aiagentindex.mit.edu/cursor-agent/} (January 2025), we filtered PRs in each repository to include only those submitted after the corresponding agent’s release. We used 20 November 2025 as the cut-off date, which corresponds to the completion of our data collection and starting of our analysis. The result of this filtration is provided in Table \ref{tab:verified_python_repos_100stars_per_agent}. 
To analyze the longitudinal behavior of agents during their adoption within repositories, we conducted a quarterly analysis of repositories. 
To perform the longitudinal analysis, similar to the methodology adopted in a previous study~\cite{agarwal2026ai}, we define the agentic activity span of each repository as the time interval between its first and most recent agentic PR. This span is then partitioned into four equal quarters, enabling a standardized temporal comparison of agent activity across repositories with different lifespans. Repositories containing only a single agentic PR have an activity span of zero, making it impossible to divide the span into meaningful temporal quarters. Consequently, a minimum of two agent-generated PRs is required to obtain a non-zero activity span suitable for quarterly analysis. Repositories with only one agent-generated PR are therefore excluded not because a single PR is considered insufficient for analysis, but because such repositories are structurally incompatible with the quarterly decomposition that forms the basis of our longitudinal study.
This selection criterion enabled us to observe the evolution of agent usage from the first recorded agentic PR to the most recent one within each repository. The resulting set of selected repositories is presented in Table \ref{tab:repos_100stars_per_agent_>=2agenttic_PRs}.
It is worth noting that we classify a PR as a human-generated PR, when its author type is not labeled as \texttt{BOT},
nor includes any of the agentic PRs' evidences.

\begin{table}[]
\centering
\caption{Distribution of verified \textit{Python} repositories with at least 100 stars and at least 1 agentic PR per each AI agent (A-PR = PR generated by agent, AM-PR = PR generated by agent which is merged, AR-PR = PR generated by agent which is rejected)}
\label{tab:verified_python_repos_100stars_per_agent}
\scriptsize
\setlength{\tabcolsep}{5pt}
\begin{tabular}{llllll}
\toprule
Agent Name & 
\# Repos with $\ge$ 1 A-PR
& \# All PRs  & \# A-PRs & \# AM-PRs & \# AR-PRs\\
\midrule
Claude Code & 50 & 54,481 & 219 & 166 & 53\\
GitHub Copilot & 123 & 66,029 & 1,259 & 874 & 385\\
Cursor & 52 & 64,173 & 541 & 291 & 250\\
Devin & 50 & 43,011 & 1,355 & 512 & 843\\
OpenAI Codex & 274 & 90,340 & 6,057 & 5,451 & 606\\
\midrule
\textbf{Total \& Unique} & \textbf{489} & \textbf{220,612} & \textbf{9,428} & \textbf{7,293} & \textbf{2,135}\\
\bottomrule
\end{tabular}
\end{table}

\begin{table}[b]
\centering
\caption{Distribution of verified \textit{Python} repositories with at least 100 stars and at least 2 agentic-PR for each AI agent (A-PR = PR generated by agent, AM-PR = PR generated by agent which is merged, AR-PR = PR generated by agent which is rejected)}
\scriptsize
\setlength{\tabcolsep}{5pt}
\label{tab:repos_100stars_per_agent_>=2agenttic_PRs}
% \tiny
\begin{tabular}{llllll}
\toprule
Agent Name & 
\# Repos with $\ge$ 2 A-PRs & \# All PRs  & \# A-PRs  & \# AM-PRs & \# AR-PRs\\
\midrule
Claude Code & 22  & 28,268 & 191 & 151 & 40\\
GitHub Copilot & 82 & 56,397  & 1,218 & 856 & 362\\
Cursor & 32  & 55,600 & 521 & 284 & 237\\
Devin & 41  & 38,960 & 1,346 & 508 & 838\\
OpenAI Codex & 171  & 72,086 & 5,954 & 5,402 & 552\\
\midrule
\textbf{Total \& Unique} & \textbf{309} & \textbf{190,766} & \textbf{9,227} & \textbf{7,200} & \textbf{2,027}\\
\bottomrule
\end{tabular}
\end{table}

\subsection{Measures}

The following subsections describe the measures used in the analysis of each research question.

\subsubsection{RQ1: Evolution of Agentic PR Mergeability} 
To investigate how the effectiveness of coding agents evolves throughout the software development lifecycle, we analyze the acceptance of agentic PRs from an agentic and a temporal perspective. These measures capture differences in the effectiveness of individual coding agents as well as changes in the merge rate of their generated PR over time.
Specifically, we consider following measures:
\begin{itemize}
    \item Agentic PR merge probability, defined as the probability that a PR generated by a particular coding agent is accepted (i.e., merged). This measure quantifies the overall success rate of each agent in terms of PRs mergeability and enables comparisons across different coding agents.
    \item Quarter-specific agentic PR merge probability, defined as the probability that an agentic PR submitted during a given quarter is accepted (i.e., merged). 
    This measure allows us to examine whether the probability of an agentic PR being merged varies across quarters of the software development lifecycle.

\end{itemize}
Together, these measures provide complementary perspectives on the longitudinal adoption and effectiveness of coding agents. The agentic PRs merge probability characterizes differences in the performance of coding agents, while the quarterly merge probability of PRs reveals trends in the acceptance of agentic contributions throughout the evolution of software systems.

\subsubsection{RQ2: Development Tasks of Agentic PRs}

RQ2 seeks to understand the types of development activities commonly performed by AI coding agents. Manually categorizing thousands of PRs is impractical and susceptible to subjective bias. 

Following prior work that has successfully applied topic modeling to analyze PR contextual information and identify common development themes~\cite{popoola2025empirical}, we employ topic modeling to automatically identify the dominant development tasks represented in agentic PRs.
In general, topic modeling is a widely used statistical technique for uncovering latent thematic structures in large-scale textual datasets~\cite{vayansky2020review, alam2025empirical}.
For each identified development task category, we report the total number of merged and rejected agentic PR and its respective normalized merge rate to assess the success of coding agents across different development activities. Furthermore, for each software development quarter, we compute the frequency and relative proportion of each task category among agentic PRs to examine how the effectiveness of agentic PRs varies over time across different development activities. 
Together, these analyses enable us to identify the tasks most frequently addressed by coding agents and determine how the mergeability of PRs associated with these tasks changes as software systems evolve.

We conduct topic modeling on the complete set of  9,428 agentic PRs extracted within previous steps (please see Table \ref{tab:verified_python_repos_100stars_per_agent}). 
We first clean and prepare the data and then apply topic modeling to it.
Following similar literature studies\cite{alam2026analyzing,alshara2022pi}, we consider the combination of PR’s title and body as the representative information of each PR. 
However, this representation introduces several challenges. First, the length of the resulting documents varies significantly across PRs, leading to an inconsistent textual representation. Second, PR titles are sometimes noisy or insufficiently descriptive, and therefore may not accurately reflect the content of the corresponding PR body.
To address these issues, we generate a concise summary for each 
agentic PR and use them as the input corpus to of topic modeling. 
To this end, we employ the "\texttt{gpt-4.1-mini}" model which is successfully applied for the PRs' contextual information summarization
\cite{taraghi2026real,li2025rise}, to produce a standardized summary of the combined PRs' title and body. 
As a result, 
the achieved summaries provide a more consistent and noise-reduced textual representation.

To ensure the accuracy, usability, and overall quality of the collected data, we apply a standard text preprocessing pipeline, following prior studies~\cite{alam2026analyzing}. This process includes removing stop words, filtering out \texttt{HTTPS} links, eliminating special characters, and applying lemmatization to normalize word forms.
These preprocessing steps are intended to reduce noise in the textual data and improve the quality and coherence of the extracted topics.

Consistent with prior works ~\cite{alam2025empirical,alam2026analyzing} showing that BERTopic is a well-suited model for analyzing the short, context-specific textual content of GitHub issues and PRs, we adopt BERTopic to identify the latent development activities in agentic PRs.
\textit{BERTopic} provides a modular architecture that enables customization of different components in the topic modeling pipeline~\cite{grootendorst2022bertopic}. In this study, we configure the pipeline by explicitly specifying the embedding model, dimensionality reduction, clustering algorithm, vectorization model, and representation model.
The embedding model transforms unstructured text into numerical vectors that capture both semantic and syntactic information~\cite{wang2019evaluating}. As embeddings are a core component of \textit{BERTopic}, we evaluate three different models, including \texttt{all-mpnet-base-v2}, \texttt{all-MiniLM-L6-v2}, and \texttt{Gemini-embedding-001}. The first two models, provided by \textit{SentenceTransformers}, are widely used general-purpose encoders trained on large-scale datasets, making them effective for semantic representation tasks~\cite{alam2025empirical,alam2026analyzing}. We also include \texttt{Gemini-embedding-001}~\cite{lee2025gemini}, a recent state-of-the-art model that ranks highly in contemporary embedding benchmarks~\cite{taraghi2026real}.

\begin{figure}
    \centering
    \includegraphics[width=0.6\linewidth]{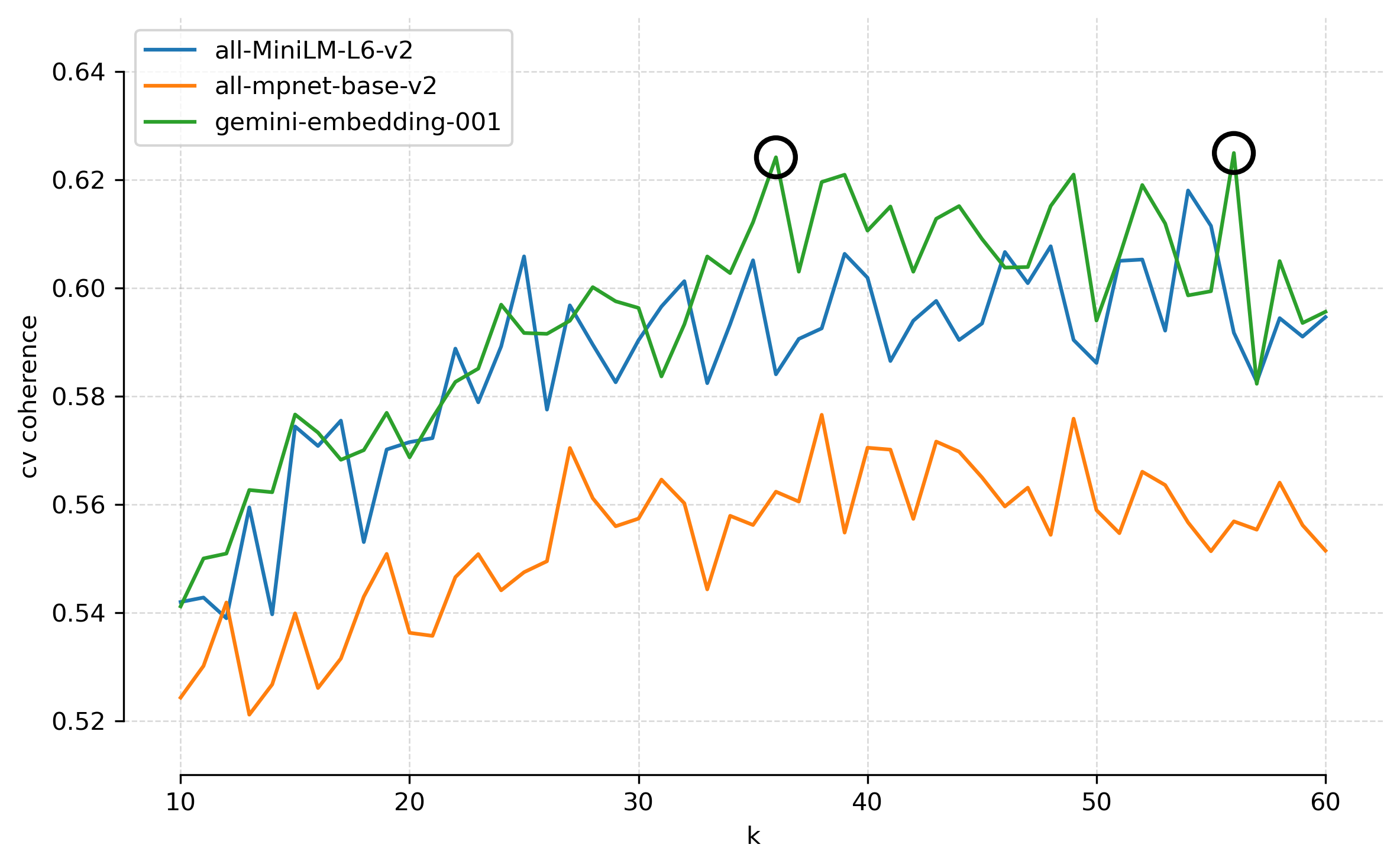}
    \caption{Comparison of $C_V$ Coherence scores across varying number of topics ($k$) for different Embedding Models.}
    \label{fig:k_vs_cv_coherence_embedding_models}
\end{figure}

To address the high dimensionality of document embeddings, we apply Uniform Manifold Approximation and Projection (UMAP) for dimensionality reduction~\cite{healy2024uniform}. For clustering, we adopt \textit{K-means}, as prior work has shown that combining \textit{BERTopic} with \textit{K-means} can improve topic coherence and produce more diverse and interpretable topics~\cite{kumi2024uncovering}. To determine the optimal number of clusters (k), we follow established practices~\cite{weng2022identification,kumi2024uncovering} by evaluating multiple values in the range of 10 to 60 and selecting the configuration that yields the highest topic coherence. Fig.\ref{fig:k_vs_cv_coherence_embedding_models} represents the $C_V$ coherence value for various \textit{k}, using three different embedding models. 
In general, higher topic coherence values indicate more semantically consistent and interpretable topics~\cite{rahimi2024contextualized}. 
Among all evaluated configurations, the highest $C_V$ coherence score ($0.624$) is achieved when $k=36$ and $k=56$ are used with the \texttt{Gemini-embedding-001} embedding model. Given that \textit{UMass} is considered another important metric to measure the quality of topic modeling~\cite{bellaouar2021topic,alagha2021topic}, we use it to decide between 36 and 56 values for \textit{k}. Generally speaking, a smaller amount of \textit{UMass} indicates a more coherent and interpretable set of topics~\cite{mimno2011optimizing}.
Since $k=36$ brings a \textit{UMass} coherence score of 
$-3.018$, outperforming $k=56$ with $-3.378$ \textit{UMass} score, we select $k=36$ for our topic modeling configuration, 

After clustering, we use a vectorization model to generate interpretable topic representations~\cite{zagatti2025investigating}. Specifically, we employ CountVectorizer~\cite{danyal2024sentiment} to compute term frequencies. To further enhance interpretability, we adopt a chained representation model~\cite{grootendorst2022bertopic} that combines part-of-speech (POS) filtering and Maximal Marginal Relevance (MMR). The POS component, implemented using the \texttt{en\_core\_web\_sm} model from \textit{SpaCy} (v3.8.11), extracts linguistically meaningful keywords, while MMR (diversity = 0.5) balances relevance and diversity, reducing redundancy among selected terms.
The detailed configuration of all components is provided in the replication package accompanying this paper~\cite{replication_package}.

\subsubsection{RQ3: Code Quality and Defect Proneness}
To examine how agentic contributions differ from human-generated contributions, we compare merged agentic and merged human-generated PRs
in terms of their characteristics and defect proneness over time. The analysis consists of two complementary parts.
In the first part, we compare several PR characteristics that serve as proxies for development complexity. Specifically, we consider: number of commits, number of contributors, number of changed files, number of comments, number of changed lines of code (LOC), and time-to-merge.
In the second part, we evaluate the long-term quality impact of contributions by comparing the bug-inducing commit of agentic and human-generated PRs. A PR is considered bug-inducing if its changes are later identified as introducing defects through our bug-inducing commit identification process. By comparing the proportion of bug-inducing PRs between the two groups, we assess whether agentic contributions differ from human-generated contributions in their propensity to introduce defects into the software system.
Together, these measures enable us to compare agentic and human-generated contributions with respect to PR complexity, review effort, collaboration, and defect proneness, providing insight into the long-term software quality implications of adopting coding agents in collaborative software development.

% ==========================================
% --------  Result Section -------------
% ==========================================
\section{Result}
\label{sec:result}

In this section, we present the results of our analysis and the key findings of this study. 
All data and materials used in this study are available in the paper’s replication package~\cite{replication_package}.

\subsection{\textbf{RQ1.} How does the mergeability of agentic PRs evolve across different stages of the software development lifecycle?}
\label{subsec:rq1}

To examine the effect of coding agents on the merge rate of PRs generated by agents throughout the software development lifecycle, we analyze the PRs extracted in the previous step (Subsection~\ref{subsec:data_collection}). To characterize agents behavior over time, we divide the agent-driven development period of each repository into four equal temporal intervals, each corresponding to one quarter of the repository’s development timeline. 
Although a repository may employ multiple coding agents concurrently, PRs are analyzed independently for each agent by including only the PRs generated by that agent. The binary response variable \textit{merged} is defined for each PR, taking the value 1 if the PR was merged and 0 when the PR was rejected.
As this study focuses exclusively on agentic PRs, we define the agentic activity span as the time interval from the first recorded PR to the most recent PR submitted by studied AI coding agents in this paper
for each repository.

To analyze the overall merge probability of PRs generated by all five agents, we employ a Generalized Linear Mixed Model (GLMM).
\textit{GLMM}s are a powerful class of statistical models that extend Generalized Linear Models (\textit{GLM}s) by incorporating mixed effects, thereby enabling the analysis of non-normally distributed response variables while accounting for correlation and heterogeneity in the data~\cite{bolker2015linear}. In particular, \textit{GLMM}s support the inclusion of both fixed and random effects, making them well suited for modeling structured and repeated observations~\cite{stroup2024generalized}. 
In this study, we fitted a binomial GLMM with a logit link function, where the unit of analysis was an individual PR. The response variable, \textit{merged}, was binary, taking the value 1 for merged PRs and 0 for rejected PRs.
The model included AI agent (\textit{agent}) and development quarter (\textit{period}) as fixed effects to evaluate their association with the probability of a PR being merged.
Repository (\textit{repo}) was included as a random intercept to account for the correlation among PRs originating from the same repository and to capture repository-specific variability in merge probabilities. Equation~\ref{equ:GLMM} presents the formal specification of the proposed \textit{GLMM}, while equation ~\ref{equ:GLMM_mathematical} provides its corresponding mathematical formulation. 
\begin{equation}
\label{equ:GLMM}
merged \sim agent + period + (1 | repo)
\end{equation}
\begin{equation}
\label{equ:GLMM_mathematical}
\text{logit}[P(merged_{ij} = 1)] = \beta_0 +\beta_1 \textit{agent}_{ij} + \beta_2 \textit{period}_{ij} + u_j
\end{equation}

where $P(merged_{ij} = 1)$ denotes the probability that PR $i$ from repository $j$ is merged, $\beta_0$ is the intercept, $\beta_1$ and $\beta_2$ represent the fixed effects of AI agent and development quarter, respectively, and  $u_j \sim \mathcal{N}(0 , \sigma_{u}^2)$ is the repository-specific random intercept. 

Estimated marginal means (EMMs) ~\cite{searle1980population}, are used to summarize the results of linear models ~\cite{lenth2023emmeans}.
In this study, EMMs were computed from the fitted binomial GLMM using the \texttt{emmeans} package in \texttt{R} ~\cite{lenth2023emmeans}. For each AI agent (or quarter), the predicted probability of a PR being merged was obtained by averaging the model's predictions over all levels of the other fixed effect, holding the repository random effect at its mean. Predictions were back-transformed from the log-odds scale to the probability scale, so the reported EMMs represent model-adjusted merge probabilities rather than raw observed proportions.

Table~\ref{tab:glmm_predict_agent} presents the EMMs of the probability of agentic PRs being merged for each AI agent.
Among the studied agents, Claude (84.3\%) and Codex (73.5\%) achieve the highest PRs merge probabilities.
It is important to note that Codex is used in a larger number of repositories (274) compared to Claude (50). However, repositories adopting Claude tend to be larger and more complex, as reflected by their higher average size, number of commits, and number of contributors. Despite this, Claude achieves a higher merge rate, suggesting stronger performance in terms of PR acceptance even in more complex development contexts.

\begin{table}[]
    \centering
    
    \caption{Estimated marginal mean (EMM) merge probabilities for each AI agent from the GLMM }
    \scriptsize
    \begin{tabular}{p{2cm}l}
        \toprule
        \textbf{Agent} & \textbf{Estimated probability}\\
        \midrule
         Claude & 84.3\% \\
         Codex & 73.5\% \\
         Cursor & 63.9\% \\
         Copilot & 59.6\% \\
         Devin & 43.0\% \\
        
        \bottomrule
    \end{tabular}
    \label{tab:glmm_predict_agent}
\end{table}

In contrast, Devin exhibits the lowest merge probability, with only 43\% of its generated PRs being accepted.
This lower performance is consistent with its usage patterns: Devin is predominantly applied in less popular repositories, as indicated by a lower average number of stars. Such repositories may have lower activity levels or different contribution standards~\cite{borges2016understanding,borges2018s}, which can influence PR acceptance rates and potentially contribute to Devin’s observed performance.
Regarding Cursor and Copilot, their performance reflects the relative maturity and stability of earlier-generation coding agents. Copilot is used across the largest number of repositories for PR generation, suggesting broad adoption and sustained reliability. Additionally, repositories employing Cursor have a higher average number of stars compared to those using other agents, indicating that Cursor is more frequently utilized in highly popular and active repositories. This observation further supports the interpretation that Cursor’s performance is associated with its maturity and acceptance within established open-source projects.

\begin{tcolorbox}[colback=blue!10!white, colframe=blue!75!black]

\textbf{Finding 1: }

Although Claude achieves the highest PR acceptance rate among the evaluated AI coding agents, this superior performance is observed despite being primarily employed in repositories that are generally larger and more complex than those using other agents.
In contrast, Codex and Copilot are used across a larger number of repositories, including more popular projects, and therefore provide a more representative baseline for evaluating the expected performance of AI coding agents in PR generation.

\end{tcolorbox}

Table~\ref{tab:glmm_predict_quarter} presents the EMMs of the probability of agentic PRs being merged for each development quarter.
The results indicate that the PRs' merge probability decreases slightly from 65.4\% in the first quarter to 63.7\% in the second quarter, increases to its highest value of 68.2\% in the third quarter, and remains similar in the fourth quarter (67.7\%). These results suggest a slight improvement in the probability of PRs being merged during the later stages of development.

\begin{table}[]
    \centering
    \caption{Estimated marginal mean (EMM) merge probabilities for each quarter from the GLMM}
    \scriptsize
    \begin{tabular}{p{2.5cm}l}
        \toprule
        \textbf{Period (Quarter)} & \textbf{Estimated probability}\\
        \midrule
         1st Q & 65.4\% \\
         2nd Q & 63.7\% \\
         3rd Q & 68.2\%  \\
         4th Q & 67.7\%  \\
        \bottomrule
    \end{tabular}
    \label{tab:glmm_predict_quarter}
\end{table}

Specifically, the third quarter exhibits the highest PR acceptance rate 
(68.2\%)
among all development quarters. This suggests that after progressing beyond the second quarter, coding agents have acquired a better understanding of the software system’s structure, which in turn leads to improved PR quality and higher acceptance rates. In addition, this effect may partly stem from increased familiarity of human developers with the style, context, and conventions of PRs generated by coding agents.
Conversely, the second quarter exhibits the lowest estimated probability of PRs being merged (63.7\%). One possible explanation is that repositories often undergo periods of active feature development and codebase evolution during the early stages of a project, which may increase the complexity of code integration and lead to more rigorous review processes. This aligns with the finding that project maturity is a significant determinant of PR review speed, with less mature projects experiencing considerably longer and more demanding review processes ~\cite{zhang2022pull}.
As a result, agentic PRs submitted during this period may require additional revisions before being accepted. Another possible explanation is that developers may still be adapting to the workflow and coding style of agentic contributions, resulting in a slightly lower merge probability.
Besides, the fourth quarter exhibits a slightly lower estimated merge probability than the third quarter, while remaining higher than those observed in the first and second quarters. One possible interpretation is that this period reflects a stabilization phase in the development process, where both the system architecture and development practices have matured ~\cite{zhang2022pull}.
Overall, although the estimated probability of PRs being merged is the highest in the third quarter, the differences across development quarters are relatively small. The slight decrease observed in the fourth quarter suggests only minor temporal variation in merge probability rather than a sustained change in the effectiveness of AI coding agents over time.

\begin{tcolorbox}[colback=blue!10!white, colframe=blue!75!black]

\textbf{Finding 2: }
The estimated marginal means (EMMs) indicate modest variation in the probability of PRs being merged across development quarters. Although the third quarter exhibits the highest merge probability (68.2\%) and the second quarter the lowest (63.7\%), none of the pairwise comparisons is statistically significant. Therefore, while these numerical differences may reflect natural fluctuations in software development, they do not provide evidence of a systematic temporal effect on the merge probability of agentic PRs.
\end{tcolorbox}

\begin{figure}[]
  \centering
    \includegraphics[width=0.55\linewidth]{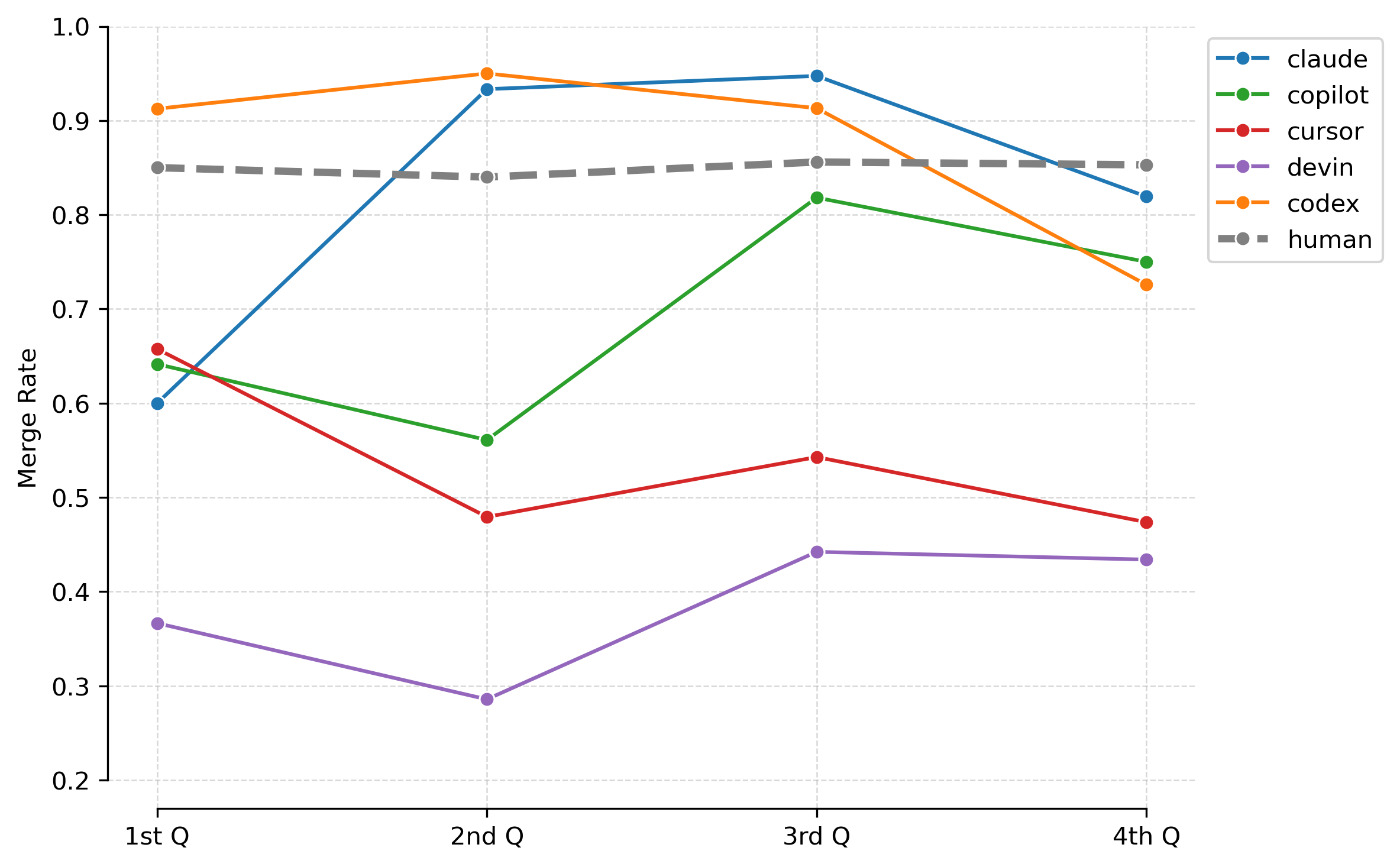}
  
  \caption{Agentic-PRs' Merge rate distributions across agents}
  \label{fig:merge_rate}
  \Description{Distribution of PRs' merge rate for all agents}
\end{figure}

Fig.~\ref{fig:merge_rate} presents the merge rates of PRs generated by different coding agents across four development quarters of the analyzed repositories. Overall, the results show that merge rates fluctuate over time, rather than following a strictly increasing or decreasing trend. This suggests that the effectiveness and adoption of coding agents evolve in a non-linear manner throughout the project lifecycle.
Notably, for three of the five studied agents (Copilot, Cursor, and Devin), the merge rate in the second quarter is lower than in the first quarter.
A similar trend is observed for human-generated PRs, although the fluctuations are less pronounced than those of agentic PRs. Specifically, the merge rate of human-generated PRs decreases slightly from 0.850 in the first quarter to 0.840 in the second, before increasing to 0.856 in the third quarter and remaining relatively stable at 0.853 in the fourth. This pattern suggests that the decline in merge rate during the second quarter is not unique to agentic PRs, but may instead reflect broader temporal or repository-level factors that influence PR acceptance throughout the software development lifecycle.
This decline may be attributed to increasing project maturity, which typically leads to stricter acceptance criteria 
and makes code changes more difficult to integrate~\cite{bird2011don,zhou2014will}. Prior studies on open-source software evolution report that code review processes become more rigorous and selective as projects mature~\cite{bacchelli2013expectations}. In addition, empirical evidence suggests that the effectiveness of AI-based coding assistants may decrease as software systems grow in size and complexity~\cite{xia2023automated,chen2021evaluating}.

Furthermore, the merge rates of PRs generated by Claude, Copilot, and Devin are higher in the fourth quarter than in the first quarter. This pattern suggests a progressive adaptation between human developers and coding agents over time. 
For example, in PR \href{https://github.com/567-labs/instructor/pull/1232}{\#1232} of the \textit{instructor} repository, generated by Devin during the first quarter of the project's development, the reviewer requested multiple revisions after evaluating the agent's proposed changes, resulting in several rounds of discussion before the PR could progress. 
After the agent modified an undesired file, the maintainer instructed,“\textit{Don't edit this file at all}”. When the subsequent revision did not fully revert those changes, the maintainer clarified, “\textit{you still edited the file, just revert the whole thing}”. The PR was merged only after the requested revisions were completed. 
In contrast, PR \href{https://github.com/567-labs/instructor/pull/1783}{\#1783}, generated by the same agent during the fourth quarter of the same repository, was approved and merged by the same reviewer with minimal discussion after successfully passing 13 of the 15 automated checks.
As development teams accumulate experience with coding agents, they become more proficient at crafting effective prompts, evaluating agent-generated code, and integrating it into established software development workflows. 
Prior research on AI-assisted programming has shown that developers’ acceptance rates and effective integration of AI-generated suggestions tend to increase as they become more familiar with the assistant’s capabilities and limitations~\cite{kumar2025intuition}. 
Moreover, by the fourth quarter, software projects often reach a more stable architectural state, which can improve coding agents’ contextual understanding of the system and reduce ambiguity in code generation.
This observation is consistent with previous empirical studies on AI coding assistants, such as GitHub Copilot, which report that developers tend to accept a higher proportion of AI-generated code as the development process progresses~\cite{dohmke2023sea}.
In contrast, Codex and Cursor achieve their lowest merge rates in the fourth quarter. This pattern may suggest comparatively weaker adaptation to evolving project contexts or limited alignment with developer expectations over time. 
An illustrative example is provided by Codex generated PR \href{https://github.com/Kiln-AI/Kiln/pull/648}{\#648} of \textit{Kiln} repository, submitted during the fourth quarter which introduced a broad change and took approximately two weeks to merge. It underwent several revisions, including a reviewer’s request to “\textit{update the return types on the concrete class implementations too}”. Another fourth quarter Codex generated contribution from the same repository, PR \href{https://github.com/Kiln-AI/Kiln/pull/693}{\#693}, remained unresolved for approximately five weeks before being closed without merging. In a follow-up commit, the developer noted that “\textit{Codex did a decent job, but I had to fix a lot, especially on UI side}”. The developer ultimately closed the PR because the implementation remained incomplete and still required token-refresh support.
By comparison, the first quarter Codex generated PR \href{https://github.com/Kiln-AI/Kiln/pull/387}{\#387} from the same repository, addressed a narrower test-configuration task and was merged within approximately one day, without recorded human-review discussion, after passing all 14 automated checks.

\begin{table}[]
    \centering
    \scriptsize
    \caption{Pairwise comparison between agents and quarters}
    \begin{tabular}{p{3cm} r r r r}
        \hline
          & \textbf{odds.ratio} & \textbf{SE} & \textit{\textbf{z.ratio}} & \textit{\textbf{p.value}}\\
        \hline
        Claude / Codex & 1.929 & 0.575 & 2.203 & 0.178 \\
        Claude / Copilot & 3.631 & 1.220 & 3.836 & \textbf{0.001} \\
        Claude / Cursor & 3.023 & 1.010 & 3.301 & \textbf{0.009} \\
        Claude / Devin & 7.098 & 2.380 & 5.846 & \textbf{<1e-3} \\
        Codex / Copilot & 1.883 & 0.379 & 3.142 & \textbf{0.015} \\
        Codex / Cursor & 1.568 & 0.314 & 2.242 & 0.164 \\
        Codex / Devin &  3.680 & 0.770 & 6.230 & \textbf{<1e-3} \\
        Copilot / Cursor & 0.833 & 0.210 & -0.726 & 0.951 \\
        Copilot / Devin & 1.955 & 0.507 & 2.586 & 0.073 \\
        Cursor / Devin & 2.348 & 0.584 & 3.432 & \textbf{0.005} \\
        \hline
        1st / 2nd Q & 1.075 & 0.110 & 0.704 & 0.896 \\
        2nd / 3rd Q &  0.819 & 0.096 & -1.698 & 0.325 \\
        3rd / 4th Q & 1.026 & 0.113 & 0.229 & 0.996 \\
        \hline
    \end{tabular}
    \label{tab:glmm_fix}
\end{table}

To compare AI agents and development quarters, we conducted pairwise comparisons of EMM using Tukey-adjusted p-values to account for multiple comparisons.
Table~\ref{tab:glmm_fix} presents the pairwise comparisons of the EMMs obtained from the fitted binomial GLMM. The upper section reports pairwise comparisons among AI coding agents, while the lower section reports pairwise comparisons among development quarters.

The pairwise comparisons indicate that PRs generated by Claude exhibit a significantly higher probability of being merged compared to those generated by Copilot, Cursor, and Devin (\textit{p}-value < 0.05), suggesting superior performance of Claude in terms of PRs merge rate. 
Although the odds ratio comparing Claude with Codex also indicates a higher merge likelihood for Claude, this difference is not statistically significant (\textit{p}-value > 0.05).
These observations are consistent with the predicted marginal effects reported in Table~\ref{tab:glmm_predict_agent}, which summarizes agent-level merge rates estimated by the GLMM.
Furthermore, the odds ratios comparing Cursor with Copilot and Devin suggest that Cursor achieves a higher PR merge probability than both Copilot and Devin, although the difference between Cursor and Copilot does not reach statistical significance (\textit{p}-value = 0.951). 
This pattern also aligns with Table~\ref{tab:glmm_predict_agent}, which identifies Cursor as the third successful agent in terms of PRs acceptance ratio, with an estimated merge probability of $63.9\%$.
Finally, Devin exhibits the lowest probability of PRs being merged among all agents, with statistically significant differences in all pairwise comparisons except with Copilot (\textit{p}-value = 0.073). This finding is corroborated by Table~\ref{tab:glmm_predict_agent}, where Devin shows the lowest estimated PR merge probability ($43.0\%$) among all studied agents.

With respect to the development quarters, none of the pairwise comparisons is statistically significant. 
Although the EMMs indicate modest variation in the probability of PRs being merged across development quarters, with the third quarter exhibiting the highest estimated merge probability and the second quarter experiencing the lowest, these differences are not statistically significant. Therefore, the fitted GLMM provides no evidence that the development quarter is associated with the probability of a PR being merged after accounting for AI agent and repository-level variability.

\begin{figure}[]
    \centering
    \includegraphics[width=0.55\linewidth]{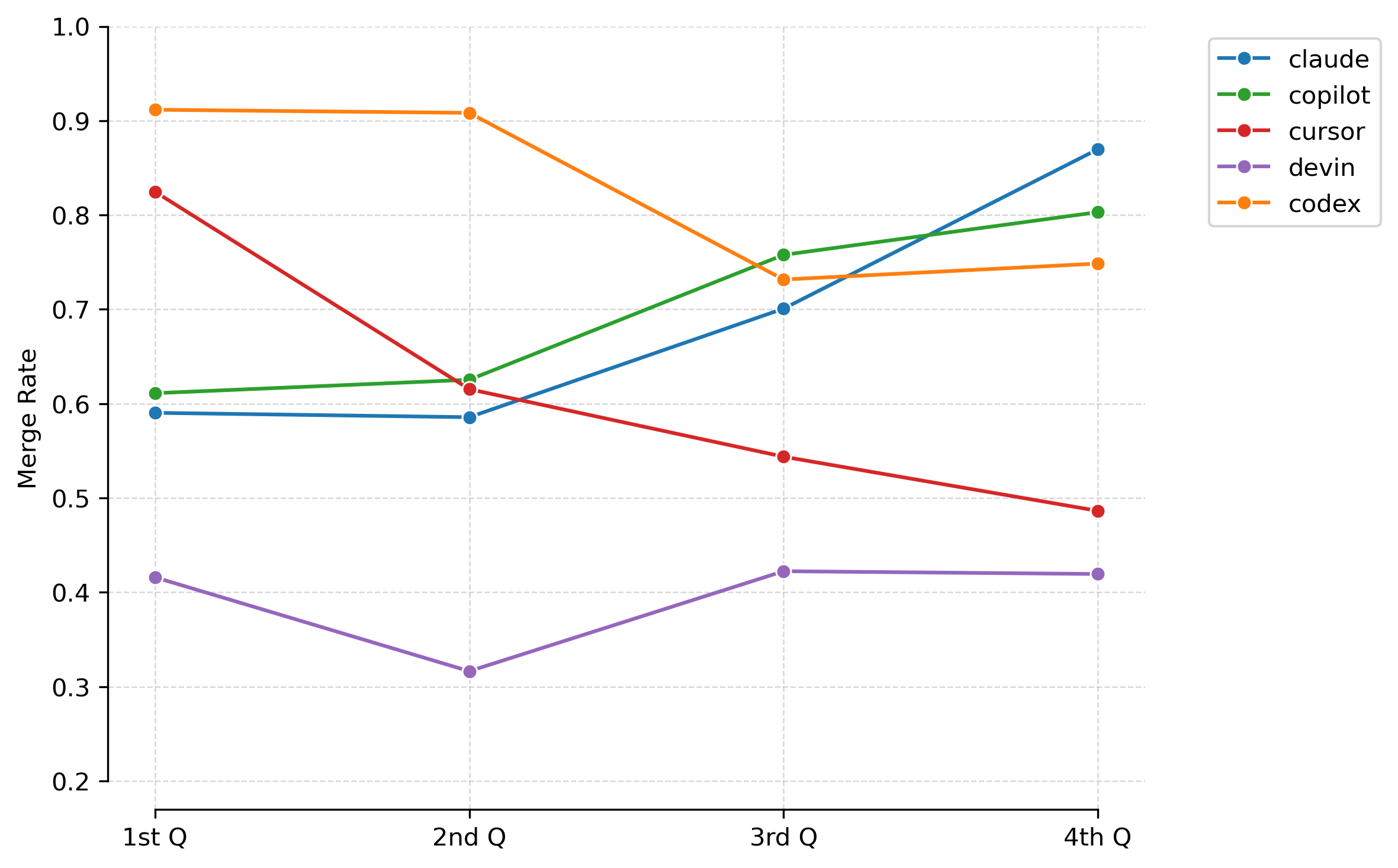}
    \caption{PRs merge rate over the whole period that coding agents have been used}
    \label{fig:agent_maturity}
\end{figure}

Given that the maturity of coding agents may influence the merge rate of PRs they generate, we analyze the temporal evolution of PR merge rates for each agent. Fig.~\ref{fig:agent_maturity} presents the proportion of merged PRs over the entire period during which each agent was used across all studied repositories. To isolate the effect of agent maturity, this analysis aggregates data across repositories and does not distinguish between individual projects.
The results indicate an increasing trend in merge rates over time for Claude and Copilot, suggesting potential improvements in agent performance. To assess the statistical significance of differences between consecutive quarters, we apply a two-proportions \textit{z}-test~\cite{tintle2018introduction}. The results show statistically significant differences only between the third and fourth quarters for Claude (\textit{p} = 0.0042) and between the second and third quarters for Copilot (\textit{p} = 0.0002).

In contrast, the merge rates for Cursor and Codex exhibit a decreasing trend over time, although statistical significance is observed only for Codex between the second and third quarters (\textit{p} = 0.00). For Devin, no consistent trend is observed. Overall, these findings do not provide strong evidence of a systematic effect of agent maturity on PR merge rates, suggesting that any such influence is limited and does not substantially affect our broader analysis.

\subsection{\textbf{RQ2.} In which types of tasks are coding agents most frequently used?}
\label{subsec:rq2}

Overall, PRs can be categorized along various dimensions, including the types of tasks they aim to perform. Prior studies have proposed various taxonomies of PRs from different perspectives~\cite{yu2018nbsl,fridh2022classification}.
Focusing on PRs generated by coding agents, Li et al.~\cite{li2025rise} introduced a categorization based on the 11 task types defined in the \textit{Conventional Commits Specification}~\cite{conventional_commits}.
To provide this taxonomy, they employ \texttt{Gemini-4.1-Mini} to automatically assign each PR to one of the predefined categories.

However, since the \textit{Conventional Commits Specification} was originally designed for human-authored commits, relying solely on this taxonomy may overlook characteristics specific to agentic PRs. To address this limitation, we adopt an unsupervised approach to characterize the tasks performed by AI agents.
Specifically, we apply topic modeling to the corpus of agentic PRs to uncover latent task categories directly from the data. This approach allows us to move beyond predefined taxonomies and identify patterns that may not be explicitly captured by existing classification schemes. As a result, our method provides a more exploratory and data-driven view of how AI agents are used in software development, revealing potentially novel or emerging task types within PRs generated by agents.

\begin{table}[]
    \centering
    \scriptsize
    \caption{Final set of clusters, topics abbreviation, and a brief description of each cluster. }
    \begin{tabular}{p{3cm} l p{10.5cm}}
        \hline
        \textbf{Label} & \textbf{Abbreviation} & \textbf{Description}\\
        \hline
         Documentation & DOC & Improving, updating, or restructuring project documentation, comments, and usage guides.\\
         LLM Integration & LLM & Changes introducing or enhancing integrations with LLMs, prompts, or AI-based functionalities.\\
         Test & TEST & Modifications focused on adding, updating, or fixing unit, integration, and end-to-end tests.\\
         JSON Template  & JSON & Updates involving JSON schemas, configuration templates, serialization, or structured data formatting.\\
         Logging & LOG & Changes related to logging mechanisms, debugging output, monitoring, or observability improvements.\\
         GitHub Actions Workflows & GIT & Modifying GitHub Actions workflows, automation pipelines, or repository-level CI scripts.\\
         Dependency & DEPs & Updates for managing, upgrading, or removing software dependencies and external libraries.\\
         Asynchronous & ASYNC & Changes introducing or refining asynchronous execution, concurrency handling, or event-driven operations.\\
         UI & UI & Focused on user interface components, layouts, styling, or user experience enhancements.\\
         Refactoring & REFACTOR & Code restructuring efforts aimed at improving maintainability, readability, or modularity without altering functionality.\\
         CI \& Build & CI & Modifications related to continuous integration pipelines, build systems, packaging, or deployment processes.\\
         Type Checking & TYPECHK & Changes involving static type analysis, type annotations, or type-safety enforcement mechanisms.\\
         Asset Management  & ASSET &  Updates managing static assets such as images, media files, resources, or frontend bundles.\\
         Hook Management & HOOK &  Introducing or modifying hooks, callbacks, middleware, or event interception mechanisms.\\
         Simulation & SIM & Changes implementing or improving simulation environments, mock systems, or synthetic execution workflows.\\
         Model Evaluation \& Datasets  & MODEL-EVAL & Updates concerning benchmarking, validation, or evaluation of machine learning models.\\
         Function Implementation & FUNC & Adding new functions, core features, or business logic implementations.\\
         Typos & TYPO & Minor corrections addressing typographical, grammatical, or formatting issues in code or documentation.\\
         AST Parsing Refactor & AST & Refactoring changes related to abstract syntax tree parsing, transformation, or compiler-related logic.\\
         Model Download \& Mirroring & MODEL-DL & Changes enabling model downloading, caching, synchronization, or repository mirroring functionalities.\\
         Cloud & CLOUD & Involving cloud infrastructure, remote services, storage, or distributed deployment configurations.\\
         Multilingual Translation & I18N & Changes supporting internationalization, localization, or multilingual translation capabilities.\\
         License Compliance & LICENSE &Updates ensuring software license compatibility, attribution, or compliance with legal requirements.\\
         Token management  & TOKEN &  Changes related to authentication tokens, API keys, credential handling, or token lifecycle management. 
         \\         
        \hline 
    \end{tabular}
    \label{tab:final_topics_label}
\end{table}

To investigate the types of tasks for which AI agents are utilized, we perform topic modeling on a cleaned dataset of $9,428$ agentic PRs (Table~\ref{tab:verified_python_repos_100stars_per_agent}). Given that \textit{k = 36} gives the best $C_V$ and \textit{UMass} scores, we set k as 36 in our topic modeling pipeline. After executing topic modeling on the whole dataset, the first two authors (one PhD candidate and one associate researcher with a PhD degree) review the topic representations collaboratively in a meeting. To this end, they examine the top 10 keywords related to each topic which are generated by \textit{BERTopic}. Moreover, they combine the topics with similar themes to make the final set of topics more concise and accurate.

To validate the accuracy and completeness of the generated clusters, we construct a validation dataset by randomly sampling 100 PRs. To ensure a balanced representation of different clusters, we employ a stratified sampling approach~\cite{meng2013scalable}, which preserves the proportional distribution of topics across the sampled PRs.
Next, the first two authors independently examine the sampled PRs along with their assigned topic labels to determine whether each label accurately represents the primary subject of the corresponding PR. In cases of disagreement, the two authors discuss their interpretations to reach a consensus. When disagreements cannot be resolved, the third author is consulted to make the final decision. To assess annotation reliability, we also compute the inter-rater agreement between the two primary annotators using  Cohen's kappa~\cite{hsu2003interrater} (using \texttt{scikit-learn} library ~\cite{pedregosa2011scikit}), giving 84.2\% which is interpreted as strong agreement~\cite{mchugh2012interrater}.

The validation process shows that 87\% of the sampled PRs are correctly associated with their corresponding topics, which is comparable to the results reported in similar prior studies~\cite{taraghi2026real}. During the validation process, the annotators also identified semantic overlap among several topics. To improve the conciseness and interpretability of the extracted themes, we merged 15 overlapping topics into 3 broader topics sharing similar underlying semantics. As a result, the final dataset consists of 24 unique topics after the validation and consolidation steps. The complete list of final topic labels is presented in Table~\ref{tab:final_topics_label}.

\begin{figure}
    \centering
    \includegraphics[width=\linewidth]{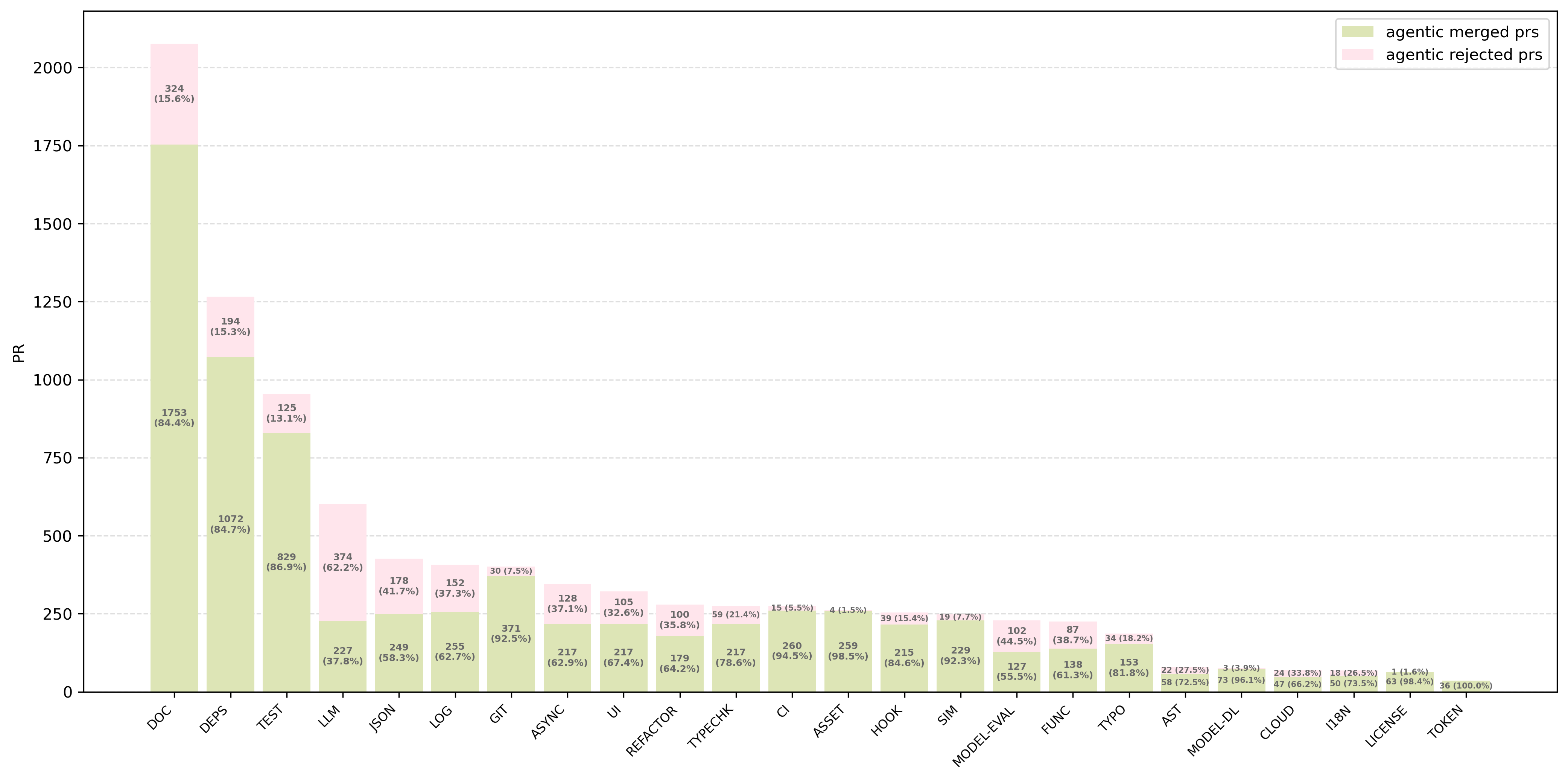}
    \caption{Merged vs rejected agentic PRs per topic}
    \label{fig:topics_to_number_of_merged_rejected_prs}
\end{figure}

Fig.~\ref{fig:topics_to_number_of_merged_rejected_prs} presents the distribution of merged and rejected agentic PRs across different development tasks. As shown in the figure, documentation, dependency management, and testing are the most frequent tasks performed by agent-generated PRs.
Among these categories, documentation-related PRs represent the largest proportion of contributions. Although prior studies on human-generated PRs have not identified documentation tasks as the most common type of contribution, these studies primarily focus on human developers rather than AI agents. One possible explanation is that documentation-related tasks are generally more straightforward, less complex, and less critical than many implementation or architectural changes~\cite{morovati2024common,taraghi2026real}. Consequently, developers may prefer to delegate such low-complexity tasks to coding agents, where the risk of introducing critical defects is relatively limited.

Dependency-related changes also appear frequently among agentic PRs. This observation is consistent with prior research showing that dependency management is one of the most common and recurring maintenance activities in modern software systems~\cite{he2023automating,decan2016github}. Many dependency updates involve repetitive and standardized modifications, such as version upgrades or configuration adjustments, making them suitable targets for automation through coding agents.
Furthermore, the high prevalence of testing-related PRs may be attributed to the structured and automatable nature of testing tasks. Automated support for testing activities has been extensively studied in AI-assisted software engineering~\cite{xue2024llm4fin}. In particular, the pattern-based structure of test code aligns well with the capabilities of modern LLM-based code generation systems, making testing activities especially amenable to automation~\cite{alshahwan2024automated,siddiq2024using}. In addition, testing-related modifications can often be validated automatically through CI pipelines and regression testing frameworks, which may increase developers’ willingness to rely on coding agents for such tasks.

In contrast, PRs related to token management and license represent the least frequent category of development tasks among agent-generated contributions. One possible explanation for this phenomenon is that token management is directly associated with operational cost and resource allocation in LLM-based software systems~\cite{artola2025your}. Modifications involving token budgeting, usage limits, or API consumption policies may have immediate financial and performance implications, particularly in environments where token usage directly affects inference cost~\cite{jiang2024d}. 
Consequently, developers may prefer to maintain tighter human control over such changes rather than delegating them to coding agents.
Furthermore, token management logic is often tightly coupled with system-level optimization policies and application-specific constraints, requiring contextual reasoning that may extend beyond pattern-based code generation~\cite{lavrentiev2026token}. Incorrect modifications in this type of tasks may lead to increased operational costs, degraded system performance, or unstable service behavior. As a result, developers may adopt a more conservative approach toward using AI agents for token-management-related tasks.
Moreover, the low frequency of agentic PRs related to repository licensing suggests that developers may be reluctant to delegate governance-sensitive tasks to AI agents. Although licensing-related modifications are typically limited in implementation scope, they can carry substantial legal, organizational, and economic implications~\cite{wolter2023open,vendome2017license}. The potentially irreversible and project-wide consequences of such changes may reduce developers’ willingness to trust AI agents for the management of license-related contributions.

\begin{tcolorbox}[colback=blue!10!white, colframe=blue!75!black]

\textbf{Finding 3: }
Developers primarily employ coding agents for repetitive, low-complexity, well-structured tasks, particularly
\textit{`documentation'}, \textit{`dependency management'}, and \textit{`testing'}, rather than governance- and cost-sensitive tasks, such as \textit{`license'} and \textit{`token management'}.

\end{tcolorbox}

\begin{figure}[b]
    \centering
    \includegraphics[width=0.95\linewidth]{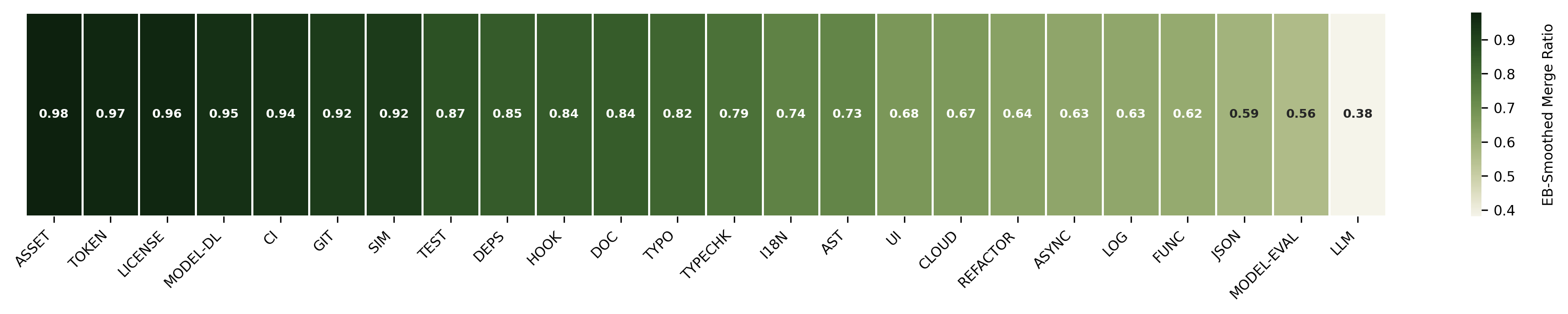}

    \caption{Agentic PRs' merge ratio per topic}
    \label{fig:EB_merge_ratio_per_topics}
\end{figure}

Fig. \ref{fig:EB_merge_ratio_per_topics} shows the merge ratio of various types of development tasks. Concerning the number of PRs in each group of development task is different from others, we first normalize the merge ratio of PRs categorized in different groups using Parametric Empirical Bayes. 
As it is obvious, a clear distinction between tasks that are comparatively easier to integrate and those that require more contextual reasoning and validation. 
Specifically, agentic PRs categorized as 
\textit{`asset management'}, \textit{`token management'}, and \textit{`license management'} show the highest merge ratios among all agentic PRs categories. One potential reason behind this phenomenon can be the fact that these tasks are generally operational in nature, often involving localized modifications with limited interaction with core application logic. 
For example, changes related to asset management and model download are frequently additive rather than transformative. 
As a result, these changes are less likely to introduce architectural inconsistencies or complex integration conflicts, increasing developers’ willingness to merge such PRs.
In contrast, categories such as \textit{`LLM integration'}, \textit{`model evaluation'}, \textit{`JSON template'}, and \textit{`function implementation'} exhibit substantially lower normalized merge ratios. These development activities often require deeper contextual understanding of system behavior, project-specific requirements, and implicit dependencies.
As an example, LLM-related tasks involve rapidly evolving frameworks, non-deterministic model behaviors, prompt sensitivity, and evaluation uncertainty, all of which complicate validation and integration~\cite{chen2025empirical}. Similarly, \textit{`JSON template'} modifications may affect schema consistency, workflow orchestration, or downstream components that rely on strict configuration compatibility, meaning that even small changes can propagate unintended system-level effects~\cite{jha2017developer}.
Moreover, the relatively low merge ratio for \textit{`function implementation'} further suggests that tasks regarding implementing functionality require semantic correctness beyond syntactic validity~\cite{herzig2013impact}. That is, function-level changes frequently interact with existing business logic, hidden assumptions, and cross-module dependencies, increasing the likelihood of integration concerns.

\begin{figure}
    \centering
    \includegraphics[width=0.6\linewidth]{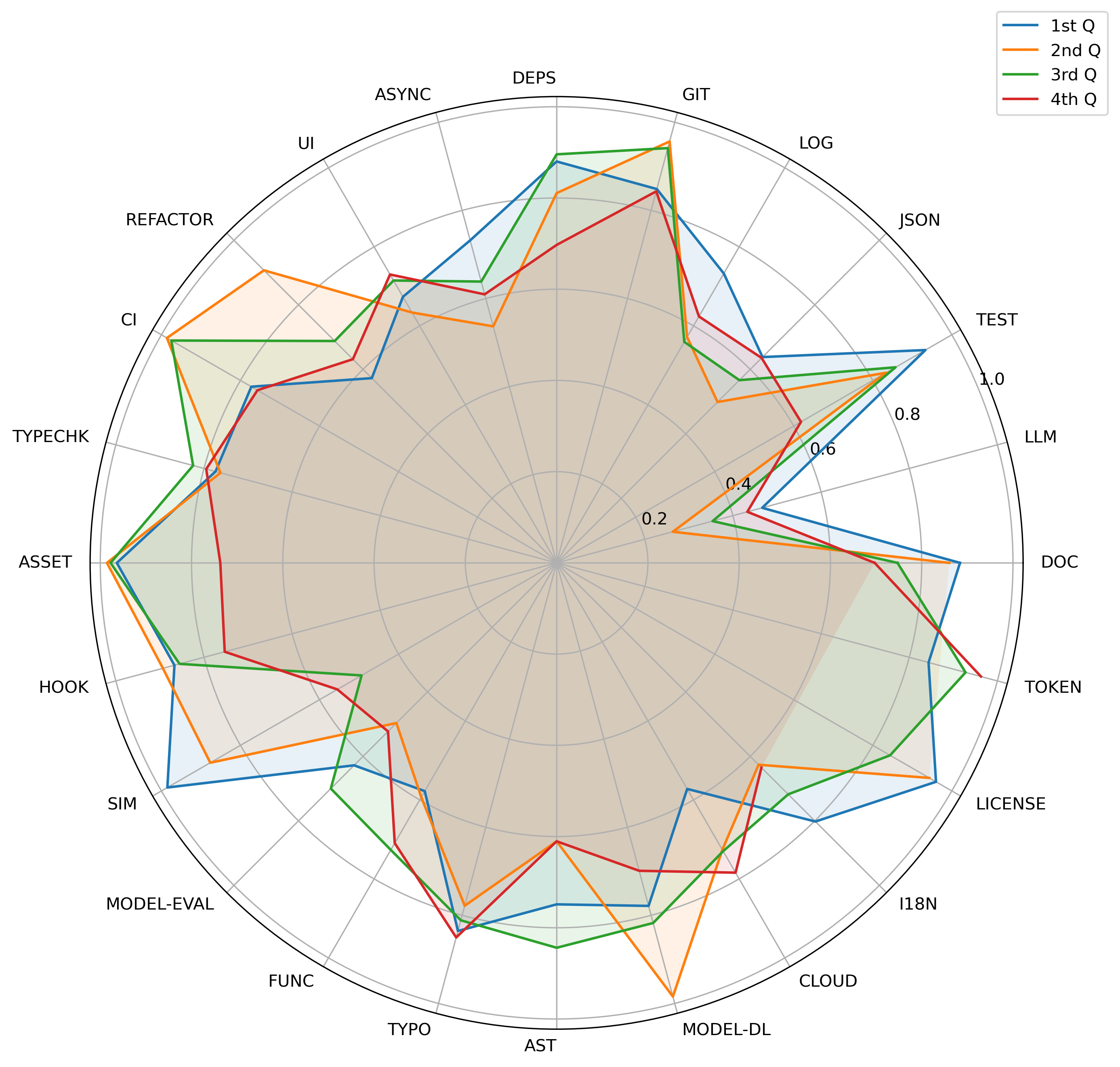}
    \caption{Distribution of  Agentic PRs' merge ratio across topics per quarter}
    \label{fig:topic_merge_ratio_per_quarter_radar_chart}
\end{figure}

Fig.\ref{fig:topic_merge_ratio_per_quarter_radar_chart} represents the activity of PRs belong to each type of development tasks in each quarter, in terms of merge rate. Given that the number of PRs in various categories are different, we need to normalize them to achieve a fair comparison. 
In other words, categories with higher number of PRs provide more reliable estimate, misleading the comparison if we only rely on the raw percentages. 
To address this issue, 
we employ a parametric Empirical Bayes approach with a Beta–Binomial model.
Parametric Empirical Bayes is one of the most popular approaches to evaluate group effects, when the effective connectivity is different between groups~\cite{ZEIDMAN201912}. 
Combining parametric Empirical Bayes and a Beta–Binomial model provides a powerful statistical framework that is generally used to estimate proportions for multiple related groups~\cite{hardcastle2013empirical}. 
Therefore, we compute the total number of PRs and the corresponding raw merge ratio for each quarter of each task type. 
To this end, we incorporate the variability of the number of PRs in each category of development task by computing the weighted mean and variance of the merge ratios, using the number of PRs as weights.

\subsubsection{Systematic Differences Across Topic Categories}
The results reveal strong heterogeneity in agentic PR merge rates across topic categories, indicating that different types of development activities exhibit systematically distinct integration behaviors.

\paragraph{\textbf{Development Activities with High Merge Ratio ($\geq$ ~0.80):}}
Activities exhibiting consistently high merge ratios reveal a distinct class of development tasks that are comparatively less complex, easier to integrate, and less susceptible to merge conflicts. In particular, infrastructure- and configuration-oriented activities such as \textit{`GitHub action workflow'}, \textit{`CI \& build'}, \textit{`asset management'}, \textit{`dependency'}, \textit{`hook management'}, \textit{`license'}, and \textit{`token management'} maintain high merge ratios across most quarters, often exceeding 0.90. This observation suggests that these categories of tasks are generally modular, well-scoped, and less entangled with concurrent development activities.
Similarly, categories such as \textit{`typos'} and \textit{`documentation'} also demonstrate high mergeability, indicating that minor textual modifications, documentation updates, and small corrective changes are typically isolated and therefore less likely to interfere with other code contributions.
It is worth noting that some categories, including \textit{`simulation'} and \textit{`model download'}, achieve very high merge ratios only during specific quarters while exhibiting noticeable fluctuations over time. Such variations may indicate that the complexity and development demand of these tasks are episodic in nature or influenced by evolving development practices and project requirements over time.
Overall, the prevalence of high merge ratios among these development activities suggests a broader pattern: declarative changes (e.g., configuration-related modifications), additive changes (e.g., assets and dependencies), and non-functional modifications (e.g., documentation and typo corrections) are substantially more likely to be integrated successfully. This finding indicates that lower coordination overhead, reduced implementation complexity, and limited interaction with concurrent development efforts increase the likelihood of successfully merging agent-generated PRs.

\paragraph{\textbf{Development Activities with Medium Merge Ratio ($\sim$0.60–0.80):}}

Categories such as \textit{`UI'}, \textit{`logging'}, \textit{`asynchronous'}, \textit{`type checking'}, \textit{`multilingual translation'}, \textit{`function implementation'}, and \textit{`cloud'} fall into a moderate acceptance range~\cite{lu2008learning}.
Changes related to these types of development tasks are neither purely isolated nor highly governance-sensitive, but typically involve moderate coordination and risk. 
That is, they often require contextual understanding of system behavior and relationship among software components, while still remaining localized enough to avoid widespread structural impact. As an example, PRs providing changes on the UI and logging depend on runtime behavior and system conventions, but do not bring high risk to the system, since they have moderate integration complexity~\cite{memon2007event,lu2008learning}. Similarly, cloud-related tasks require alignment with external systems correctness rather than purely syntactic validity~\cite{cotroneo2019bad}. 
It is also proven by previous studies that such cross-cutting and context-dependent changes tend to increase reasoning effort and integration uncertainty, while not necessarily inducing large-scale architectural modifications~\cite{herzig2013impact}.

\paragraph{\textbf{Development Activities with Low Merge Ratio ($\leq$ $\sim$0.60 in at least one quartile):}}

PRs belong to categories such as \textit{`LLM integration'}, \textit{`model evaluation'}, 
and \textit{`JSON template'} exhibit lower merge ratio or highly fluctuated for various quarters. This phenomenon can be emerged from the fact that these categories often involve context-sensitive, specification-dependent, and experimentally driven development tasks. 
In particular, PRs categorized as \textit{`LLM integration'} and \textit{`model evaluation'} are associated with rapidly evolving development practices, unstable APIs, prompt-specific behaviors, and evaluation uncertainty, all of which can increase implementation challenges and integration risk~\cite{chen2025empirical}.
In the case of PRs related to \textit{`JSON template'}, the relatively low merge ratio may stem from the structural and dependency-sensitive nature of configurations regarding template. Although these modifications are often syntactically small, they can propagate unintended effects across pipelines, services, prompts, workflows, or downstream components that rely on strict schema compatibility and configuration consistency~\cite{liu2024rethinking,huang2015confvalley}. Consequently, errors in template definitions may lead to cascading failures that are difficult to detect during code review, reducing developers’ willingness to merge agent-generated contributions in this category.
Besides, modern JSON Schema introduces substantial challenges in schema interpretation and formal validation due to features such as dynamic references and annotation-dependent validation, increasing JSON data validation complexity from polynomial time in Classical JSON Schema to PSPACE complexity~\cite{attouche2024validation}.
Accordingly, high amount of risks and uncertainty related to these types of PRs often face more challenges and relatively lower acceptance rates.

\subsubsection{Temporal Trends Across Quarters}
Since Fig.~\ref{fig:topic_merge_ratio_per_quarter_radar_chart} presents the merge ratio of agentic PRs across different development task categories and quarters, we further investigate how the merging behavior of PRs evolves over time within each category. To facilitate this analysis, we group development tasks into three major categories based on their temporal merge-ratio trends including increasing, decreasing, and stable over time, as described below. 
It should also be noted that, for some development categories, certain quarters do not contain any PRs to report merge or rejection measurements. For example, no merge/rejection observations are available for \textit{`token management'} during Q2.

\paragraph{\textbf{Increasing merge ratios over time:}}
Interestingly, several development categories, including \textit{`function implementation'}, \textit{`cloud'}, \textit{`token management'}, and \textit{`UI'}, exhibit progressively increasing merge ratios over time. One plausible explanation is that both developers and AI-assisted development workflows gradually adapt to the characteristics and constraints of these tasks over time. As repositories accumulate experience with agent-generated contributions, developers may become more familiar with reviewing, validating, and integrating such changes, leading to increased trust and reduced integration friction. Besides, improved understanding of prompting strategies, agent capabilities, repository-specific conventions, and development tooling may further enhance the quality and consistency of agentic PRs over time.
For example, categories such as \textit{`cloud'} often rely on recurring design patterns, reusable templates, and standardized workflows, which may become easier for agents to reproduce as development practices mature~\cite{gunawi2014bugs}. Similarly, the increasing merge ratio of \textit{`function implementation'} PRs  over time may indicate improved alignment between generated code and repository-specific implementation patterns, reducing semantic inconsistencies during review.

\paragraph{\textbf{Decreasing merge ratios over time}:}
several development categories exhibit progressively decreasing merge ratios over time, potentially indicating increasing integration complexity, stricter review practices, or evolving quality expectations for agent-generated contributions. While categories such as \textit{`dependency'} and \textit{`documentation'} demonstrate a relatively gradual decline in mergeability across quarters, other categories including \textit{`asset management'}, \textit{`test'}, and \textit{`simulation'} experience substantially sharper reductions. 
One plausible explanation is that, as repositories accumulate more experience with agent-generated PRs, developers adopt more conservative review practices for categories that may introduce hidden dependencies, maintenance overhead, or reproducibility concerns. 
For example, changes related to dependency may increasingly require careful compatibility verification as projects evolve~\cite{he2023automating}, while contributions on documentation may face stricter expectations regarding consistency, accuracy, and alignment with rapidly changing codebases over time~\cite{wu2024comprehensive}.
Besides, the sharp decline observed in categories such as \textit{asset management'}, \textit{test'}, and \textit{`simulation'} may stem from the growing complexity and validation burden associated with these development tasks. As an example, test-related changes require semantic correctness and adequate behavioral coverage beyond syntactic validity~\cite{zhang2026enhancing}. Similarly, PRs work on \textit{`asset management'} modifications can affect resource organization, build pipelines, or deployment artifacts~\cite{zheng2026github}.

\paragraph{\textbf{Stable Merge Ratios}:}
PRs categorized under development activities such as \textit{`type checking'}, \textit{`hook management'}, \textit{`typos'}, and \textit{`GitHub action workflow'} exhibit relatively stable merge ratios or only limited fluctuations over time. In particular, the merge ratio of \textit{`type checking'} PRs consistently remains within a narrow range (approximately 76\%–82\%), while the remaining categories demonstrate slightly broader yet still stable variation patterns. One plausible explanation is that these development activities are governed by comparatively stable validation criteria, standardized workflows, and well-established repository conventions.
For example, modifications related to \textit{`type checking'} are often supported by deterministic static analysis and compiler feedback, whereas \textit{`GitHub workflow'} and \textit{`hook management'} changes typically rely on structured automation pipelines and reusable configuration patterns~\cite{da2024chronicles,rzig2024empirical}. 
Similarly, typo-related PRs generally involve localized and syntactically constrained modifications that can be verified with relatively low semantic ambiguity~\cite{zheng2026github}. 
Consequently, the review and integration processes for these categories may remain comparatively consistent over time, resulting in stable mergeability patterns for agentic PRs.

\begin{tcolorbox}[colback=blue!10!white, colframe=blue!75!black]

\textbf{Finding 4: }
Tasks with localized scope and well-defined semantics (e.g., \textit{`GitHub action workflow'},\textit{`Dependency management'}, \textit{`Documentation'}, and \textit{`CI \& build'}) consistently achieve high merge rates, while tasks requiring extensive contextual understanding or involving evolving specifications (e.g., \textit{`LLM integration'}, \textit{`model evaluation'}, and \textit{`JSON template'}) exhibit lower or less stable merge rates.

\end{tcolorbox}

\subsubsection{Quartile-Specific Insights}
This section compares the merge ratios of PRs associated with various development tasks from the perspective of individual quarters in isolation. Specifically, we investigate how the behavior of coding agents varies across different quarters with respect to the mergeability of PRs generated for different categories of development activities. The following highlights the corresponding merge rate patterns across the various development tasks.

\paragraph{\textbf{1st Quartile:} }
Generally, categories such as \textit{`simulation'}, \textit{`asset management'}, \textit{`license management'}, and \textit{`test'} have the highest merge ratio in the first quarter. Although changes related to these tasks may cause some risks to the repositories, they are mostly highly structured, easy to verify (manually or by CI/CD pipeline), and less challenging in the early stage of the software development process. Besides, these changes rarely affect core execution logic of the software system.  
In contrast, PRs associated with development activities such as \textit{`function implementation'}, \textit{`LLM integration'}, and \textit{`refactoring'} exhibit comparatively low merge ratios, even during the early stages of AI-assisted software development adoption. One possible explanation is that these tasks generally require a deeper understanding of system behavior, architectural constraints, and project-specific requirements~\cite{sillito2006questions}. Unlike operational or configuration-oriented modifications, these activities often involve substantial semantic reasoning and may affect multiple interacting components. Consequently, validating the correctness and broader impact of such changes requires greater review effort, which may reduce developers' willingness to merge agentic PRs belong to these categories.

\paragraph{\textbf{2nd Quartile:} }
The second quartile shows a high merge rate for infrastructure-related and automation-oriented tasks such as \textit{`CI \& build'}, \textit{`model download'}, \textit{`asset management'}, and \textit{`GitHub actions workflows'}. This phenomenon can be interpreted as the fact that reliance on AI agents for these types of development tasks increase in the second quarter. On the other hand, developers become more conservative for tasks which are directly related to AI integration like \textit{`LLM integration'} and \textit{`model evaluation'} involving semantic correctness. 
It is also note mentioning that some categories of tasks such as \textit{`refactoring'}, and \textit{`model download'} reach their peak in this quarter. 
However, concerning the decline in merge ratio of PRs work on \textit{`refactoring'}, and \textit{`model download'} tasks in the next quarters, this pattern likely reflects quarter-specific development dynamics rather than a persistent characteristic of these task categories.
Overall, while merge rate of PRs working on well-structured or infrastructure-related changes maintain high in the second quarter, more complex or uncertain domains tend to exhibit reduced merge rate, reflecting increasingly differentiated evaluation criteria.

\paragraph{\textbf{3rd Quartile:}}
The third quartile exhibits among the highest variability across topics, with merge ratios spanning a wide range from approximately 0.35 to 0.97. Interestingly, several development tasks experience substantial declines, including 
\textit{`simulation'}, and \textit{`documentation'}. One possible reason behind this phenomenon can be the fact that simulation-related tasks become more challenging by increasing the size and accordingly, the complexity of the software systems. In contrast, some other categories such as \textit{`function implementation'}, \textit{`type checking'}, and \textit{`dependency'} experience a higher merge rate of related PRs. It can refer to the fact that repositories may become more adopted with integrating the functional-related changed generated by AI agents over time, because of better understanding of agents about the project structure and logic.

\paragraph{\textbf{4th Quartile:} }
In the fourth quarter, the distribution of merge ratios shifts toward a more moderate range, with many development categories converging between approximately 0.60 and 0.75. Notably, categories such as \textit{`test'}, \textit{`dependency'}, and \textit{`asset management'}, which exhibited consistently high merge ratios in earlier quarters, experience a noticeable decline. One plausible explanation is that, as repositories become mature, larger, and more complicated, modifications in these areas require more extensive validation to ensure compatibility, maintainability, and system stability.
At the same time, several development tasks, including \textit{`token management'}, \textit{`typo'}, and \textit{`cloud'}, achieve comparatively higher merge ratios. The high mergeability of PRs categorized as \textit{typo'} is unsurprising, as such changes are typically localized, easy to verify, and involve limited semantic ambiguity. 
Similarly, \textit{token management'} modifications often follow well-defined operational rules, enabling developers to evaluate related changes with relatively low effort. The increasing merge ratio of \textit{cloud'}-related PRs may reflect the growing standardization of cloud deployment workflows and automation practices, which can reduce integration uncertainty and facilitate review.

\begin{tcolorbox}[colback=blue!10!white, colframe=blue!75!black]

\textbf{Finding 5: }
The mergeability of agentic PRs is strongly task-dependent and evolves across repository maturity. Structured and less challenging tasks (e.g., \textit{`asset management'}, \textit{`CI \& build'}, and \textit{`GitHub actions workflows'}) consistently achieve higher merge rates during the early and middle quarters, whereas semantically complex tasks (e.g., \textit{`function implementation'} and \textit{`LLM integration'}) exhibit persistently lower mergeability.
In the fourth quarter, merge rates become more uniform across task categories, indicating that the differences in mergeability between structured and complex development tasks diminish as repositories evolve. 
\end{tcolorbox}

\subsection{\textbf{RQ3.} 
How do agentic PRs differ from human generated PRs in terms of code quality and defect proneness?
}
\label{subsec:rq3}

In this RQ, we compare the characteristics of agentic PRs with those provided by human developers. To this end, we investigate differences in the complexity of code changes introduced through agentic and human-generated PRs.
To characterize PRs' complexity, we employ a set of well-established software engineering metrics that have been widely used in prior studies~\cite{morovati2024bug,kononenko2018studying}, including the number of commits, the number of modified files, the number of changed lines of code, the number of comments associated with the PR, the number of contributors participating in the PR and its discussions, and the duration (in days) that a PR remains open before being merged.
In the second part, we compare the bug-inducing commit rates associated with merged agentic and human-generated PRs to better understand their relative impact on software quality.

Following the sampling strategy used in prior work~\cite{watanabe2025use}, we construct balanced samples of agentic and human generated PRs (as described in Section~\ref{subsec:data_collection}) for each repository quarter. Specifically, for every development quarter, we randomly sample PRs from the larger group so that the numbers of agentic and human generated PRs are equal. This approach mitigates potential biases caused by differences in the volume of contributions across the two groups and enables a fair comparison. Quarters containing PRs from only one group (i.e., either agentic or human generated PRs) are excluded from the analysis because they do not permit direct comparison.
The resulting matched sample consists of 2,275 merged agentic PRs and 2,275 merged human-generated PRs.

\begin{figure}[]
    \centering

    \begin{subfigure}{0.48\textwidth}
        \centering
        \includegraphics[width=\linewidth]{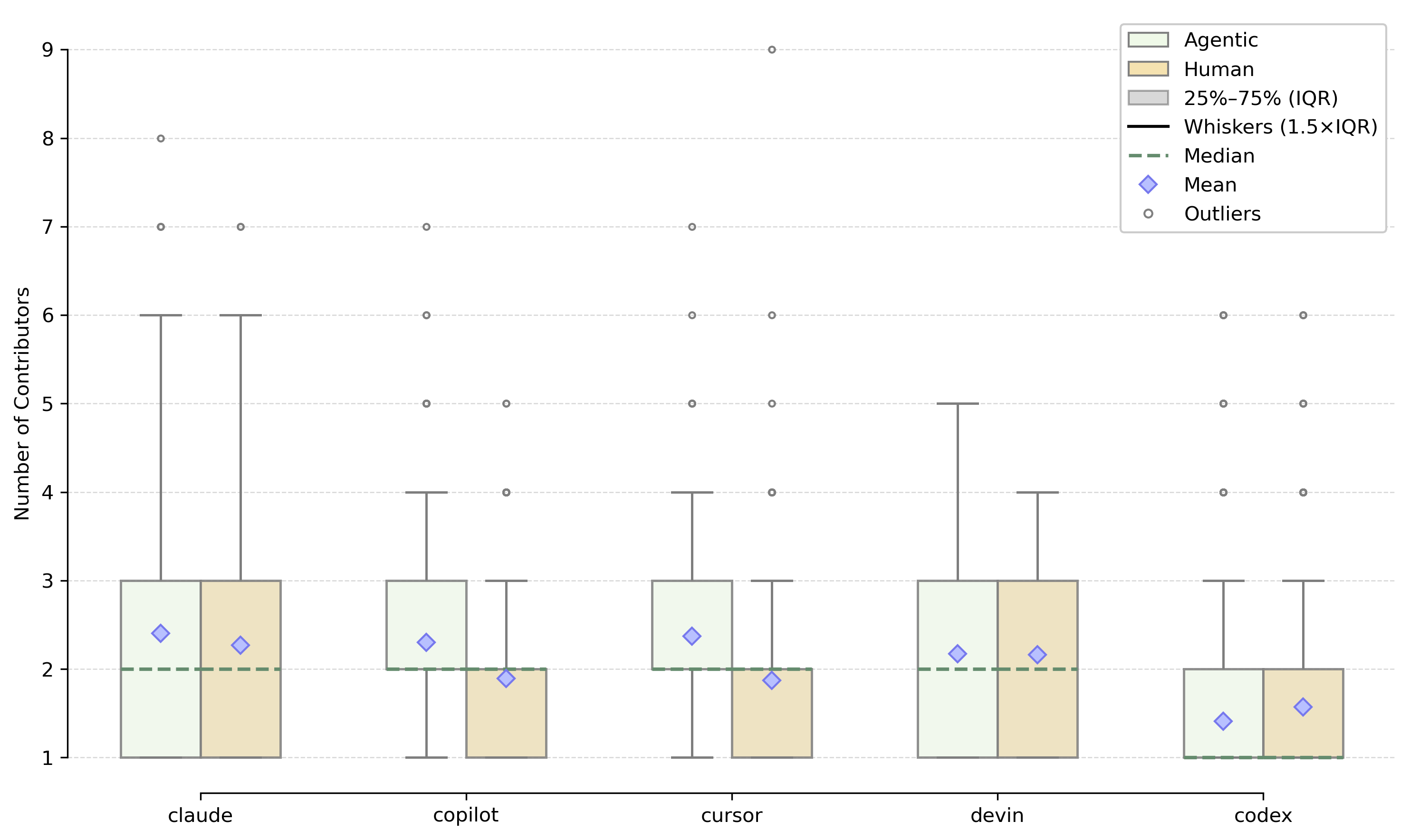}
        \caption{Num of Contributors}
        \label{fig:metrics_contributors}
    \end{subfigure}
    \hfill
    \begin{subfigure}{0.48\textwidth}
        \centering
        \includegraphics[width=\linewidth]{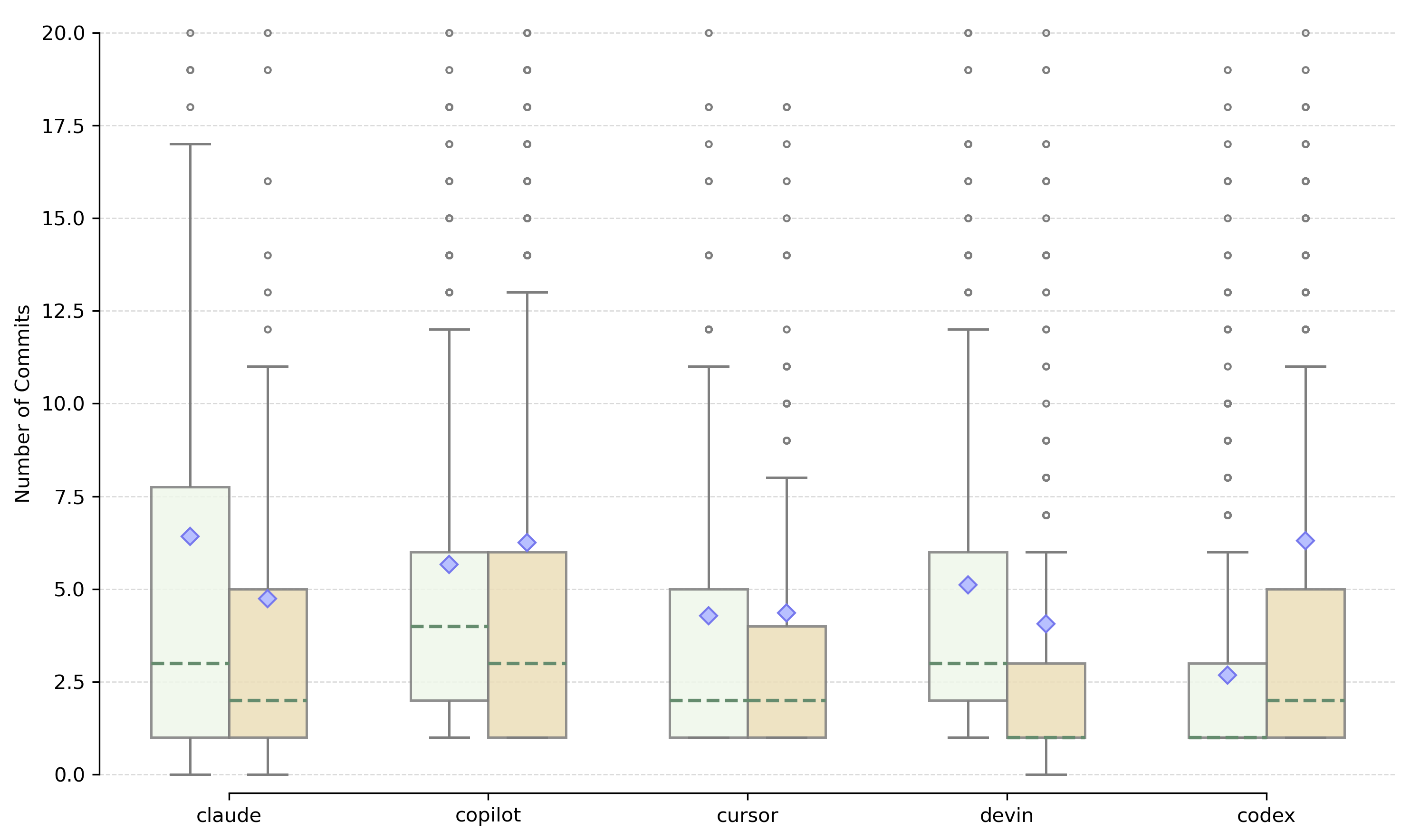}
        \caption{Num of Commits}
        \label{subfig:metrics_commits}
    \end{subfigure}

    \vspace{0.5cm}

    \begin{subfigure}{0.48\textwidth}
        \centering
        \includegraphics[width=\linewidth]{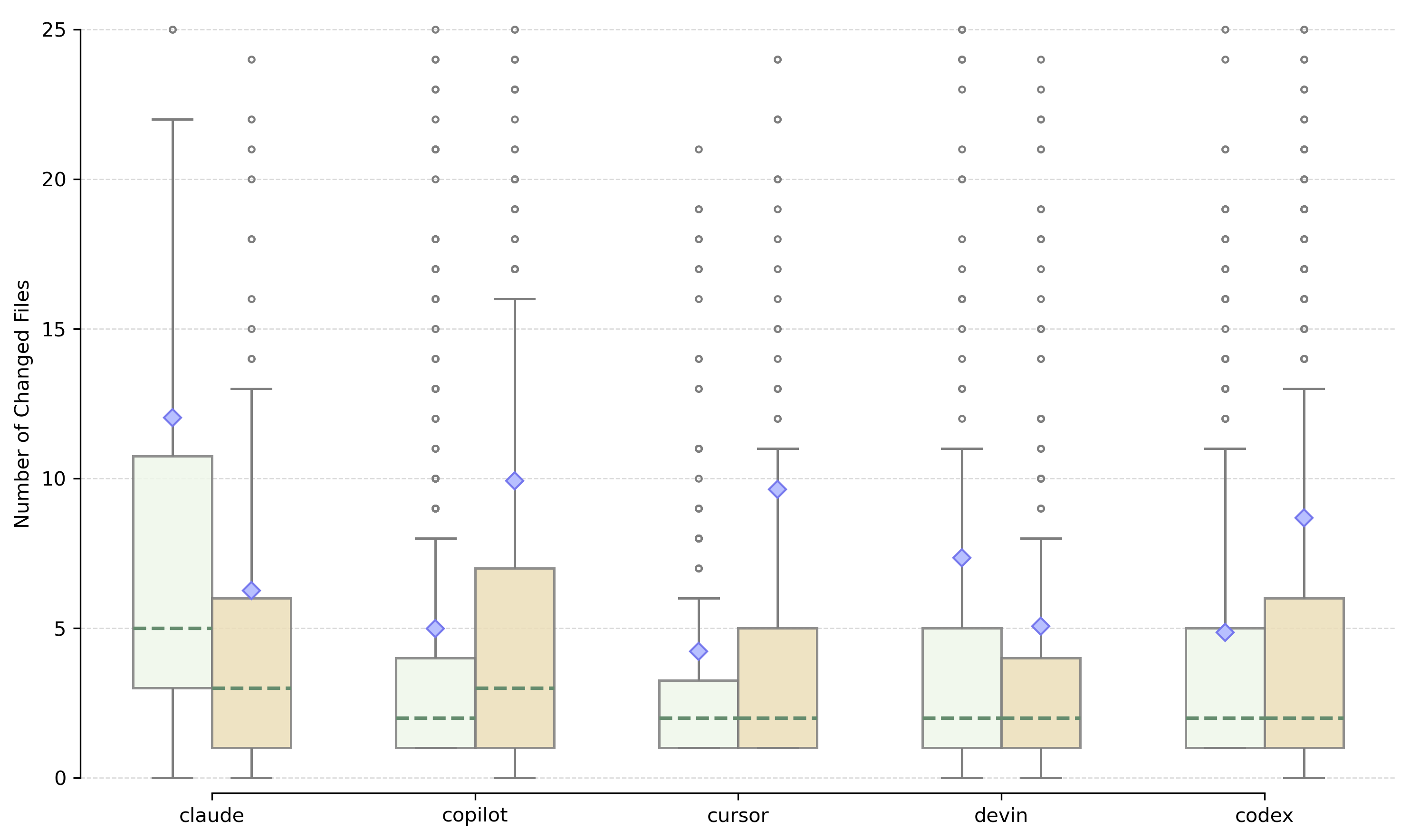}
        \caption{Num of Changed Files}
        \label{subfig:metric_file}
    \end{subfigure}
    \hfill
    \begin{subfigure}{0.48\textwidth}
        \centering
        \includegraphics[width=\linewidth]{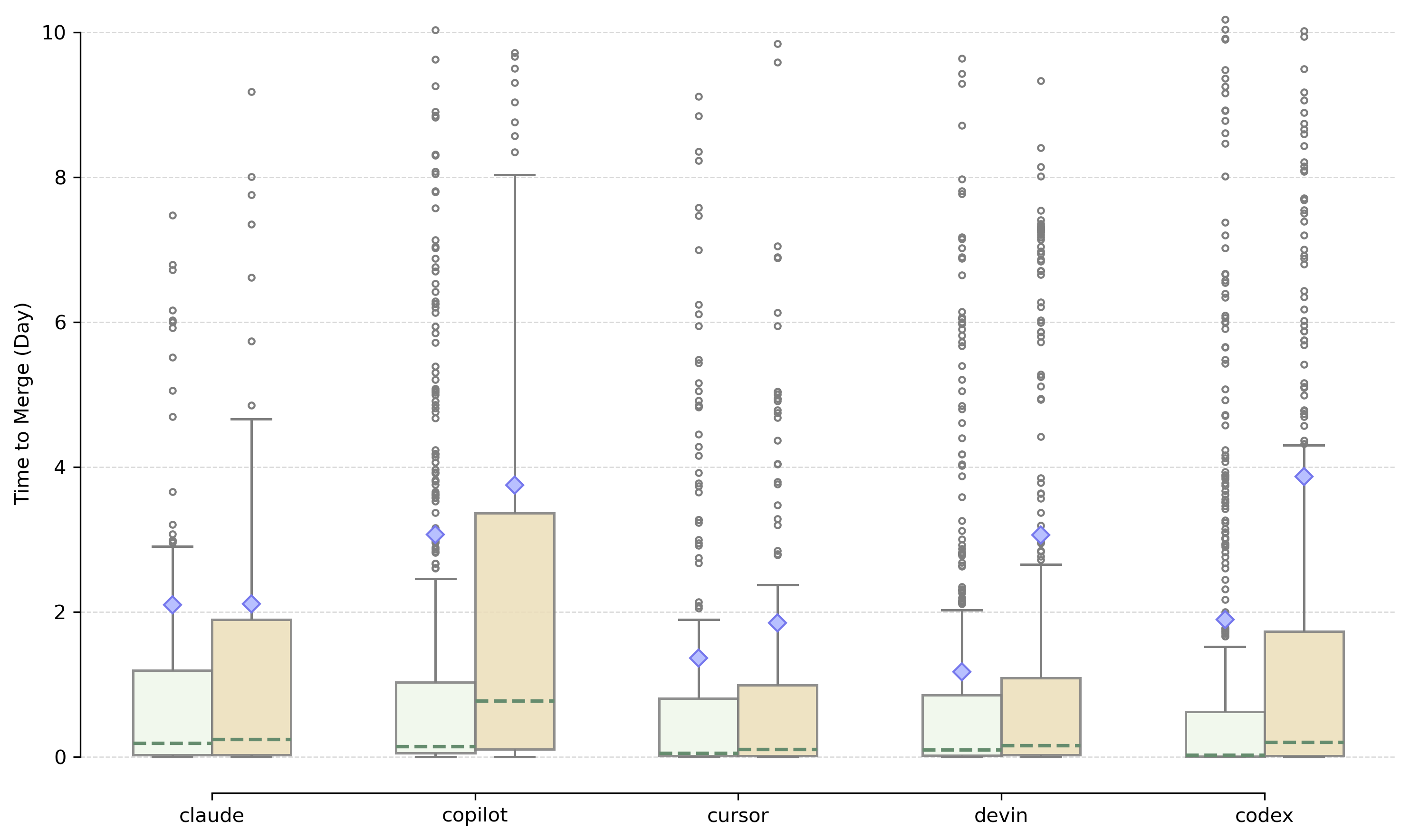}
        \caption{Time to Merge}
        \label{subfig:metric_duration_merge}
    \end{subfigure}
    \vspace{0.5cm}

    \begin{subfigure}{0.48\textwidth}
        \centering
        \includegraphics[width=\linewidth]{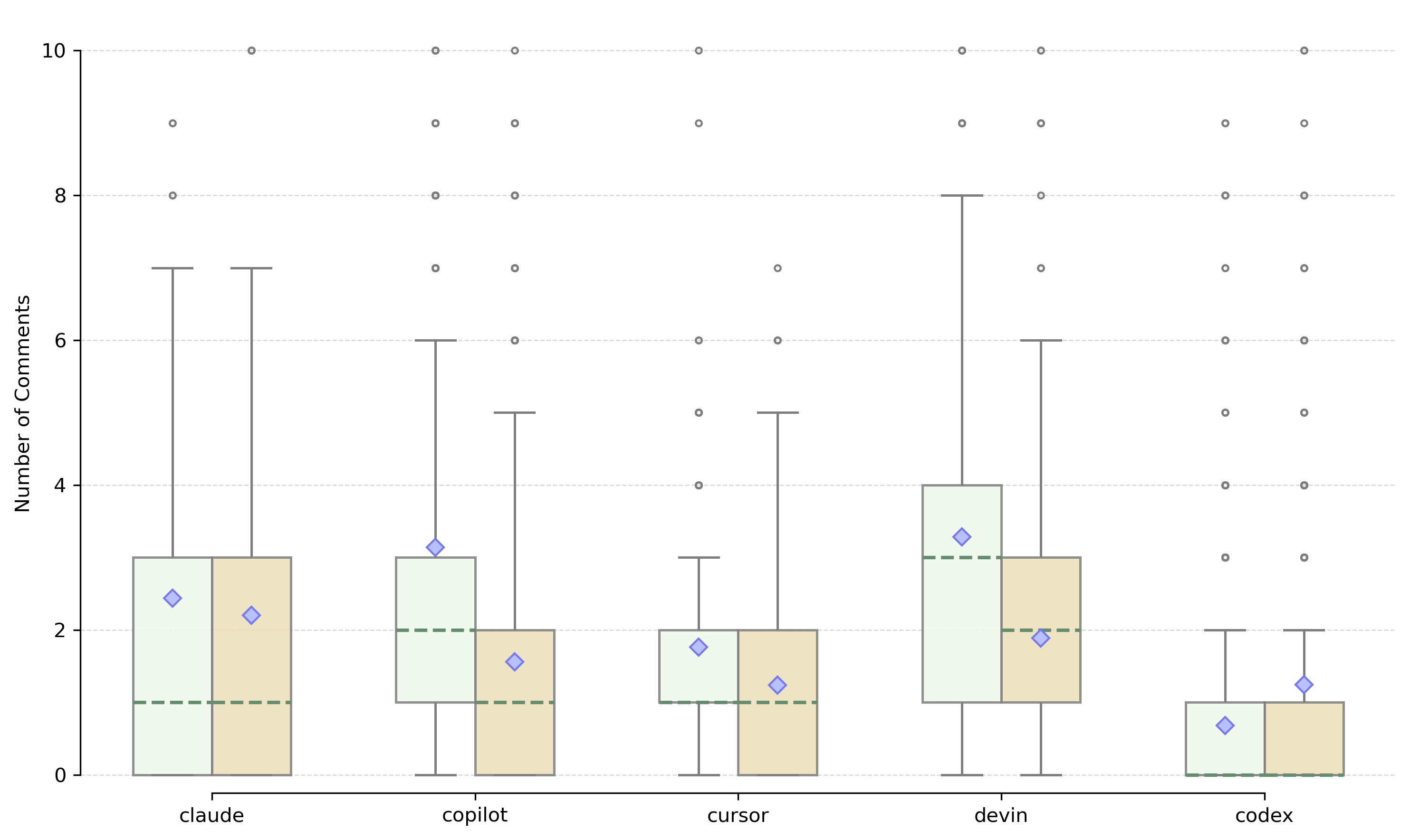}
        \caption{Num of Comments}
        \label{subfig:metric_comment}
    \end{subfigure}
    \hfill
    \begin{subfigure}{0.48\textwidth}
        \centering
        \includegraphics[width=\linewidth]{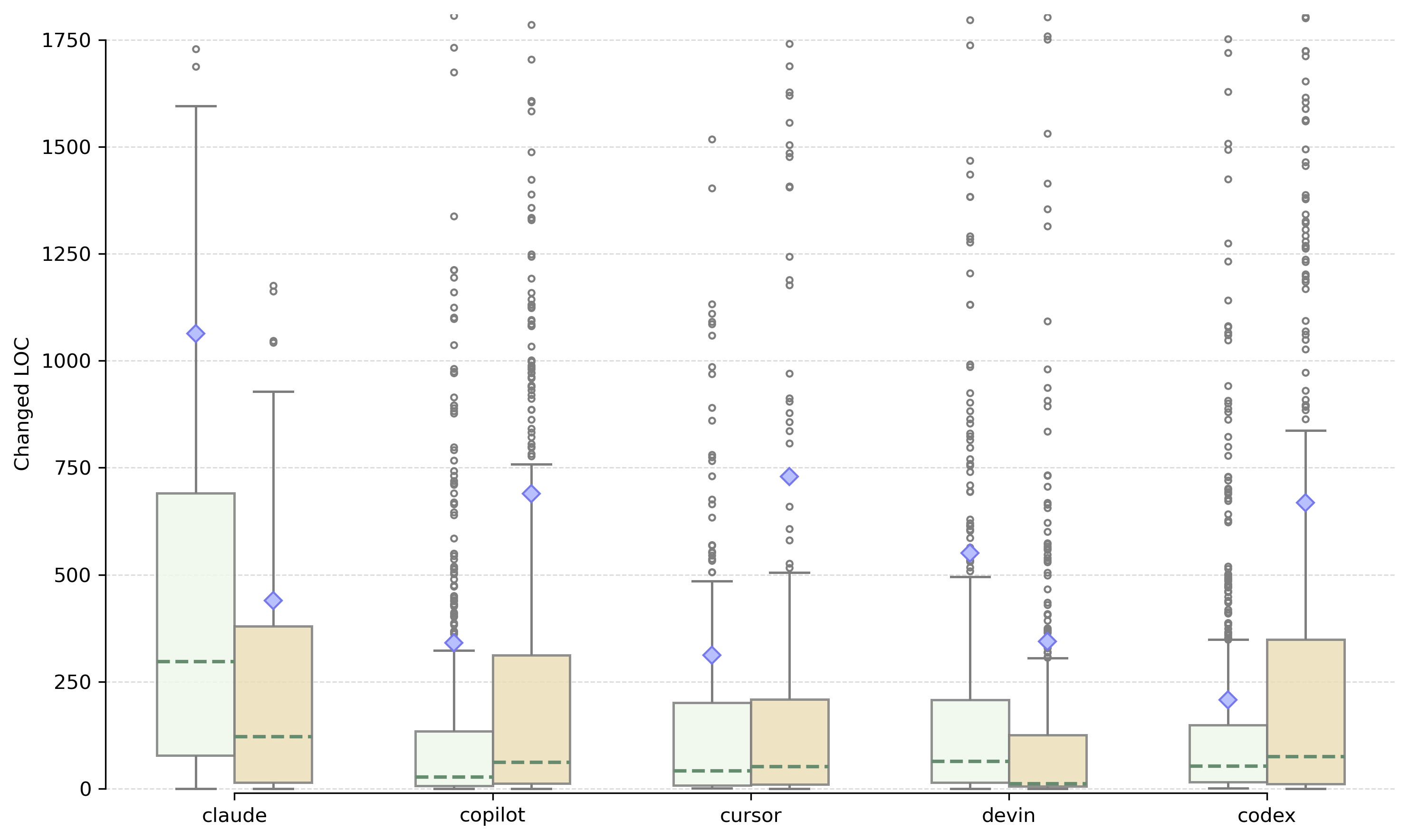}
        \caption{Num of Changed LOC}
        \label{subfig:metric_LOC}
    \end{subfigure}

    \caption{Boxplot distributions of the complexity metrics for A-PRs H-PRs across different agents}
    \label{fig:Compexity_metric_APRs_vs_HPRs_per_agents}
\end{figure}

\subsubsection{Complexity Analysis of Agentic and Human-generated PRs}

Following literature study~\cite{morovati2024bug}, 
We use six metrics including number of commits, number of contributors, number of changes files, number of comments, number of changed Line Of Code (LOC), and time to merge
to measure the complexity of PRs.
Fig. ~\ref{fig:Compexity_metric_APRs_vs_HPRs_per_agents} represents the comparison of various metrics between agentic and human-generated PRs, for five AI agents separately. To ensure the significancy of the difference between agentic and human-generated PRs, as well as between various AI agents, we employed the \textit{Mann–Whitney U test} ~\cite{arcuri2014hitchhiker} as the statistical test, and \textit{Cliff’s Delta} ~\cite{macbeth2011cliff} to measure effect size to evaluate the magnitude of the differences, since relying solely on the \textit{p}-value may be misleading~\cite{kampenes2007systematic}. 
Table ~\ref{tab:stat_test_agent_human_metric}
summarizes the results of the statistical test, along with the effect size of comparing characteristics of agentic and human-generated PRs, over various metrics. 
Table \ref{tab:Statistical_tests_comparing_AM-PRs_of_different_agents_across_metrics} also reports the statistical comparisons among various agents over metrics.

\paragraph{\textbf{Number of Contributors}} Regarding the number of developers who collaborate on the PRs, as it is understandable from Fig.\ref{fig:metrics_contributors}, for 4 out of 5 it is higher in agentic PRs in average. 
The contributors to an agentic PR include the PR author (agent) and the developers (human or agent) who participate in its discussion. 
Considering related \textit{p}-value and effect size represented in Table ~\ref{tab:stat_test_agent_human_metric}, the differences between agentic and human generated PRs are significant only for Copilot, Cursor, and Codex ($p<0.001$). Regarding Copilot and Cursor showing higher number of contributors 
in average for agentic PRs, one possible reason can be the fact that the code generated by these agents revise multiple times during the review process. In contrast, PRs generated by Codex use fewer number of contributors, compare to ones generated by humans. 
Overall, the results suggest that agentic workflows do not universally alter collaboration breadth, but some agents—particularly Copilot and Cursor—are associated with moderately more collaborative PR processes.
Considering the comparison among the number of developers contribute on the PRs generated by various agents, as shown in Table \ref{tab:Statistical_tests_comparing_AM-PRs_of_different_agents_across_metrics}, the number of contributors for Codex is significantly lower, compared to other agents. 
Despite the \textit{p}-value of comparing Devin with Copilot and Cursor is less that 0.05 (significant difference), we can consider them as insignificant differences, based on the negligible related effect sizes.

\paragraph{\textbf{Number of Commits}} 
Another metric commonly used in prior studies to characterize the complexity of the task addressed by a PR, as well as a factor influencing PR merge outcomes, is the number of commits contained in the PR~\cite{morovati2024bug,kononenko2018studying}. As shown in Fig.~\ref{subfig:metrics_commits}, the average number of commits in PRs generated by Devin is significantly higher than that of the corresponding human generated PRs. In contrast, PRs generated by Codex contain significantly fewer commits on average than their human-generated counterparts. Moreover, the observed effect sizes indicate that these differences are non-negligible (Table \ref{tab:stat_test_agent_human_metric}).
Although the average number of commits in Copilot generated PRs is lower than that of human generated PRs, the median number of commits is higher for PRs generated by Copilot, and the corresponding effect size is positive (Cliff's $\delta$ = 0.171). These results suggest that Copilot generated PRs tend to contain 
more (small effect size) commits than human generated PRs in typical cases. The apparent discrepancy between the mean and median values may be explained by the presence of a small number of human generated PRs with exceptionally high number of commit, which increase the average while having a limited effect on the median. 
One possible explanation for the higher number of commits in PRs generated by Devin and Copilot is that developers may need to make additional revisions to the agent's proposed changes, potentially to address issues in the generated code. Conversely, the lower number of commits observed in Codex generated PRs may suggest that Codex produce changes that require less human intervention. This could indicate a better understanding of project context, code structure, or development objectives, enabling their generated solutions to align more closely with developers' expectations.

Furthermore, a comparison of the number of commits across PRs generated by various agents reveals that PRs generated by Copilot and Devin involve the highest average numbers of commits. The observed differences are also statistically significant (\textit{p}-value < 0.05). In contrast, Codex generated PRs tend to contain fewer commits than those generated by the other agents.
Notably, the comparison between Codex and Copilot yields a large effect size (Cliff's $\delta$ = -0.580), indicating a substantial difference in their commit count distributions. This result suggests that PR generated by Copilot generally contain considerably more commits than Codex generated PRs, highlighting distinct patterns in how these agents contribute changes to software projects.

\begin{table}[t]
\centering
\caption{Statistical tests comparing agentic and human-generated PRs across various metrics (n = negligible, s = small, m = medium, l = large)}
\scriptsize
\setlength{\tabcolsep}{4pt}
\begin{tabular}{lcccccccccccc}
\toprule
\multirow{2}{*}{\textbf{Agent}} &
\multicolumn{2}{c}{\textbf{Contributors}} &
\multicolumn{2}{c}{\textbf{Commits}} &
\multicolumn{2}{c}{\textbf{Changed Files}} &
\multicolumn{2}{c}{\textbf{Time to Merge}} &
\multicolumn{2}{c}{\textbf{Comments}} &
\multicolumn{2}{c}{\textbf{Changed LOC}} \\
\cmidrule(lr){2-3}
\cmidrule(lr){4-5}
\cmidrule(lr){6-7}
\cmidrule(lr){8-9}
\cmidrule(lr){10-11}
\cmidrule(lr){12-13}
&
\textit{p}-value & Cliff's $\delta$ &
\textit{p}-value & Cliff's $\delta$ &
\textit{p}-value & Cliff's $\delta$ &
\textit{p}-value & Cliff's $\delta$ &
\textit{p}-value & Cliff's $\delta$ &
\textit{p}-value & Cliff's $\delta$ \\
\midrule
Claude  & 0.374    & 0.058  (n)  & 0.267    & 0.074 (n)  & \textbf{<1e-3}    & 0.255 (s)  & 0.779    & -0.019 (n)  & 0.472    & 0.048 (n)  & \textbf{<1e-3}     & 0.287 (s)  \\
Copilot & \textbf{<1e-3 }   & 0.281 (s)  & \textbf{<1e-3}    & 0.171 (s)  & \textbf{<1e-3}    & -0.187 (s)  & \textbf{<1e-3}    & -0.204 (s)  & \textbf{<1e-3}    & 0.281 (s)  & \textbf{<1e-3}    & -0.174 (s)  \\
Cursor  & \textbf{<1e-3}    & 0.279  (s)  & 0.824    & 0.010 (n)  & \textbf{0.009}    & -0.127 (n)  & 0.143   & -0.074 (n)  & \textbf{<1e-3}    & 0.239 (s)  & 0.375    & -0.044 (n)  \\
Devin   & 0.675 & -0.014 (n)  & \textbf{<1e-3} & 0.344 (m)  & 0.393 & 0.031 (n)  & \textbf{0.020} & -0.087 (n)  & \textbf{<1e-3} & 0.365 (m)  & \textbf{<1e-3}  & 0.269 (s)  \\
Codex   & \textbf{<1e-3}  & -0.121 (n)  & \textbf{<1e-3}   & -0.255 (s)  & 0.192   & -0.040 (n)  & \textbf{<1e-3}   & -0.210 (s)  & \textbf{<1e-3}   & -0.148 (s)  & \textbf{0.001}   & -0.101 (n)  \\

\bottomrule
\end{tabular}

\label{tab:stat_test_agent_human_metric}
\end{table}

\begin{table}[b]
\centering
\caption{Statistical tests comparing agentic PRs of various agents across metrics}
\scriptsize
\setlength{\tabcolsep}{4pt}
\begin{tabular}{lcccccccccccc}
\toprule
\multirow{2}{*}{\textbf{Agent}} &
\multicolumn{2}{c}{\textbf{Contributors}} &
\multicolumn{2}{c}{\textbf{Commits}} &
\multicolumn{2}{c}{\textbf{Changed Files}} &
\multicolumn{2}{c}{\textbf{Time to Merge}} &
\multicolumn{2}{c}{\textbf{Comments}} &
\multicolumn{2}{c}{\textbf{Changed LOC}} \\
\cmidrule(lr){2-3}
\cmidrule(lr){4-5}
\cmidrule(lr){6-7}
\cmidrule(lr){8-9}
\cmidrule(lr){10-11}
\cmidrule(lr){12-13}
&
\textit{p}-value & Cliff's $\delta$ &
\textit{p}-value & Cliff's $\delta$ &
\textit{p}-value & Cliff's $\delta$ &
\textit{p}-value & Cliff's $\delta$ &
\textit{p}-value & Cliff's $\delta$ &
\textit{p}-value & Cliff's $\delta$ \\
\midrule
Codex vs Claude  & \textbf{<1e-3}   &  -0.467 (m)  & \textbf{<1e-3}   &  -0.292 (s)  & \textbf{<1e-3}   &  -0.357 (m)  & \textbf{<1e-3}   &  -0.285 (s)  & \textbf{<1e-3}   &  -0.475 (l)  & \textbf{<1e-3}   &  -0.471 (m)  \\
Codex vs Copilot & \textbf{<1e-3}   &  -0.633 (l)  & \textbf{<1e-3}   &  -0.580 (l)  & \textbf{<1e-3}   &  0.112 (n)  & \textbf{<1e-3}   &  -0.351 (m)  & \textbf{<1e-3}   &  -0.647 (l)  & \textbf{<1e-3}   &  0.130 (n)  \\
Codex vs Cursor  & \textbf{<1e-3}   &  -0.546 (l)  & \textbf{<1e-3}   &  -0.211 (s)  & \textbf{0.002}   &  0.123 (n)  & \textbf{0.004}   &  -0.120 (n)  & \textbf{<1e-3}   &  -0.511 (l)  & 0.316   &  0.042 (n)  \\
Codex vs Devin   & \textbf{<1e-3}   &  -0.467 (m)  & \textbf{<1e-3}   &  -0.388 (m)  & 0.506   &  0.022 (n)  & \textbf{<1e-3}   &  -0.228 (s)  & \textbf{<1e-3}   &  -0.794 (l)  & 0.133   &  -0.052 (n)  \\
Copilot vs Claude   & \textbf{0.178}   &  0.065 (n)  & \textbf{<1e-3}   &  0.182 (s)  & \textbf{<1e-3}   &  -0.431 (m)  & 0.294   &  0.056 (n)  & \textbf{0.035}   &  0.109 (n)  & \textbf{<1e-3}    &  -0.513 (l)  \\
Copilot vs Cursor   & 0.610   &  -0.019 (n)  & \textbf{<1e-3}   &  0.317 (s)  & 0.853   &  0.007 (n)  & \textbf{<1e-3}    &  0.270 (s)   & \textbf{<1e-3}   &  0.168 (s)  & 0.100    &  -0.069 (n)  \\
Copilot vs Devin   & \textbf{0.003}   &  0.094 (n)  & \textbf{<1e-3}   &  0.154 (s)  & \textbf{0.006}   &  -0.090 (n)  & \textbf{<1e-3}    &  0.173 (s)  & \textbf{<1e-3}   &  -0.238 (s)  & \textbf{<1e-3}    &  -0.170 (s)  \\
Devin vs Claude   & 0.522   & -0.033 (n)  & 0.453   & 0.0408 (n)  & \textbf{<1e-3}   & -0.369 (m)  & 0.064   & -0.102 (n)  & \textbf{<1e-3}   & 0.305 (s)  & \textbf{<1e-3}   & -0.400 (m)  \\
Devin vs Cursor   & \textbf{0.021}   &  -0.098 (n)  & \textbf{<1e-3}   &  0.157 (s)  & \textbf{0.020}   &  0.100 (n)  & \textbf{0.014}  &  0.110 (n)  & \textbf{<1e-3}   &  0.429 (m)  & \textbf{0.047}   &  0.089 (n)  \\
Cursor vs Claude   & 0.336   &  0.055 (n)  & 0.091   &  -0.099 (n)  & \textbf{<1e-3}   &  -0.450 (m)  & \textbf{0.001}   &  -0.193 (s)  & 0.348  &  -0.055 (n)  & \textbf{<1e-3}   &  -0.438 (m)  \\

\bottomrule
\end{tabular}

\label{tab:Statistical_tests_comparing_AM-PRs_of_different_agents_across_metrics}
\end{table}

\paragraph{\textbf{Number of Changed Files}}
Regarding the number of files modified within PRs, Fig.~\ref{subfig:metric_file} shows that PRs generated by Claude and Devin tend to affect a larger number of files than those generated by human developers. 
However, the statistical analysis (Table~\ref{tab:stat_test_agent_human_metric}) indicates that only the difference between Claude generated and human generated PRs is statistically significant. In contrast, Copilot and Cursor exhibit the opposite pattern, with their PRs modifying significantly fewer files than human generated PRs. It is worth noting that the corresponding effect size for Cursor is negligible, suggesting that the practical difference is limited despite statistical significance.
One possible explanation is that PRs generated by Claude are associated with broader scope of changes. Prior research has identified the number of modified files as a proxy for change complexity, as changes spanning multiple files often require greater coordination and may involve multiple interacting components~\cite{badampudi2023modern}. Therefore, the observed pattern may indicate that Claude is frequently applied to development tasks involving more extensive code modifications than those addressed by humans.
A comparison among agents further reveals that Claude and Codex tend to modify significantly more files than the other agents, with Claude exhibiting a 
stronger effect (Cliff's $\delta$ = -0.357). This finding suggests that these agents are more commonly associated with PRs involving broader code changes. Alternatively, developers may be more willing to integrate larger scope modifications generated by Claude and Codex, reflecting greater confidence in the quality or usefulness of their proposed changes. Nevertheless, additional investigation would be required to establish whether this phenomenon is driven by task selection, agent capability, developer trust, or a combination of these factors.

\paragraph{\textbf{Merge Duration}}
As shown in Fig.~\ref{subfig:metric_duration_merge}, PRs generated by all examined agents are merged faster, on average, than human generated PRs, which is align with previous studies~\cite{gao2026autopilot}. 
However, the observed differences are statistically significant only for Copilot and Codex, with the small corresponding effect sizes (Table~\ref{tab:stat_test_agent_human_metric}). In the case of Devin, the difference is not practically meaningful, as the associated effect size is negligible (Cliff's $\delta$ = -0.087).
One possible explanation for the shorter merge time of agentic PRs is that these PRs may be more focused and limited in scope, allowing reviewers to assess and integrate them more quickly. Another potential reason is that agentic PRs are often submitted for relatively well-defined development tasks, such as documentation updates (Fig. \ref{fig:topics_to_number_of_merged_rejected_prs}), which may require less coordination before merging. 
With respect to merge duration across different agents, the statistical test results (Table~\ref{tab:Statistical_tests_comparing_AM-PRs_of_different_agents_across_metrics}) indicate that Codex generated PRs are merged significantly faster, on average, than PRs generated by the other agents. One possible explanation is that the changes proposed by Codex are easier for reviewers to understand, validate, and integrate into the target project. Alternatively, this finding may reflect differences in the types of tasks for which Codex is employed. In particular, Codex may be used more frequently for relatively well-defined or less complex tasks, which naturally require less review effort and can therefore be merged more quickly.
In contrast, PRs generated by Copilot and Claude exhibit longer merge durations. A plausible explanation is that these agents are more often utilized for complex development tasks involving larger or more intricate code changes. Such changes typically require more extensive review, discussion, testing, and validation before they can be safely integrated. This interpretation is consistent with our earlier findings, which suggest that  Claude generated PRs tend to be associated with tasks of greater complexity.

\paragraph{\textbf{Number of Comments}}

As shown in Fig.~\ref{subfig:metric_comment}, PRs generated by four of the five studied agents receive, on average, more comments than human generated PRs. According to the statistical test results (Table~\ref{tab:Statistical_tests_comparing_AM-PRs_of_different_agents_across_metrics}), these differences are statistically significant only for Copilot, Cursor, and Devin. Notably, PRs generated by Devin exhibit the largest difference in the number of comments compared with their human generated counterparts corresponding to a medium effect size (Cliff's $\delta$ = 0.365).
A higher number of comments generally suggests increased reviewer discussion during the review process. This phenomenon may arise for several reasons. For instance, reviewers may need additional clarification regarding the generated changes, identify issues that require revision, or discuss alternative implementation strategies~\cite{zampetti2017developers}. Moreover, PRs associated with more complex development tasks often require more extensive communication among contributors before they can be approved and merged~\cite{morovati2024bug}.

When comparing different agents, the statistical test results indicate that Codex generated PRs receive significantly fewer comments, on average, than those generated by the other agents. In contrast, Devin generated PRs attract substantially more discussion during the review process. 
One possible explanation is that Codex tends to generate changes that are easier to review and integrate, thereby reducing the need for extensive reviewer interaction. 
Alternatively, Codex may be applied more frequently to well-scoped tasks with lower review complexity. In contrast, the larger number of comments associated with PRs generated by Devin may indicate that these PRs require additional revisions or prompt more discussion regarding the correctness and suitability of the proposed changes.
Overall, the higher volume of comments observed for agentic PRs suggests that reviewers engage more actively with these contributions than with human generated PRs. Consistent with prior work~\cite{morovati2024bug}, this increased discussion may reflect several factors, including the quality of the generated changes, the complexity of the underlying tasks, the need for clarification or refinement, and the level of reviewer confidence in the proposed modifications.

\paragraph{\textbf{Number of Changed LOC}}

The number of changed lines of code (LOC) is another commonly used indicator of change complexity and development effort, and has been shown to influence the mergeability of PRs~\cite{rastogi2018relationship}. As illustrated in Fig.~\ref{subfig:metric_LOC}, no uniform pattern emerges when comparing the number of changed LOC in agentic and human generated PRs across different coding agents.
According to the results of the statistical analysis (Table~\ref{tab:Statistical_tests_comparing_AM-PRs_of_different_agents_across_metrics}), PRs generated by Claude and devin modify significantly more lines of code, on average, than those generated by human developers. One possible explanation is that these agents are associated with broader scope code changes that require modifications across larger portions of the codebase. Given that change size is frequently considered a proxy for implementation effort and change complexity~\cite{rastogi2018relationship}, this finding is consistent with our earlier observations suggesting that Claude generated PRs are often associated with development activities involving more extensive code modifications.
In contrast, PRs generated by Copilot exhibit significantly smaller code changes than their human generated counterparts. This result may indicate that these agents are more frequently employed for localized modifications or narrowly defined development tasks. Finally, the number of changed LOC in PRs generated by Cursor does not differ significantly from that of human generated PRs (Cliff's $\delta$ = 0.375).

By comparing agentic PRs, the results of the statistical analysis reported in Table~\ref{tab:Statistical_tests_comparing_AM-PRs_of_different_agents_across_metrics} indicate that PRs generated by Claude modify significantly more LOC, on average, than those generated by the other studied agents. One possible explanation is that Claude is more frequently associated with development activities involving more complicated code changes, which typically require modifications to larger portions of the code. This interpretation is consistent with our earlier findings showing that Claude generated PRs are often linked to development tasks characterized by higher change complexity.
In contrast, the differences in the number of changed LOC among the remaining agents are either statistically insignificant or associated with negligible effect sizes. This suggests that, despite minor variations, the overall scope of code changes proposed by these agents is largely comparable.

\begin{figure}[]
    \centering

    \begin{subfigure}{0.30\textwidth}
        \centering
        \includegraphics[width=\linewidth]{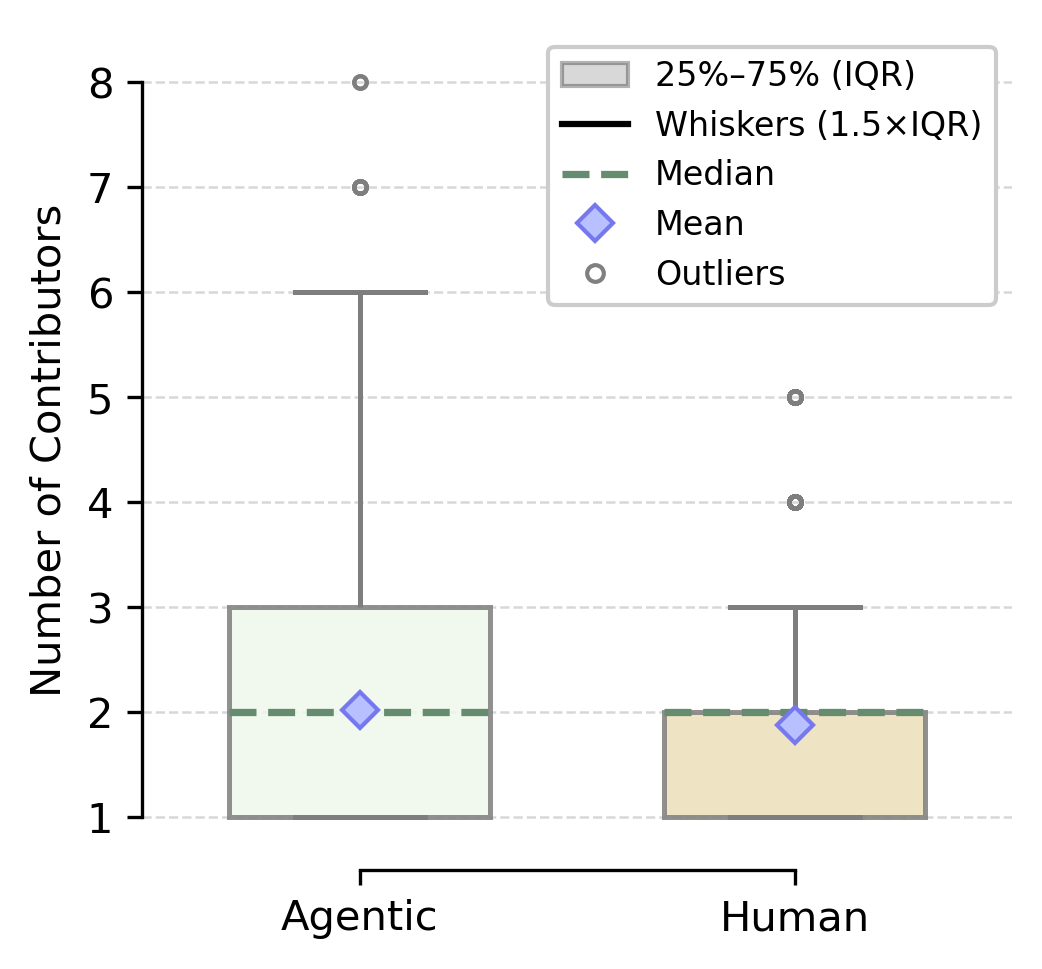}
        \caption{Num of Contributors}
    \end{subfigure}
    \hfill
    \begin{subfigure}{0.30\textwidth}
        \centering
        \includegraphics[width=\linewidth]{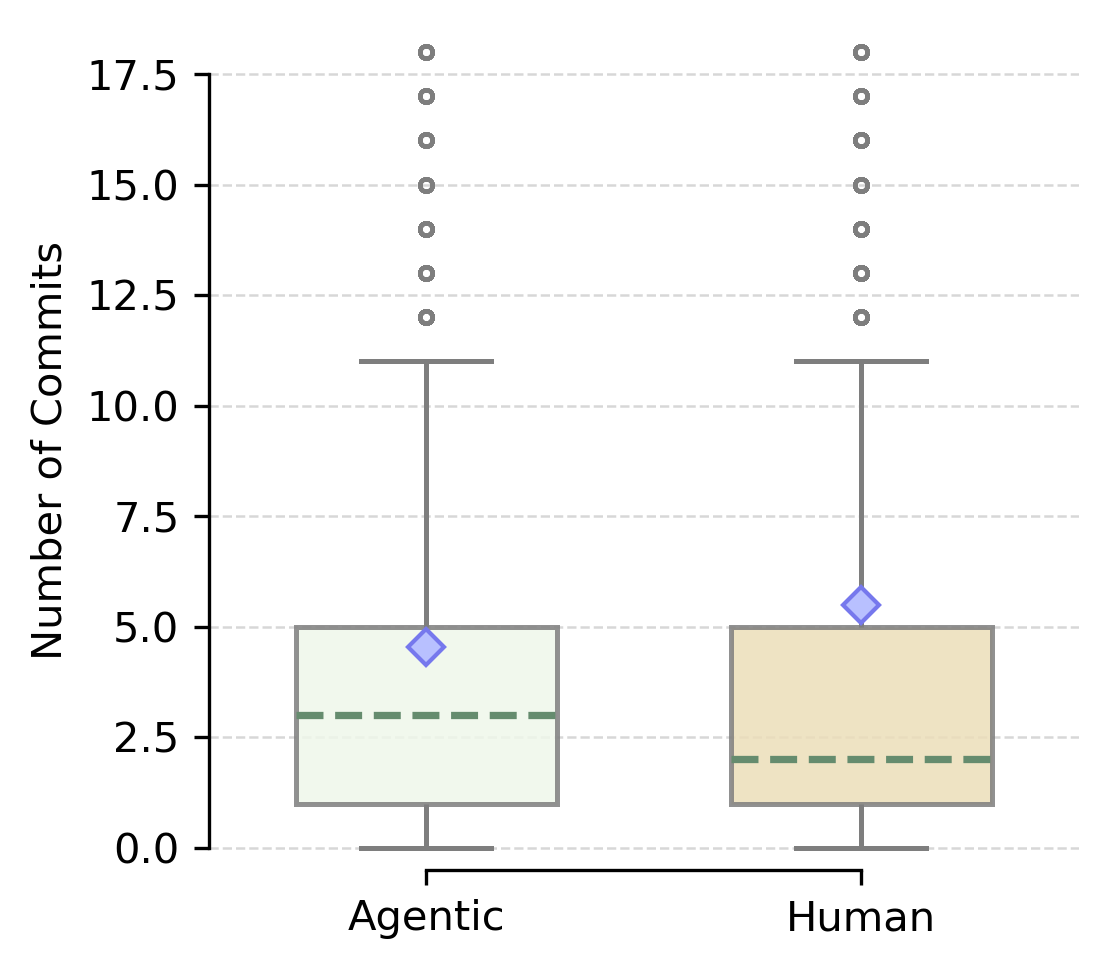}
        \caption{Num of Commits}
    \end{subfigure}
    \hfill
    \begin{subfigure}{0.30\textwidth}
        \centering
        \includegraphics[width=\linewidth]{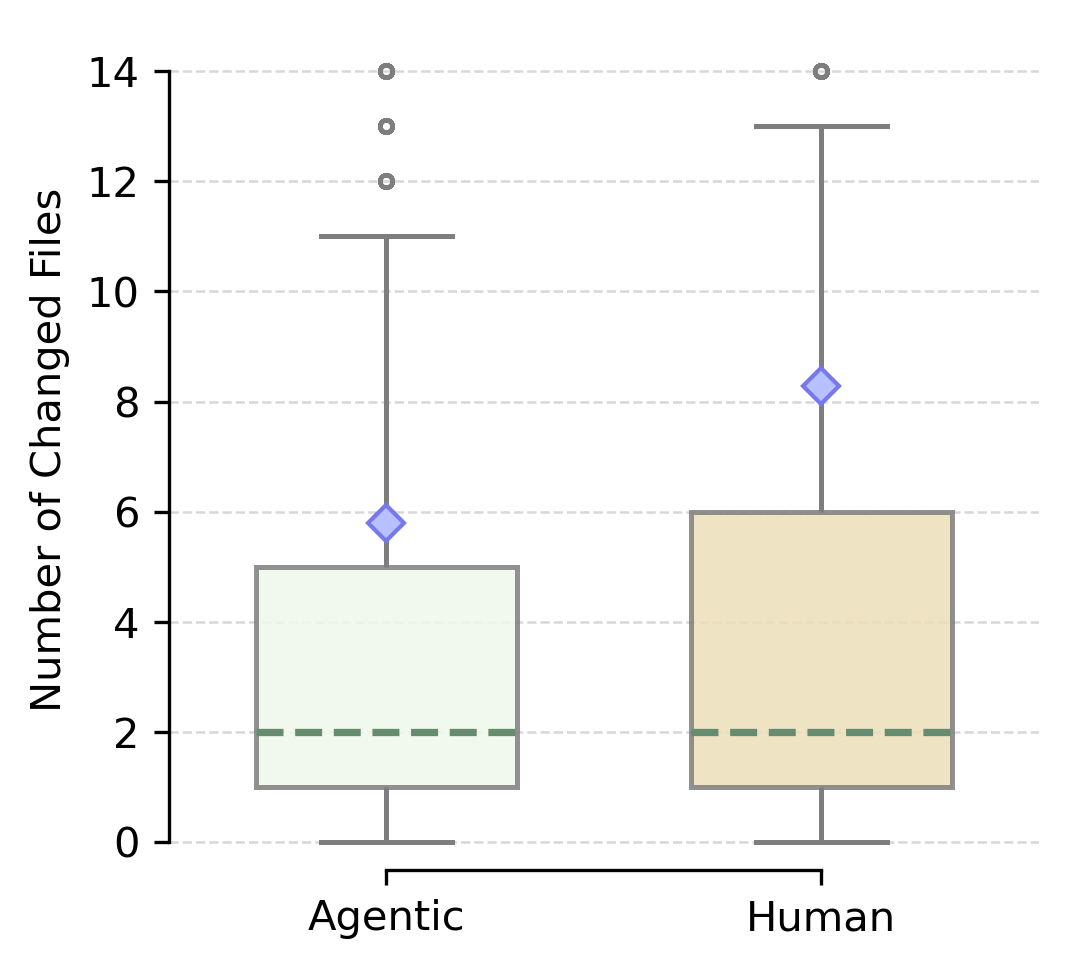}
        \caption{Num of Changed Files}
    \end{subfigure}
    \vspace{0.5cm}

    \begin{subfigure}{0.30\textwidth}
        \centering
        \includegraphics[width=\linewidth]{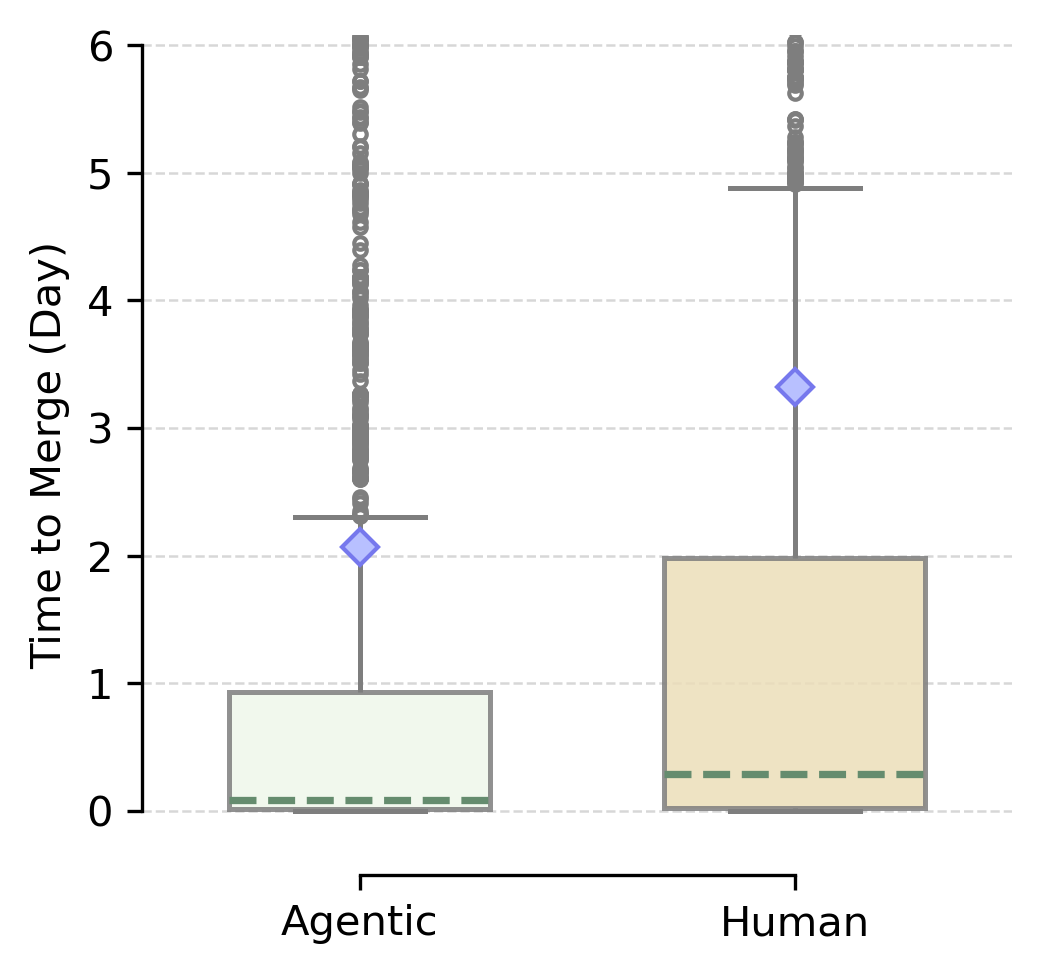}
        \caption{Time to Merge}
        \label{subfig:agent_metric_mergeTime}
    \end{subfigure}
    \hfill
    \begin{subfigure}{0.30\textwidth}
        \centering
        \includegraphics[width=\linewidth]{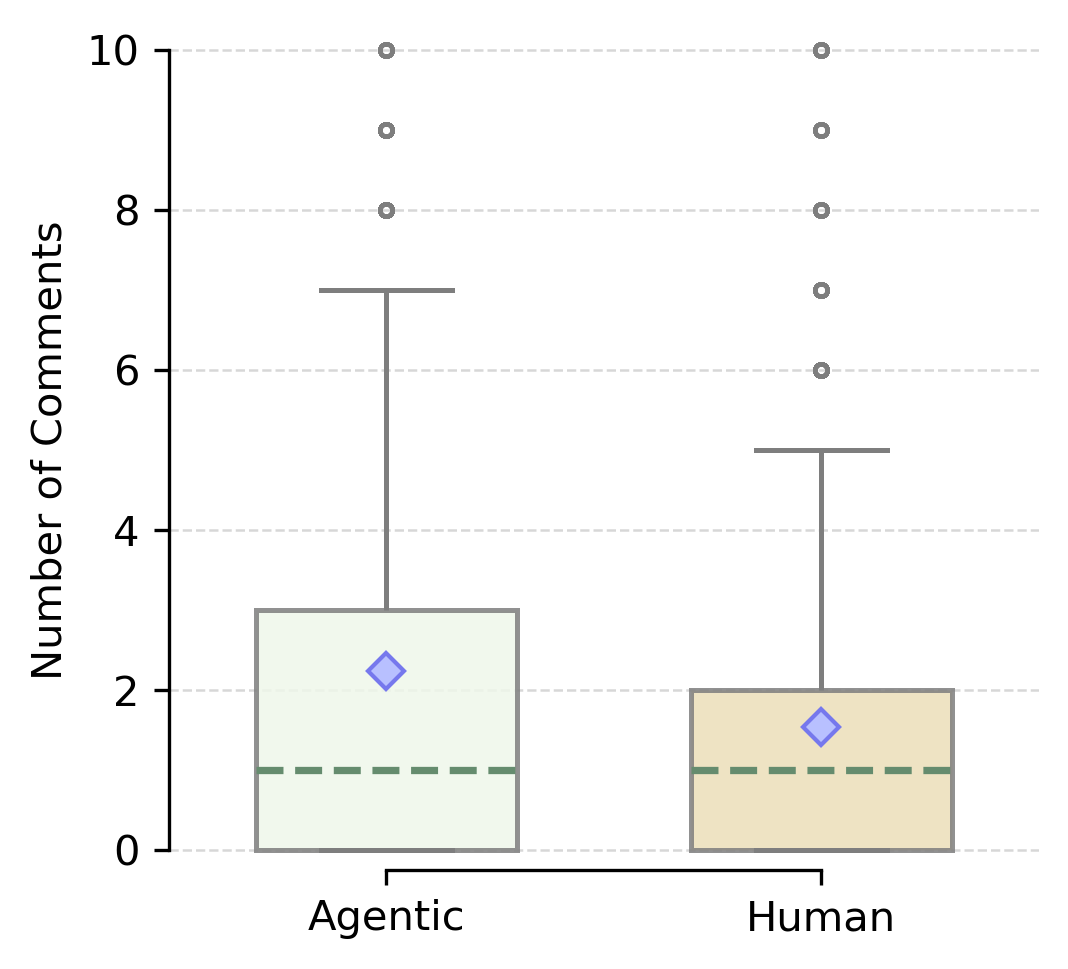}
        \caption{Num of Comments}
        \label{subfig:agent_metric_comment}
    \end{subfigure}
    \hfill
    \begin{subfigure}{0.30\textwidth}
        \centering
        \includegraphics[width=\linewidth]{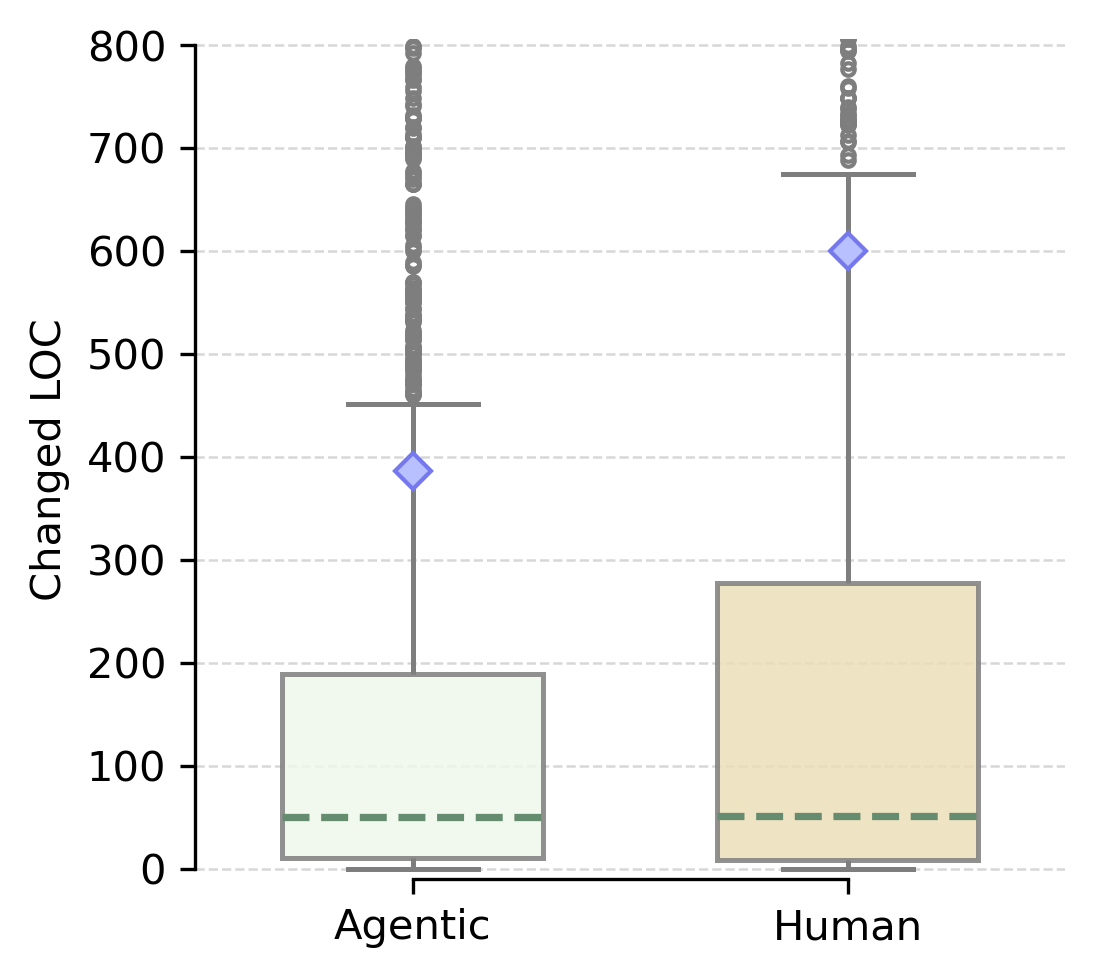}
        \caption{Num of Changed LOC}
    \end{subfigure}

    \caption{Complexity metrics comparing aggregated merged agentic PRs with human generated PRs}
    \label{fig:Compexity_metric_APRs_vs_HPRs_all_agent}
\end{figure}

\subsubsection{Overall Complexity Analysis}

We aggregate agentic and human generated PRs across all agents and compare the two groups without controlling for the effect of individual agents. 
Figure \ref{fig:Compexity_metric_APRs_vs_HPRs_all_agent} presents the boxplot distributions of the evaluated complexity metrics comparing all agentic PRs with all human PRs. Table \ref{tab:Statistical_tests_comparing_Overall_AM-PRs_and_HM-PRs_across_metrics} also summarizes the 
result of \textit{Mann–Whitney U} test ~\cite{arcuri2014hitchhiker} and Cliff's $\delta$ ~\cite{macbeth2011cliff} as the effect size of the comparison between aggregated agentic PRs and human PRs across the evaluated metrics.
Across five complexity metrics (Number of Contributors, Number of Commits, Number of Changed Files, Time to Merge, Number of Comments), we observe statistically significant differences between agentic PRs and human PRs, although the magnitude of these differences is consistently negligible according to related effect sizes.

\begin{table}[]
\centering
\caption{Statistical tests comparing overall AM-PRs and HM-PRs across metrics}
\scriptsize
\setlength{\tabcolsep}{5pt}

\begin{tabular}{llll}
\toprule
Metric & \textit{p}-value & Cliff's $\delta$  \\
\midrule
Num of Contributors  & \textbf{<1e-3} & 0.075 (n)  \\
Num of Commits & \textbf{<1e-3} & 0.060 (n)  \\
Num of Changed Files  & \textbf{<1e-3} & -0.066 (n)  \\
Time to Merge   & \textbf{<1e-3} & -0.146 (n)  \\
Num of Comments   & \textbf{<1e-3} & 0.137 (n)  \\
Num of Changed LOC   & 0.296 & -0.018 (n)  \\
\bottomrule
\end{tabular}

\label{tab:Statistical_tests_comparing_Overall_AM-PRs_and_HM-PRs_across_metrics}
\end{table}

The largest differences are observed for the \textit{time to merge} and \textit{number of comments} metrics. With respect to \textit{time to merge}, Fig.~\ref{subfig:agent_metric_mergeTime} shows that agentic PRs tend to be merged slightly faster, on average, than human-generated PRs. One possible explanation is that agentic PRs are often associated with more structured and well-defined development tasks, which may require less effort to review, validate, and integrate. In addition, developers may devote greater attention to the validation of agentic contributions, enabling potential issues to be identified and resolved earlier in the review process. Another plausible explanation is that agentic PRs are typically narrower in scope, making them easier to assess and merge than broader human-generated changes.
Nevertheless, given the relatively small magnitude of the observed differences, these findings should be interpreted as indicating a modest tendency toward faster integration of agent-generated PRs rather than a substantial reduction in review effort.
Regarding the \textit{Number of Comments} metric, as shown in Fig. \ref{subfig:agent_metric_comment}, both groups share an identical median of 1 comment, while agentic PRs exhibit a higher mean (2.24 vs. 1.54) and a larger maximum value (90 vs. 29), yet close faster, potentially reflecting the well-scoped, more self-contained scope of agentic changes ~\cite{farrag2026productivity}. The higher average appears to be driven by a small subset of highly commented agentic PRs. 

\subsubsection{Overall Complexity per Quarter}

\begin{figure}[]
    \centering
   
    \begin{subfigure}{0.30\textwidth}
        \centering
        \includegraphics[width=\linewidth]{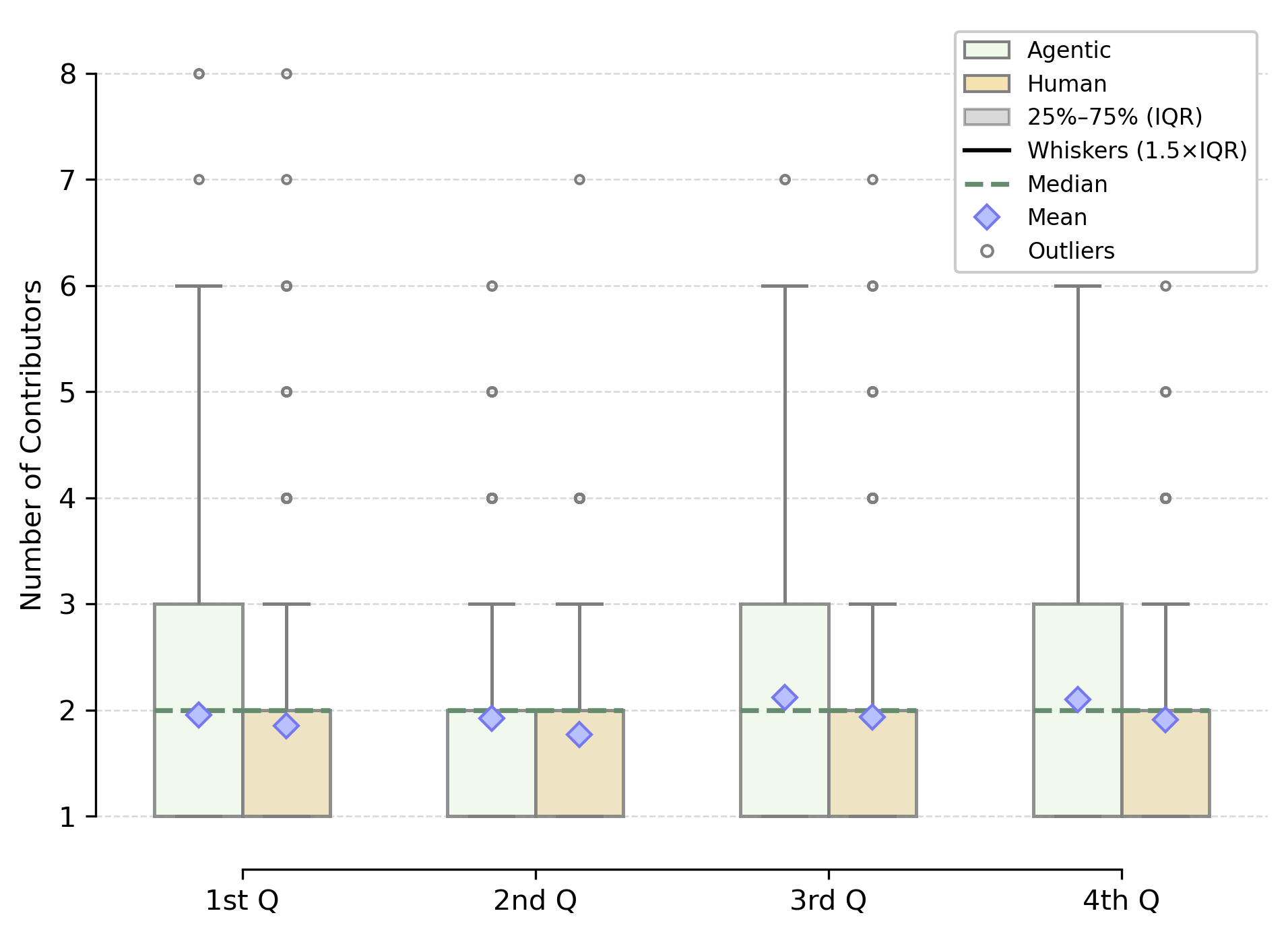}
        \caption{Num of Contributors}
        \label{subfig:agg_contribute}
    \end{subfigure}
    \hfill
    \begin{subfigure}{0.30\textwidth}
        \centering
        \includegraphics[width=\linewidth]{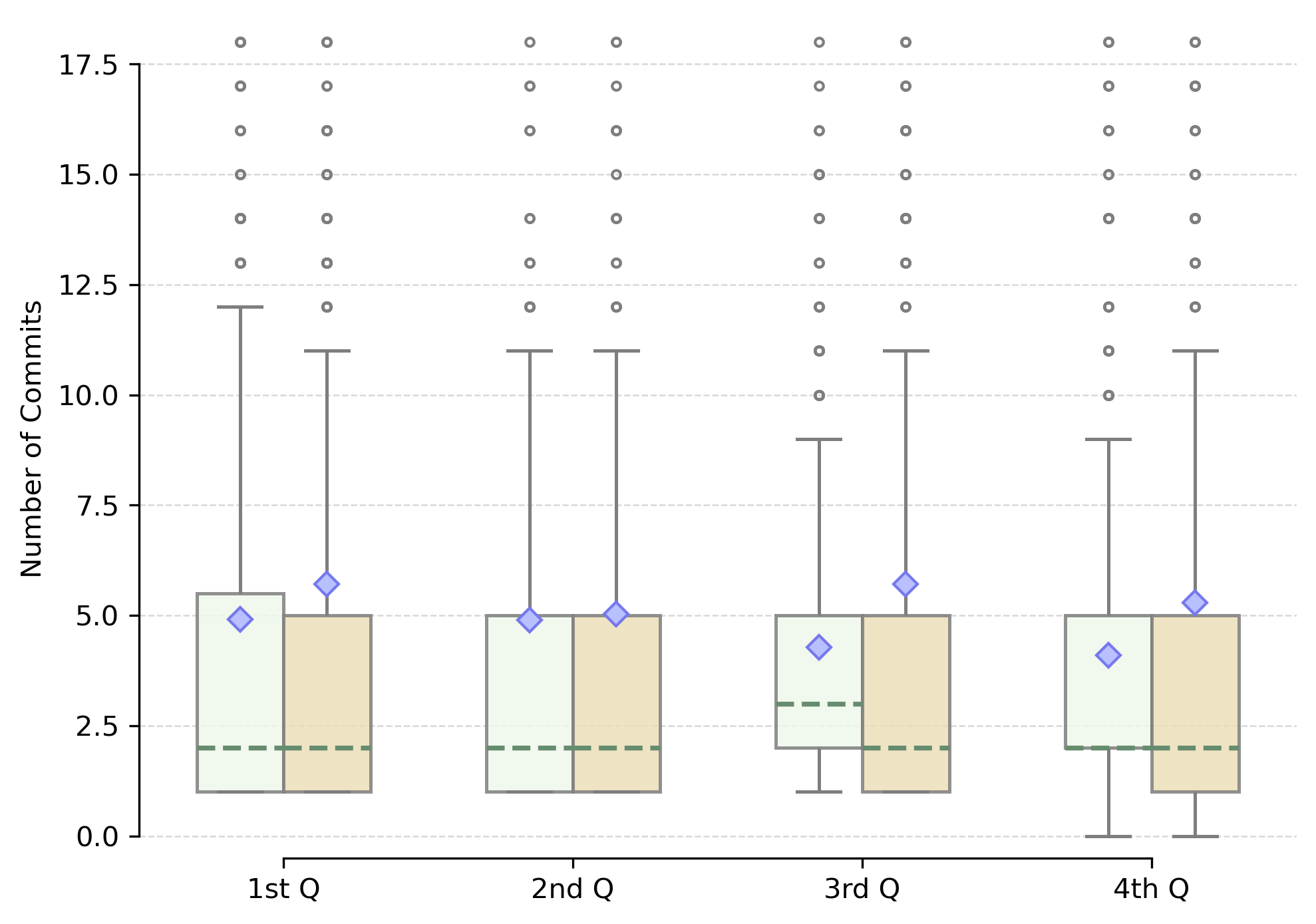}
        \caption{Num of Commits}
        \label{subfig:agg_commits}
    \end{subfigure}
    \hfill
    \begin{subfigure}{0.30\textwidth}
        \centering
        \includegraphics[width=\linewidth]{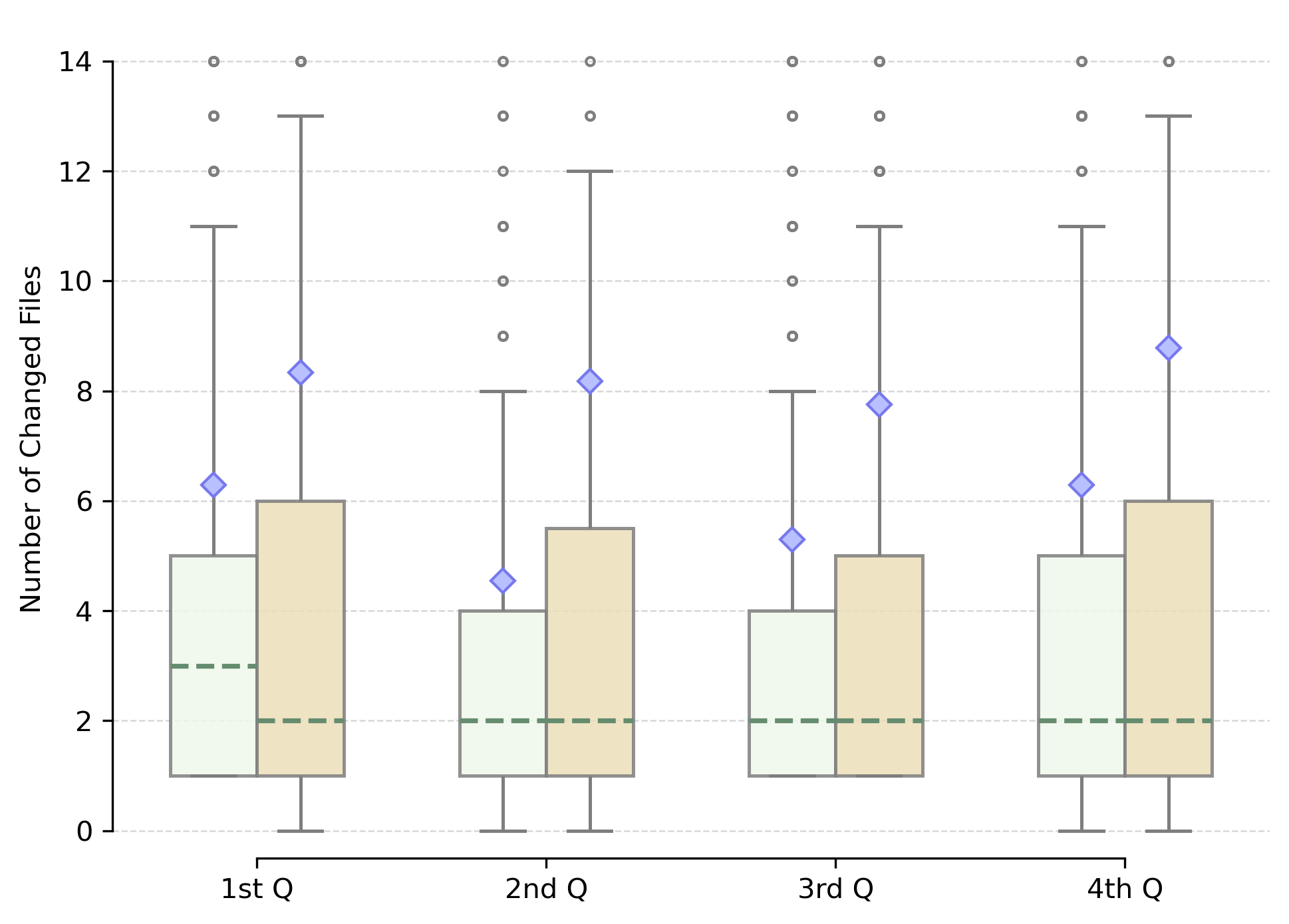}
        \caption{Num of Changed Files}
        \label{subfug:agg_changedFiles}
    \end{subfigure}
    \vspace{0.5cm}
    \begin{subfigure}{0.30\textwidth}
        \centering
        \includegraphics[width=\linewidth]{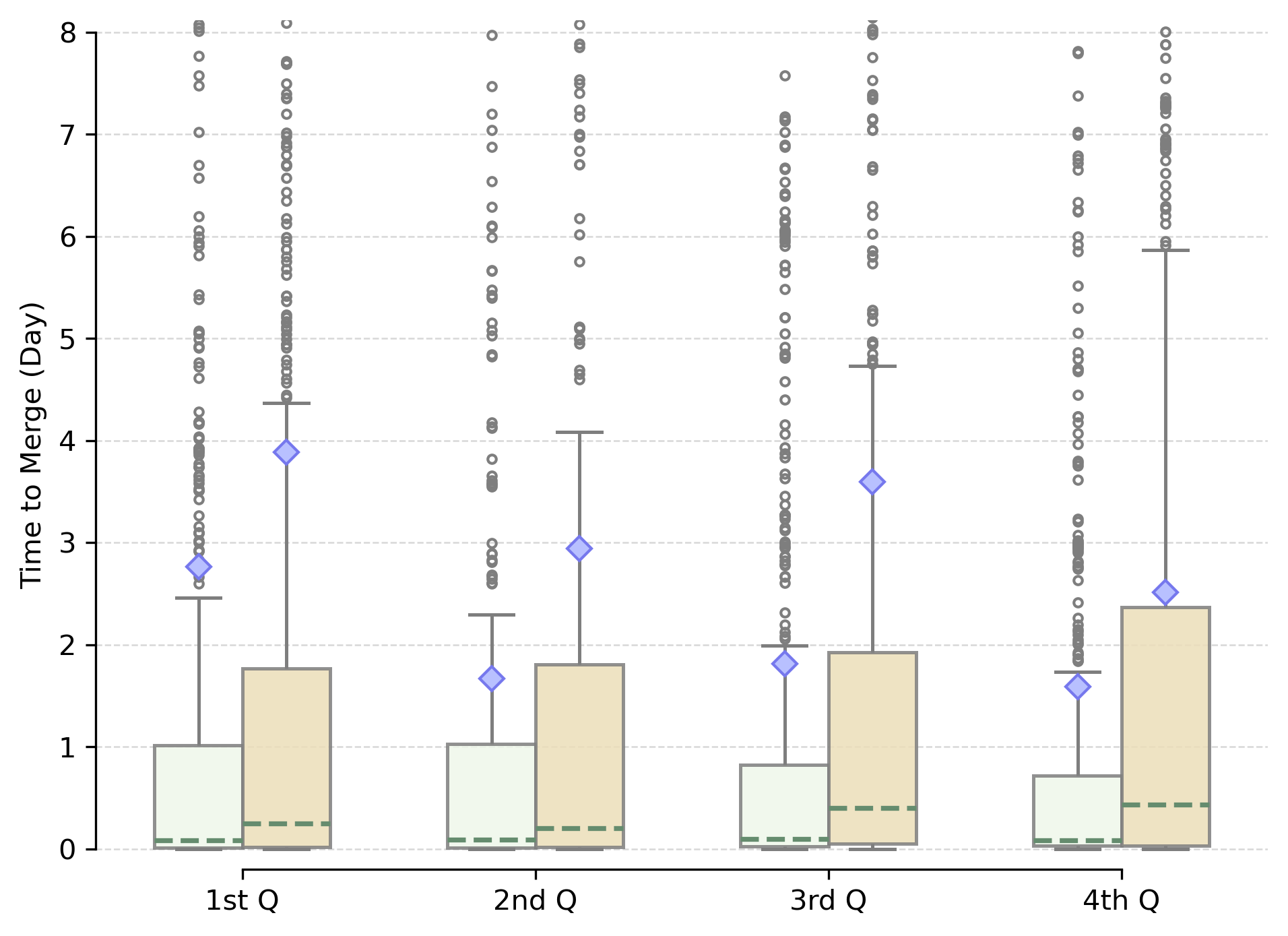}
        \caption{Time to Merge}
        \label{subfug:agg_mergeTime}
    \end{subfigure}
    \hfill
    \begin{subfigure}{0.30\textwidth}
        \centering
        \includegraphics[width=\linewidth]{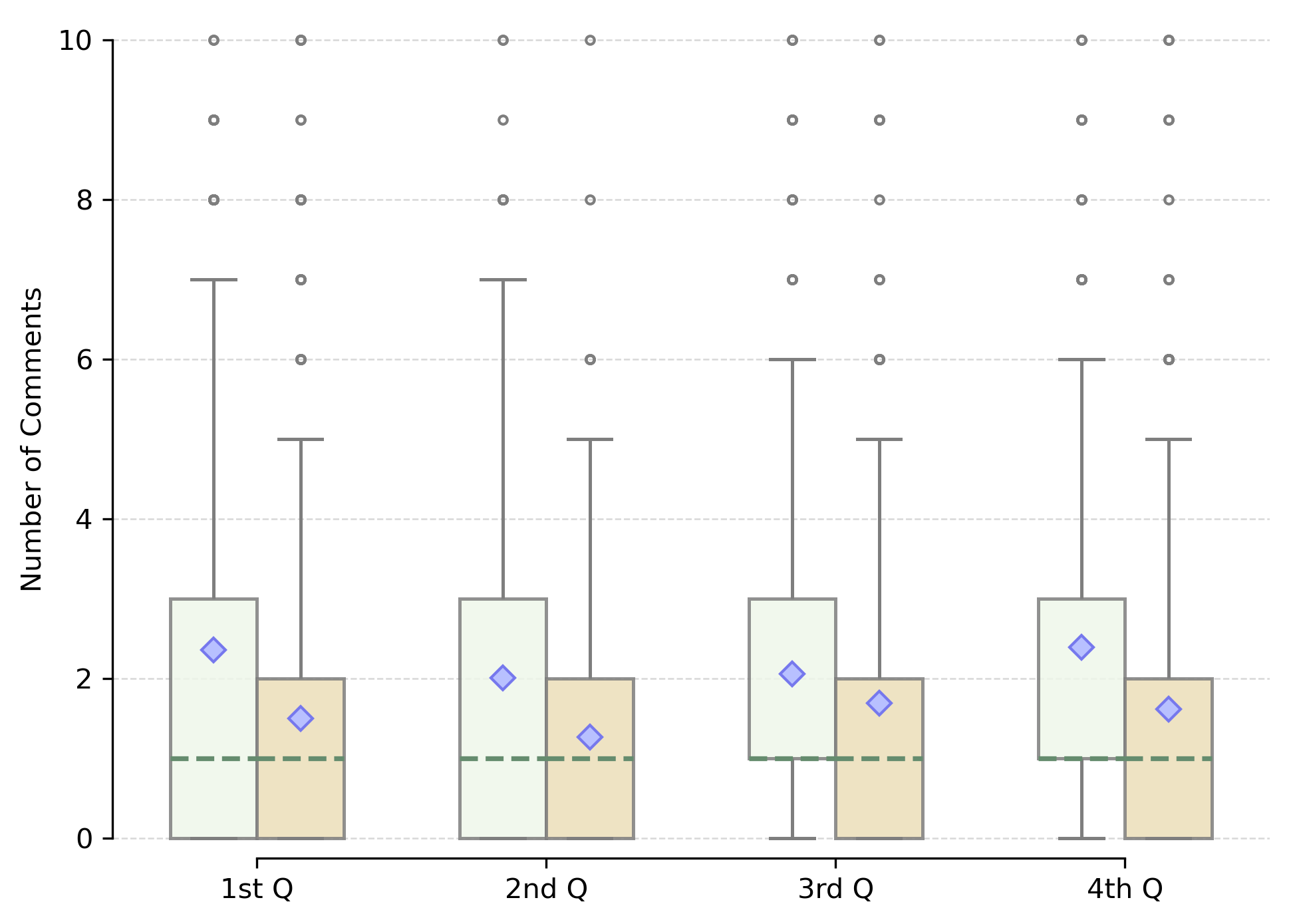}
        \caption{Num of Comments}
        \label{subfig:agg_commets}
    \end{subfigure}
    \hfill
    \begin{subfigure}{0.30\textwidth}
        \centering
        \includegraphics[width=\linewidth]{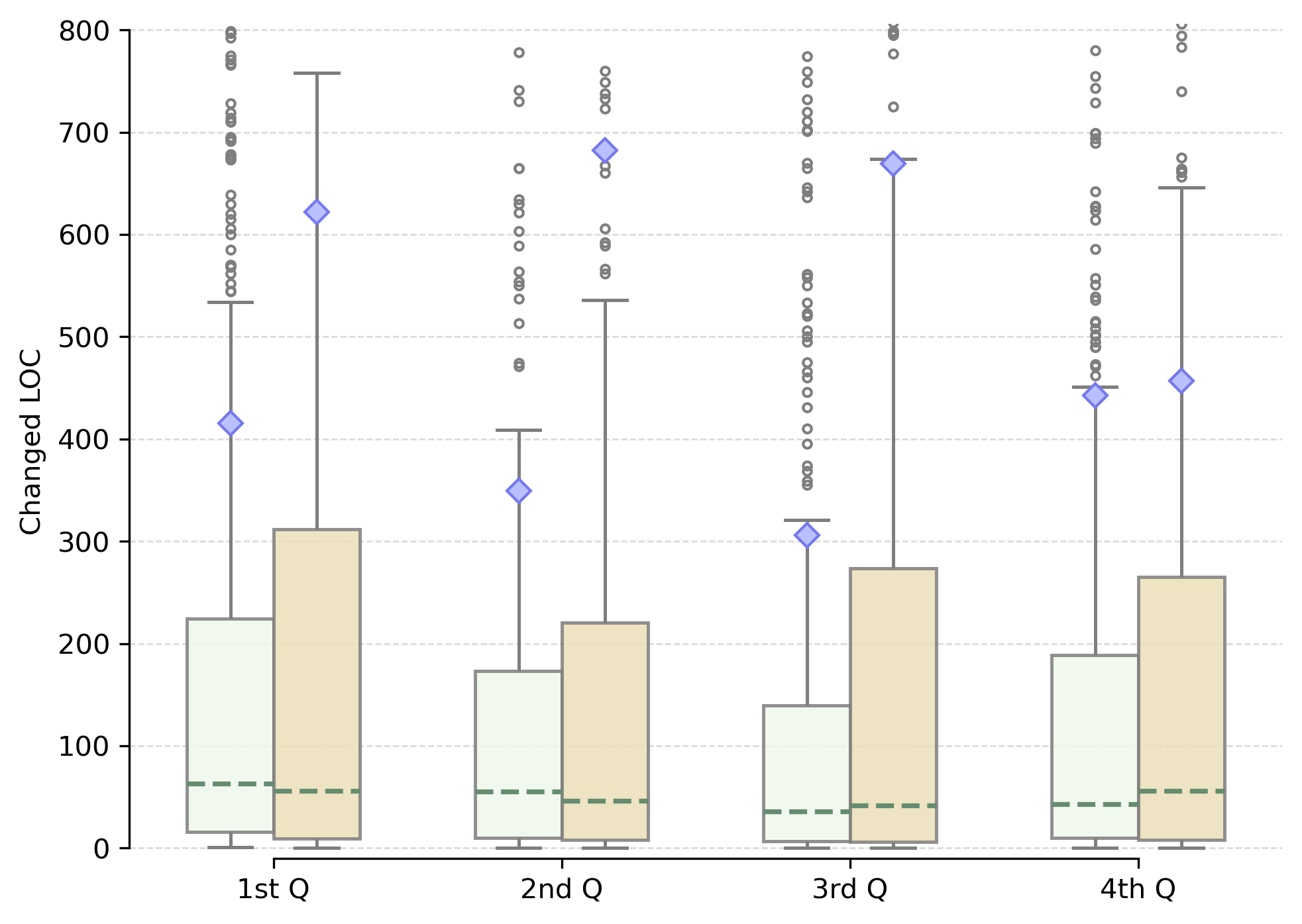}
        \caption{Num of Changed LOC}
        \label{subfig:agg_loc}
    \end{subfigure}

    \caption{Complexity metrics comparing aggregated merged agentic PRs with human generated PRs per quarter} 
    \label{fig:Compexity_metric_APRs_vs_HPRs_all_agent_per_quarter}
\end{figure}

To examine how the complexity of agentic PRs has evolved relative to human generated PRs over time, we compare the aggregated agentic PRs with the aggregated human generated PRs across different development quarters. Fig.~\ref{fig:Compexity_metric_APRs_vs_HPRs_all_agent_per_quarter} presents the quarterly comparison of agentic and human generated PRs with respect to the six complexity metrics.
To assess whether the observed differences are statistically significant, we employ the \textit{Mann–Whitney U} test ~\cite{arcuri2014hitchhiker} and use Cliff's $\delta$ ~\cite{macbeth2011cliff} to quantify the effect size. Table~\ref{tab:Statistical_tests_comparing_AM-PRs_and_HM-PRs_across_metrics_quarter} summarizes the results of the statistical comparisons between agentic and human generated PRs for each quarter. In addition, Table~\ref{tab:Statistical_tests comparing_AM_PRs_in_different_quarters_across_metrics} reports the results of pairwise comparisons among different quarters of agentic PRs, enabling us to investigate how the characteristics of agentic contributions have changed over time.

\paragraph{\textbf{Number of Contributors}}
As shown in Fig.~\ref{subfig:agg_contribute}, agentic PRs consistently involve slightly more contributors than human generated PRs across all quarters. According to the statistical test results reported in Table~\ref{tab:Statistical_tests_comparing_AM-PRs_and_HM-PRs_across_metrics_quarter}, no statistically significant difference is observed during the first quarter. However, statistically significant differences emerge from the second quarter onward, with the strongest level of significance observed in the fourth quarter.
Despite these statistically significant results, the corresponding effect sizes remain below 0.11 in all cases, indicating negligible practical differences between agentic and human generated PRs. Therefore, while agentic PRs appear to involve slightly broader collaboration, the magnitude of this difference is limited and unlikely to be practically meaningful.
A comparison of agentic PRs across quarters further indicates that the number of contributors remains largely stable over time. Although a statistically significant difference is observed between the second and third quarters (Table~\ref{tab:Statistical_tests comparing_AM_PRs_in_different_quarters_across_metrics}), the associated effect size is negligible (Cliff's $\delta=-0.115$). This result suggests that the contributor count of agentic PRs remains relatively consistent throughout the studied period, with only minor fluctuations.

\paragraph{\textbf{Number of Commits}}

With respect to the number of commits, Fig.~\ref{subfig:agg_commits} indicates only limited differences between agentic  and human generated PRs across the examined quarters. According to the statistical test results presented in Table~\ref{tab:Statistical_tests_comparing_AM-PRs_and_HM-PRs_across_metrics_quarter}, statistically significant differences are observed only during the third and fourth development quarters. However, the corresponding effect sizes are negligible, suggesting that the practical significance of these differences is limited.
Nevertheless, the results indicate a slight tendency for agentic PRs to contain more commits than human generated PRs during the second half of the development period. One possible explanation is that, as developers gain confidence in agentic changes and agents become better integrated into the development workflow, developers may increasingly employ these agents for more complex tasks that require higher level of commit activity in later stages of the project development. 
However, given the negligible effect sizes, this interpretation should be treated as a tentative observation rather than strong evidence of a meaningful trend.
Comparing agentic PRs across different quarters reveals a largely stable pattern in the number of commits. The only statistically significant difference is observed between the second and third quarters (Table~\ref{tab:Statistical_tests comparing_AM_PRs_in_different_quarters_across_metrics}). Yet, the associated effect size is also negligible, indicating that the observed difference has little practical importance. Overall, these findings suggest that commit activity within agentic PRs remains relatively stable throughout the project lifecycle, with only minor fluctuations over time.

\paragraph{\textbf{Number of Changed Files}}
Fig.~\ref{subfug:agg_changedFiles} compares agentic and human generated PRs with respect to the number of files modified across different development quarters. As shown in the figure, agentic PRs consistently modify fewer files, on average, than human generated PRs throughout all quarters. According to the statistical test results reported in Table~\ref{tab:Statistical_tests_comparing_AM-PRs_and_HM-PRs_across_metrics_quarter}, the differences are statistically significant in all quarters except the first. However, the corresponding effect sizes are negligible (Cliff's $\delta \leq 0.144$), indicating that the practical magnitude of these differences is limited.
These results suggest that agentic PRs tend to be slightly more localized in terms of file modifications, potentially reflecting a tendency to address more focused development tasks. Nevertheless, the negligible effect sizes indicate that the practical distinction between agentic and human generated PRs remains small.
When comparing agentic PRs across quarters, Table~\ref{tab:Statistical_tests comparing_AM_PRs_in_different_quarters_across_metrics} shows statistically significant differences between the first two quarters and the last two quarters (\textit{p}-value < 0.05). However, the associated Cliff's $\delta$ values remain negligible, suggesting that these differences have little practical importance. Overall, the results indicate minor variations in the scope of agentic PRs over time, but no substantial changes in the extent of file modifications throughout the project lifecycle.

\begin{table}[t]
\centering
\caption{Statistical tests comparing agentic and human generated PRs across metrics per quarter (n = negligible, s = small, m = medium, l = large)}
\scriptsize
\setlength{\tabcolsep}{4pt}
\begin{tabular}{lcccccccccccc}
\toprule
\multirow{2}{*}{\textbf{Quarter}} &
\multicolumn{2}{c}{\textbf{Contributors}} &
\multicolumn{2}{c}{\textbf{Commits}} &
\multicolumn{2}{c}{\textbf{Changed Files}} &
\multicolumn{2}{c}{\textbf{Time to Merge}} &
\multicolumn{2}{c}{\textbf{Comments}} &
\multicolumn{2}{c}{\textbf{Changed LOC}} \\
\cmidrule(lr){2-3}
\cmidrule(lr){4-5}
\cmidrule(lr){6-7}
\cmidrule(lr){8-9}
\cmidrule(lr){10-11}
\cmidrule(lr){12-13}
&
\textit{p}-value & Cliff's $\delta$ &
\textit{p}-value & Cliff's $\delta$ &
\textit{p}-value & Cliff's $\delta$ &
\textit{p}-value & Cliff's $\delta$ &
\textit{p}-value & Cliff's $\delta$ &
\textit{p}-value & Cliff's $\delta$ \\
\midrule
1st  & 0.102    & 0.044(n)   & 0.333    & 0.027 (n)   & 0.846    & -0.005 (n)   & \textbf{<1e-3}    & -0.123 (n)   & \textbf{<1e-3}     & 0.114 (n)   & 0.424    & 0.023 (n)   \\
2nd  & \textbf{0.037}    & 0.085 (n)   & 0.337    & 0.041 (n)   &  \textbf{0.046}  & -0.085 (n)   & \textbf{0.011}    & -0.118 (n)   & \textbf{<1e-3}    & 0.176 (s)   & 0.945    & -0.003 (n)   \\
3rd  & \textbf{0.008}    & 0.088 (n)   & \textbf{0.009}    & 0.091 (n)   & \textbf{<1e-3}     & -0.144 (n)   & \textbf{<1e-3}    & -0.171 (s)   & \textbf{<1e-3}    & 0.135 (n)   & \textbf{0.023}    & -0.080 (n)   \\
4th  & \textbf{<1e-3} & 0.110 (n)   & \textbf{0.005} & 0.093 (n)   & \textbf{0.044} & -0.068 (n)   & \textbf{<1e-3} & -0.183 (s)   & \textbf{<1e-3} & 0.145 (n)   & 0.460 & -0.025 (n)   \\

\bottomrule
\end{tabular}

\label{tab:Statistical_tests_comparing_AM-PRs_and_HM-PRs_across_metrics_quarter}
\end{table}

\paragraph{\textbf{Time to Merge}}
As illustrated in Fig.~\ref{subfug:agg_mergeTime}, agentic PRs consistently require less time to be merged than human generated PRs across all development quarters. The statistical test results reported in Table~\ref{tab:Statistical_tests_comparing_AM-PRs_and_HM-PRs_across_metrics_quarter} further indicate that these differences are statistically significant throughout the entire study period. However, the corresponding effect sizes suggest that the differences are practically meaningful only during the third and fourth quarters, where small effect sizes are observed.
One possible explanation is that, over time, both developers and AI agents become better integrated into the development workflow. As developers gain experience in effectively prompting and utilizing AI agents, and as agents are increasingly applied to tasks that align with their capabilities, the resulting PRs may require fewer revisions and less review effort before integration. Another plausible explanation is that developers become more familiar with evaluating agentic contributions, enabling them to review and merge such changes more rapidly.
When comparing merge duration across different quarters of agentic PRs, the results indicate a largely stable pattern over time. As shown in Table~\ref{tab:Statistical_tests comparing_AM_PRs_in_different_quarters_across_metrics}, no statistically significant differences are observed between quarters (\textit{p}-value > 0.05). This finding suggests that, despite the growing adoption of agentic development practices, the time required for agentic PRs to be merged remains relatively consistent throughout the project lifecycle. 

\paragraph{\textbf{Number of Comments}}

As shown in Fig.~\ref{subfig:agg_commets}, agenticPRs receive a higher average number of comments than human generated PRs across all development quarters. 
Furthermore, the statistical test results presented in Table~\ref{tab:Statistical_tests_comparing_AM-PRs_and_HM-PRs_across_metrics_quarter} indicate that these differences are statistically significant in every quarter. However, only the difference observed in the second quarter reaches practical significance, with a small effect size.
One possible explanation is that the complexity of tasks assigned to agents increases as projects evolve, leading to more extensive discussions and revisions during the review process. This trend appears to peak in the second quarter, where the largest difference between agentic and human generated PRs is observed. As development progresses, contributors may develop a better understanding of the capabilities and limitations of AI agents, while agents may be employed in contexts where their strengths are better understood. This increased familiarity could reduce the need for extensive review discussions, contributing to the decline in the number of comments observed during the third and fourth quarters.
In addition, Table~\ref{tab:Statistical_tests comparing_AM_PRs_in_different_quarters_across_metrics} shows that comparisons of agentic PRs' comment counts across development quarters reveal a statistically significant difference only between the second and third quarters. Nevertheless, the corresponding effect size is negligible, indicating that the practical magnitude of this difference is very small. Therefore, although the average number of comments exhibits some fluctuations throughout the project lifecycle, these variations are not practically meaningful. Overall, the results suggest that agentic PRs consistently attract a relatively stable level of review discussion across different development stages.

\paragraph{\textbf{Number of Changed LOC}}

As shown in Fig.~\ref{subfig:agg_loc}, agentic PRs modify fewer lines of code (LOC), on average, than human generated PRs across all development quarters. However, the statistical test results reported in Table~\ref{tab:Statistical_tests_comparing_AM-PRs_and_HM-PRs_across_metrics_quarter} indicate that a statistically significant difference is observed only in the third quarter (\textit{p}-value = 0.023). Importantly, the corresponding effect size is negligible, suggesting that the practical difference between agentic and human generated PRs with respect to changed LOC is minimal. Therefore, although a statistical difference is detected, its magnitude is too small to indicate a meaningful distinction in practice.
Furthermore, the comparison of agentic PRs across various development quarters, reveals statistically significant differences between all pairs of quarters (presented in Table~\ref{tab:Statistical_tests comparing_AM_PRs_in_different_quarters_across_metrics}). Nevertheless, all corresponding effect sizes are negligible, indicating that the observed variations are minor from a practical perspective. Taken together, these results suggest that the amount of code modified by agentic PRs remains relatively stable throughout the project lifecycle.
A closer examination of the quarterly trends shows a modest increase in the number of changed LOC during the first three quarters, followed by a slight decline in the fourth quarter. One possible explanation is that, as projects evolve, contributors gradually assign agents to tasks involving broader code modifications, leading to a moderate increase in the size of agentic changes. The subsequent decline in the final quarter may reflect a shift in development priorities from feature implementation toward refinement, maintenance, and quality improvement activities, which often require smaller and more targeted code modifications. This interpretation is consistent with the typical evolution of software projects, where development activity transitions from functionality expansion to stabilization and optimization as the project mature~\cite{bennett2000software}.

\begin{tcolorbox}[colback=blue!10!white, colframe=blue!75!black]

\textbf{Finding 6: }
Comparisons of PRs characteristics at agent level reveals no consistent differences between various agents, with meaningful effects limited to specific agents and metrics. At the agents aggregated level, several metrics (contributors, commits, changed files, time to merge, and comments) differ statistically significantly between the agentic and human generated PRs, but with negligible effect sizes. Across development quarters, agentic PRs generally tend to be merged faster and receive more comments than human generated PRs.

\end{tcolorbox}

\begin{table}[t]
\centering
\caption{Statistical tests comparing agentic PRs in different quarters across metrics (n = negligible, s = small, m = medium, l = large)}
\scriptsize
\setlength{\tabcolsep}{4pt}
\begin{tabular}{lcccccccccccc}
\toprule
\multirow{2}{*}{\textbf{Quarters}} &
\multicolumn{2}{c}{\textbf{Contributors}} &
\multicolumn{2}{c}{\textbf{Commits}} &
\multicolumn{2}{c}{\textbf{Changed Files}} &
\multicolumn{2}{c}{\textbf{Time to Merge}} &
\multicolumn{2}{c}{\textbf{Comments}} &
\multicolumn{2}{c}{\textbf{Changed LOC}} \\
\cmidrule(lr){2-3}
\cmidrule(lr){4-5}
\cmidrule(lr){6-7}
\cmidrule(lr){8-9}
\cmidrule(lr){10-11}
\cmidrule(lr){12-13}
&
\textit{p}-value & Cliff's $\delta$ &
\textit{p}-value & Cliff's $\delta$ &
\textit{p}-value & Cliff's $\delta$ &
\textit{p}-value & Cliff's $\delta$ &
\textit{p}-value & Cliff's $\delta$ &
\textit{p}-value & Cliff's $\delta$ \\
\midrule
1st vs 2nd   & 0.853   & 0.007 (n)   & 0.992   & -0.000 (n)   & \textbf{0.006}   & 0.099 (n)    & 0.740   & -0.012 (n)   & 0.677   & -0.015 (n)   & \textbf{0.025}   & 0.083 (n)  \\
2nd vs 3rd    & \textbf{0.002}   & -0.115 (n)   & \textbf{0.032}   & -0.084 (n)   & 0.141   & 0.057 (n)    & 0.078   & -0.070 (n)   & \textbf{0.043}   & -0.078 (n)   & \textbf{0.044}   & 0.080 (n)  \\
3rd vs 4th   & 0.849    & -0.006 (n)   & 0.076    & 0.060 (n)   & \textbf{0.005}    & -0.094 (n)    & 0.321    & 0.034 (n)   & 0.393    & -0.029 (n)   & \textbf{0.015}    & -0.083 (n)  \\

\bottomrule
\end{tabular}

\label{tab:Statistical_tests comparing_AM_PRs_in_different_quarters_across_metrics}
\end{table}

\subsubsection{Bug Inducing Analysis of Agentic and Human Generated PRs}

Bug-inducing commits are commits whose changes subsequently lead to software defects later in the software lifecycle~\cite{tang2024enhancing}. 
In this section, we compare agentic and human generated PRs, with respect to their association with bug-inducing commits. Specifically, we investigate the extent to which contributions produced by AI agents differ from human contributions in their potential impact on the long-term defect proneness of software systems.
To conduct this analysis, we use the dataset of merged agentic and human generated PRs collected in the previous stages of the study.

To study bug-inducing changes, we first identified bug-fix commits. Next, we applied a bug-inducing commit detection algorithm to trace the origins of bugs, and finally determined whether the identified bug-inducing commits were associated with human generated or agentic PRs. 
Similar to prior studies ~\cite{abidi2021multi,morovati2023bugs}, we identified bug-fixing commits using a keyword-based approach. Commits whose messages contained predefined bug-related keywords (\textit{bug}, \textit{fail}, \textit{crash}, \textit{fix}, \textit{resolve}, \textit{failure}, \textit{broken}, etc.), were classified as bug-fixing commits. Subsequently, Following the methodology adopted in previous studies ~\cite{abidi2021multi, morovati2023bugs}, we used PyDriller~\cite{spadini2018pydriller} for repository mining. To identify bug-inducing commits, we employed the implementation of the SZZ ~\cite{sliwerski2005changes} provided by PyDriller,  consistent with previous studies ~\cite{yu2024studying, abidi2021multi}.

\begin{figure}
    \centering
    \includegraphics[width=0.6\linewidth]{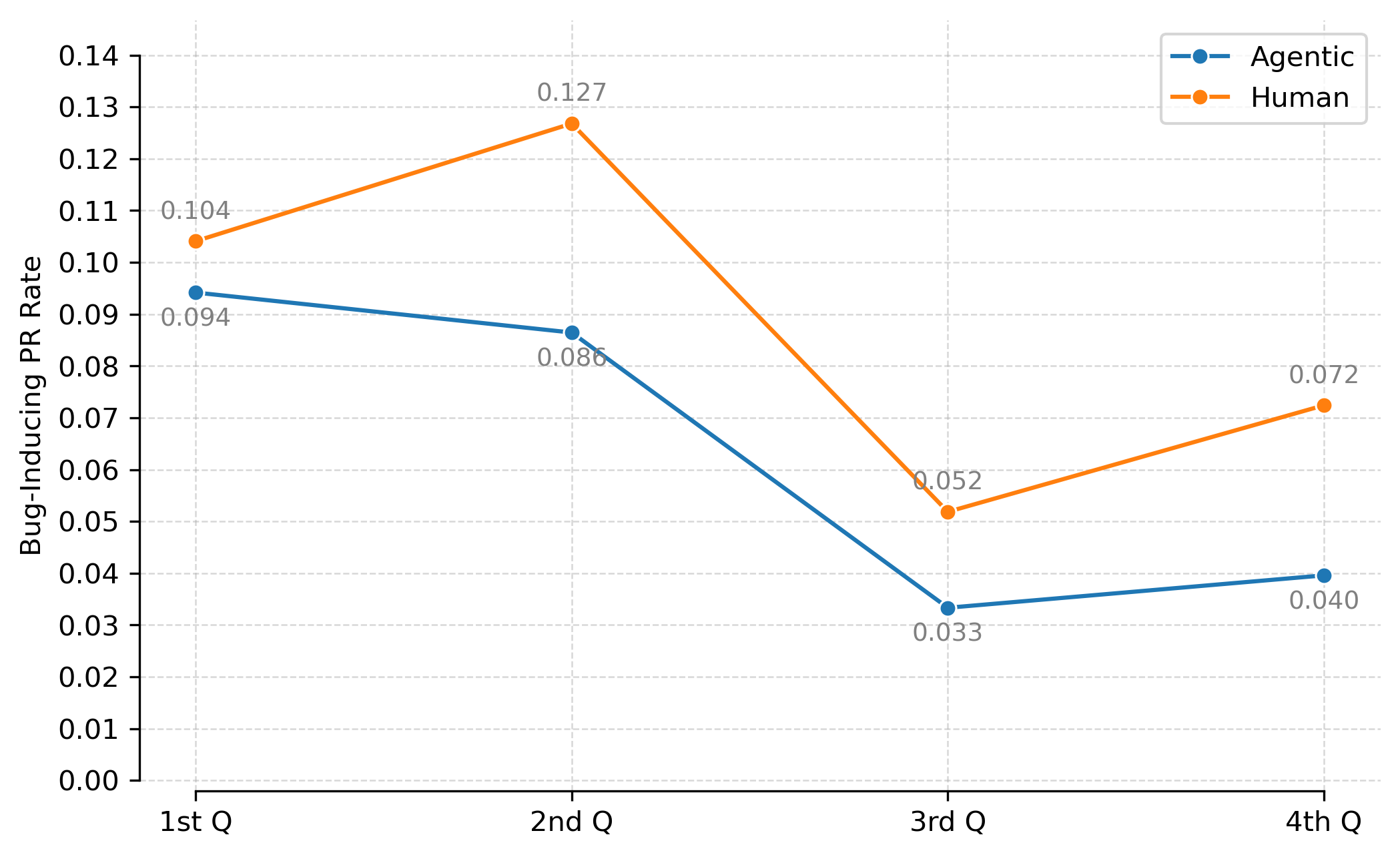}
    \caption{Bug-Inducing behavior of  agentic vs human generated PRs over time}
    \label{fig:bug_induce_PR}
\end{figure}

Figure~\ref{fig:bug_induce_PR} presents the proportion of agentic and human generated PRs that contain at least one bug-inducing commit across different development quarters. 
As shown in the figure, agentic PRs consistently exhibit a lower proportion of bug-inducing commits than human generated PRs throughout all stages of the development lifecycle. 
To assess whether these differences are statistically significant, we perform a \textit{Chi-squared} test ~\cite{mchugh2013chi} and compute \textit{Cramér's V} ~\cite{cramer1999mathematical} as a measure of effect size. According to the results reported in Table~\ref{tab:bug_induce_PR}, only the fourth quarter exhibits a statistically significant difference between agentic and human generated PRs. 
However, the corresponding effect size is negligible, indicating that the practical magnitude of the observed difference is very limited. Overall, these results suggest that although human generated PRs are slightly more likely to contain bug-inducing commits than agentic PRs, the difference remains small across all development stages.

A comparison across development quarters further reveals a decreasing trend in the proportion of agentic PRs associated with bug-inducing commits, with the most pronounced reduction occurring between the second and third quarters. 
The statistical test results reported in Table~\ref{tab:agentic_bug_induce} indicate that only this comparison yields a statistically significant difference, accompanied by a small effect size. One possible explanation is that, as projects mature, both developers and reviewers become more familiar with the strengths and limitations of AI generated contributions, leading to improved validation and integration practices. Additionally, the observed decline may reflect improvements in the underlying AI models and coding agents themselves, which can generate more reliable code as the technology evolves. 
Collectively, the results suggest that agent-assisted contributions are not associated with increased defect risk; rather, they may contribute to more stable and less bug-prone changes as the development process matures.

\begin{table}[b]
\centering
\caption{Statistical analysis (\textit{Chi Square} test and \textit{Cramér’s V} as effect size) of bug inducing commits in agentic and human generated PRs}
\label{tab:bug_inducing_analysis}

\begin{subtable}[t]{0.48\textwidth}
\centering
\scriptsize
\caption{Agentic vs. human generated PRs over time, in terms of their association with bug inducing commits}
\label{tab:bug_induce_PR}

\setlength{\tabcolsep}{5pt}
\begin{tabular}{lll}
\toprule
Quarter & \textit{p}-value & Effect size \\
\midrule
1st & 0.560 & 0.015 (n) \\
2nd & 0.110 & 0.060 (n) \\
3rd & 0.175 & 0.041 (n) \\
4th & \textbf{0.021} & 0.068 (n) \\
\bottomrule
\end{tabular}
\end{subtable}
\hfill
\begin{subtable}[t]{0.48\textwidth}
\centering
\scriptsize
\caption{
Comparison of various development quarters for the ratio of agentic PRs associated with bug inducing commits.}
\label{tab:agentic_bug_induce}

\setlength{\tabcolsep}{5pt}
\begin{tabular}{lll}
\toprule
Quarter & \textit{p}-value & Effect size \\
\midrule
1st vs 2nd & 0.760 & 0.009 (n) \\
2nd vs 3rd & \textbf{0.001} & 0.109 (s) \\
3rd vs 4th & 0.691 & 0.012 (n) \\
\bottomrule
\end{tabular}
\end{subtable}

\end{table}

\begin{tcolorbox}[colback=blue!10!white, colframe=blue!75!black]

\textbf{Finding 7: }
Agentic PRs maintain bug-inducing rates comparable to, and often lower than, human-generated PRs throughout the development lifecycle. The observed differences are generally not statistically significant and exhibit negligible practical effects, while the bug-inducing rate of agentic PRs shows a significant decline between the second and third development quarters.

\end{tcolorbox} 

\section{Related Work}\label{sec:related_work}

\subsection{Effect of AI coding Agents on Software Development}

The growing adoption of AI coding agents throughout the software development lifecycle has sparked increasing interest in understanding their impact on software engineering practices and outcomes. 
Dakhel et al.~\cite{dakhel2023github} investigated the effectiveness of Copilot by comparing its generated solutions with those produced by human developers on a dataset of \textit{Python} programming tasks, including several fundamental algorithmic problems such as sorting. 
Their results showed that Copilot was able to generate solutions for all considered tasks, although some of the generated solutions contained bugs or exhibited reproducibility issues. 
The authors further reported that human generated solutions achieved higher correctness rates than those produced by Copilot. At the same time, Copilot-generated code tended to exhibit lower complexity than human generated code, suggesting that the effort required to identify and fix defects in Copilot generated scripts is lower than that for human generated code.
Although this study provides valuable insights into the capabilities and limitations of Copilot as a widely adopted AI coding assistant, its scope is restricted to a predefined set of programming tasks evaluated in a controlled setting. Consequently, it does not examine how the effectiveness of AI coding agents evolves over time in real-world software projects, nor does it investigate whether their impact varies across different stages of the software development lifecycle. 

Hora et al.~\cite{hora2026coding} investigated the role of AI coding agents, including Claude, Copilot, and Cursor, in supporting software testing activities. To this end, they analyzed more than 48,000 agentic commits across projects developed in multiple programming languages such as \textit{Python}, \textit{JavaScript}, and \textit{TypeScript}. Their findings reveal that approximately 60\% of repositories adopting AI coding agents contain agentic contributions related to testing activities. Furthermore, they report that a larger proportion of agentic commits modify testing files (23\%) compared to human generated commits (13\%). 
Similarly, 36\% of agentic commits introduce or modify test mocks, whereas the corresponding proportion for human generated commits is 26\%.
While these findings provide valuable insights into the growing role of AI coding agents in software testing, the study does not examine potential differences among individual agents with respect to their effectiveness or testing-related contributions. Instead, all agentic activities are treated as a single group, potentially masking substantial variations across agents. In addition, the study does not investigate how the effectiveness and usage patterns of AI coding agents evolve over time. Given the rapid advancement of agent capabilities and the increasing familiarity of developers with these tools, understanding temporal trends is essential for obtaining a more comprehensive view of their impact on software development practices.

Tambon et al.~\cite{tambon2025bugs} investigated the types of bugs that may occur in \textit{Python} code generated by AI coding agents. To this end, they manually analyzed 333 bugs identified in code produced by three widely used AI coding agents (such as Codex and CodeGen) and derived a taxonomy consisting of ten distinct bug patterns, e.g. misinterpretation, syntax errors, prompt-biased code generation, and etc. 
Their results indicate that misinterpretation and missing corner cases are the two most prevalent bug categories observed in agentic code. Furthermore, based on a survey of software developers, they found that while most identified bug patterns can be detected and fixed with reasonable effort, they still require manual intervention. In particular, bugs related to misinterpretation and missing corner cases were perceived as the most difficult and time-consuming to identify and resolve.
The findings of this study shed a light on the impact of AI coding agents on software quality by highlighting the types of defects that commonly emerge in agentic code. However, the analysis is limited to a static view of agent behavior and does not examine how bug patterns may evolve over time. Given the rapid advancement of AI coding agents and their increasing adoption in software development, investigating temporal changes in the quality of generated code could provide a more comprehensive understanding of their long-term effectiveness and limitations.

Robbes et al.~\cite{robbes2026agentic} investigated the adoption of AI coding agents in GitHub projects and reported that, as of February 2026, approximately 28.66\% of GitHub repositories had incorporated AI coding agents into their development processes. Their study further revealed that repositories associated with Microsoft, Google, and Apache account for the largest numbers of projects utilizing AI coding agents.
While these findings provide valuable insights into the prevalence of AI coding agents in open-source software development, they primarily focus on adoption trends and do not highlight the effect of the agentic contributions of the software development process.

\subsection{Effect of Agentic PRs of Software Development}

As AI coding agents have become increasingly integrated into open-source software development, several studies have investigated agentic PRs from different perspectives. For example, Li et al.~\cite{li2025rise} collected data from GitHub repositories to analyze the activities of five popular AI coding agents and constructed a dataset of real-world agentic contributions (AIDev), which also serves as the primary data source for our study. Using this dataset, they examined the acceptance rates of PRs generated by different agents and compared them with those of human generated PRs. Their findings indicate that human generated PRs generally achieve higher merge rates. They also observed that AI coding agents tend to perform better on relatively simple and low-risk tasks, such as documentation related changes. Furthermore, their results suggest that coding agents can resolve issues more quickly than human contributors and that their effectiveness varies across programming languages, potentially reflecting domain specific strengths and limitations.
While this study provides a valuable dataset and offers important insights into the characteristics of agentic contributions, several research gaps remain. First, the analysis provides a largely static view of agent behavior and does not investigate how the activities and effectiveness of AI coding agents evolve over time. Given the rapid advancement of these technologies, understanding temporal trends is essential for assessing their long-term impact on software development practices. Second, although the study examines contribution outcomes such as merge rates, it does not explore the implications of agentic contributions for software quality, compared to those generated by humans.

Peralta et al.~\cite{peralta2026agentic} investigated the factors influencing the acceptance and rejection of agentic PRs. Using the AIDev dataset, they constructed a sample of 717 randomly selected agentic PRs, including both merged and rejected contributions. They then performed a manual analysis of these PRs to identify the underlying reasons for their outcomes. Their findings reveal that only 35.7\% of rejected agentic PRs were rejected due to the agent's failure in correctly completing the assigned task. Furthermore, they reported that only 15.4\% of merged agentic PRs required human intervention through feedback loops or reviewer commits before being integrated.
This study provides valuable insights into the factors affecting the acceptance of agentic contributions and offers evidence regarding the extent of human involvement required to integrate such contributions into software projects. However, the analysis does not consider potential differences across programming languages or individual AI coding agents, despite the possibility that agent capabilities and effectiveness may vary substantially across these dimensions. In addition, the study adopts a static perspective and does not investigate how the characteristics, outcomes, and review processes of agentic PRs evolve over time.

Popescu et al.~\cite{popescu2026investigating} conducted a large-scale empirical study comparing the characteristics of contributions generated by five popular AI coding agents, including Codex, Claude, Copilot, Jules, and Devin. Their analysis was based on a dataset of approximately 110,000 GitHub PRs, allowing them to compare both agentic and human generated contributions. Their findings indicate that agentic PRs tend to modify a larger volume of code than human generated PRs, while also exhibiting shorter merge times on average. In addition, they observed that agentic contributions target \textit{JavaScript} and \textit{TypeScript} files more frequently, compared with files written in other programming languages. Finally, their results suggest that agentic code is retained less frequently than human generated code and is more susceptible to subsequent code churn.
Similar to other previous studies, this paper also 
does not explicitly differentiate contributions across programming languages. Given that the effectiveness and usage patterns of AI coding agents may vary substantially depending on language specific ecosystems, tooling support, and code characteristics, such differences may influence the observed results. 
Furthermore, although the study considers temporal aspects of agentic contributions, it does not examine how agent behavior and contribution characteristics evolve throughout the project development lifecycle.

Huang et al.~\cite{huang2026more} investigated the quality of code generated by AI coding agents using the AIDev dataset. Their analysis revealed that agentic code tends to contain more redundant code than human generated contributions. Interestingly, despite this increased redundancy, human reviewers expressed relatively limited negative sentiment toward such issues during the review process. 
Similarly, Agarwal et al.~\cite{agarwal2026ai} leveraged the AIDev dataset to examine the broader impact of AI coding agents on software development. Their results indicate that the use of AI coding agents is associated with increased code complexity, suggesting that agentic contributions may introduce additional maintenance challenges despite their potential productivity benefits.
These two studies provide important insights into the quality characteristics of code produced by AI coding agents and their implications for software development. However, both studies treat agentic contributions as a homogeneous group and do not investigate potential differences among individual coding agents. Furthermore, neither of these two studies examines how the characteristics and quality of agentic contributions evolve over time.

\subsection{Position of this Study}

This study aims to address several gaps in the existing literature on AI coding agents and to complement prior findings regarding their impact on software development processes. Although our analysis is based on the same dataset used in many previous studies, we adopt a broader and more fine-grained perspective. Specifically, we compare the characteristics of contributions generated by different AI coding agents, allowing us to identify variations in their behavior, effectiveness, and contribution patterns. In addition, we compare agentic contributions with human generated contributions to better understand the similarities and differences between the two groups.
Furthermore, unlike prior work, which largely provides a static view of agentic contributions, we investigate how agent behavior evolves throughout the software development lifecycle. To this end, we analyze contribution characteristics across different development quarters, using project lifecycle stages as a proxy for temporal evolution. This longitudinal perspective enables us to examine whether the role, effectiveness, and impact of AI coding agents change as projects mature, thereby providing a more comprehensive understanding of their adoption and influence in real-world software development.

\section{Threats to Validity}
\label{sec:threats}

\paragraph{\textbf{Construct Validity}}

Potential threats to construct validity stem from four primary sources: (1) the selection of software engineering agents and the PRs generated by them, (2) the repository inclusion criteria, specifically the restriction to repositories with more than 100 GitHub stars, (3) our operationalization of repository lifecycle stages, (4) the exclusion of open PRs from the analysis, and (5) the use of SZZ to identify bug-inducing commits. The remaining components of our methodology, including PR clustering and the identification of bug-inducing commits, are based on established techniques that have been widely adopted in prior software engineering research.
First, our findings may be influenced by the specific agents represented in the dataset. To mitigate this threat, we relied on the AIDev dataset, which contains PRs generated by five widely adopted software engineering agents. Although these agents cover a substantial portion of current agent-assisted development practices, they may not fully represent the broader and rapidly evolving ecosystem of software engineering agents. Consequently, caution should be exercised when generalizing our findings to agents not included in the dataset.
Second, the repository selection criterion may affect the representativeness of our sample. Following the filtered AIDev dataset, we restricted our analysis to repositories with more than 100 GitHub stars. This threshold has been widely used in prior software engineering studies to select repositories with sufficient popularity and visibility ~\cite{moriconi2025ghalogs,tang2022towards,romeo2025uml}. This threshold helps ensure that the studied repositories have sufficient activity and community engagement ~\cite{sakib2025understanding}, reducing noise from inactive or experimental projects. However, repositories with fewer stars may exhibit different development practices, maintenance processes, and adoption patterns of AI agents. Therefore, our results may not generalize to less popular or newly created repositories.
Another threat concerns our definition of repository lifecycle stages. Repository evolution is inherently continuous~\cite{joshi2025swe,zheng2025humanevo}. To operationalize this evolution for empirical analysis, we define the lifecycle of each agent--repository pair as the period between the first PR generated by the target agent in the repository and the last agent-generated PR observed before our data collection cutoff date. This definition enables consistent comparisons across repositories and agents but may not capture all dimensions of project maturity or evolution. Alternative lifecycle definitions could potentially yield different observations.
Another threat arises from restricting our analysis to closed PRs and excluded PRs that remained open at the time of data collection. This choice was necessary because several outcome measures considered in our study, including merge status and post-merge quality indicators, cannot be reliably determined for unresolved PRs. However, open PRs may differ systematically from closed PRs in terms of complexity, review effort, or eventual outcomes, which may introduce bias into the observed results.

Finally, identifying bug-inducing commits using SZZ poses a threat to construct validity.
To reduce implementation-related errors, we relied on the official SZZ implementation provided by PyDriller~\cite{spadini2018pydriller}
rather than developing a custom implementation. 
However, Rosa et al.~\cite{rosa2023comprehensive}
reported that the PyDriller implementation achieved a recall of 0.67, a precision of 0.39, and an F1-score of 0.49 without issue-date filtering. 
These values indicate that, although PyDriller identifies a substantial proportion of known bug-inducing commits, it may also produce false positives and miss some relevant commits. Because a developer-validated ground truth is unavailable for our repositories, we could not directly measure SZZ accuracy on our dataset. Consequently, errors in identifying bug-inducing commits may affect the estimated defect-proneness of both agentic and human-generated PRs, and our findings should be interpreted with this limitation in mind.

\paragraph{\textbf{Internal Validity}}
Potential threats to internal validity primarily arise from: (1) the manual validation of topic labels, 
and (2) the sampling strategy used to construct the comparison set of human-generated PRs. 
First, although BERTopic automatically generates topic clusters, the interpretation and assignment of topic labels involve human judgment and may introduce subjective bias. To mitigate this threat, the first two authors—one PhD candidate and one associate researcher with a PhD degree—independently reviewed a randomly selected sample of 100 agent-generated PRs spanning all identified topics and verified the assigned labels. 
To assess annotation reliability, we computed Cohen's kappa ~\cite{hsu2003interrater}. The resulting kappa score of 0.842 indicates strong agreement ~\cite{mchugh2012interrater}, suggesting that the assigned labels are reliable. Any disagreements were subsequently discussed until consensus was reached on the final labeling decisions. 
Second, our analysis in RQ3 compares agentic PRs with human-generated PRs. A potential threat is that differences between the two groups may be confounded by temporal variations in repository activity, development practices, or project evolution. To reduce this threat, we sampled human-generated PRs from the same repositories and the same calendar quarter as the corresponding agentic PRs. In addition, we excluded all PRs associated with any of the five coding agents (like prior studies ~\cite{li2025rise}) considered in this study and removed PRs whose authors were identified as \texttt{BOT}s. These criteria increase our confidence that the selected PRs represent human-generated contributions and that the compared PRs were created under similar project conditions and development contexts. Nevertheless, some agent-assisted contributions may not be explicitly identifiable through repository metadata, and other unobserved confounding factors may still influence the observed differences between agent-generated and human-generated PRs.

\paragraph{\textbf{External Validity}}
Our study focuses exclusively on \textit{Python}-based repositories associated with the selected software engineering agents, as \textit{Python} is currently one of the most widely adopted programming languages ~\cite{{jimenez2023swe,morovati2024bug}}, and it is  well-representated in AIDev dataset ~\cite{li2025rise}. Although our analysis is limited to \textit{Python} projects, we expect that similar patterns and observations may generalize to agent-generated PRs in other programming languages, given that the underlying development workflows and agentic contribution processes are likely to remain comparable.
Another potential threat to external validity concerns the generalizability of our findings, as the AIDev dataset relies solely on GitHub for data collection. Nevertheless, GitHub represents the largest and most prominent platform for hosting and maintaining open-source software projects ~\cite{li2020exploratory,aghili2023studying}, making it a representative and appropriate source for large-scale empirical software engineering studies.

\section{Conclusion and Future Work}
\label{sec:conclusion}

Recent dramatic advances in LLMs have accelerated the adoption of AI-powered tools throughout almost all aspects of software engineering process. In particular, the emergence of AI coding agents has transformed the way software is developed, making automated coding assistance increasingly common in both industrial and open-source projects. While these agents are capable of addressing a wide range of development tasks, their contribution patterns and impact on software quality may vary throughout the software development lifecycle. Understanding these dynamics is therefore essential for assessing the benefits and limitations of AI coding agents in real-world software development.
In this study, we leveraged the AIDev dataset to investigate the characteristics of agenic PRs, compare them with human-generated contributions, and examine how agents' behavior evolves across different stages of project development. Our findings show that Claude generated PRs achieve the highest merge rates among the studied agents. We also observed that PRs submitted during the third development quarter are more likely to be merged than those submitted during other stages of the project lifecycle. Furthermore, our analysis reveals that \textit{`documentation'}, \textit{`dependency management'}, and \textit{`testing'} are the development tasks most frequently supported by AI coding agents. Finally, our results show 
that agentic contributions differ from human-generated contributions across several quality-related and process-related metrics (e.g., number of commits, modified files, changed lines of code, review discussions, etc.).
As future work, our analysis could be extended beyond the PR level to investigate agent behavior at the commit level, enabling a more fine-grained understanding of how AI coding agents contribute to software evolution. 
We hope that the findings of this study contribute to a deeper understanding of the evolving role of AI coding agents and support both researchers and practitioners in making informed decisions regarding their adoption and use throughout the software development lifecycle.

\bibliographystyle{ACM-Reference-Format}
\bibliography{bibliography}

@article{li2025rise,
  title={The rise of ai teammates in software engineering (se) 3.0: How autonomous coding agents are reshaping software engineering},
  author={Li, Hao and Zhang, Haoxiang and Hassan, Ahmed E},
  journal={arXiv preprint arXiv:2507.15003},
  year={2025}
}

@article{watanabe2025use,
  title={On the use of agentic coding: An empirical study of pull requests on github},
  author={Watanabe, Miku and Li, Hao and Kashiwa, Yutaro and Reid, Brittany and Iida, Hajimu and Hassan, Ahmed E},
  journal={arXiv preprint arXiv:2509.14745},
  year={2025}
}

@misc{openai_codex,
    author = {OpenAI Codex},
    title = {OpenAI Codex official website},
    howpublished = {\url{https://openai.com/codex/}},
    year={2025},
    note = {Accessed: 2025-11-15}
}

@misc{github_copilot,
    author = {GitHub Copilot},
    title = {GitHub Copilot official website},
    howpublished = {\url{https://github.com/copilot}},
    year={2025},
    note = {Accessed: 2025-11-15}
}

@misc{cursor,
    author = {Cursor},
    title = {Cursor official website},
    howpublished = {\url{https://cursor.com/}},
    year={2025},
    note = {Accessed: 2025-11-15}
}

@misc{devin,
    author = {Devin},
    title = {Devin official website},
    howpublished = {\url{https://devin.ai/}},
    year={2025},
    note = {Accessed: 2025-11-15}
}

@misc{claude_code,
    author = {Claude CodeDevin},
    title = {Claude Code official website},
    howpublished = {\url{https://www.claude.com/product/claude-code}},
    year={2025},
    note = {Accessed: 2025-11-15}
}

@article{chen2021evaluating,
  title={Evaluating large language models trained on code},
  author={Chen, Mark},
  journal={arXiv preprint arXiv:2107.03374},
  year={2021}
}

@article{jimenez2023swe,
  title={Swe-bench: Can language models resolve real-world github issues?},
  author={Jimenez, Carlos E and Yang, John and Wettig, Alexander and Yao, Shunyu and Pei, Kexin and Press, Ofir and Narasimhan, Karthik},
  journal={arXiv preprint arXiv:2310.06770},
  year={2023}
}

@misc{github_api_v3,
  author = {GitHub developer guideline documentation},
  title = {GitHub REST API},
  howpublished = {\url{https://developer.github.com/v3/}},
  year={2021},
  note = {Accessed: 2025-09-05}
}

@inproceedings{bird2011don,
  title={Don't touch my code! Examining the effects of ownership on software quality},
  author={Bird, Christian and Nagappan, Nachiappan and Murphy, Brendan and Gall, Harald and Devanbu, Premkumar},
  booktitle={Proceedings of the 19th ACM SIGSOFT symposium and the 13th European conference on Foundations of software engineering},
  pages={4--14},
  year={2011}
}

@article{zhou2014will,
  title={Who will stay in the floss community? modeling participant’s initial behavior},
  author={Zhou, Minghui and Mockus, Audris},
  journal={IEEE Transactions on Software Engineering},
  volume={41},
  number={1},
  pages={82--99},
  year={2014},
  publisher={IEEE}
}

@inproceedings{bacchelli2013expectations,
  title={Expectations, outcomes, and challenges of modern code review},
  author={Bacchelli, Alberto and Bird, Christian},
  booktitle={2013 35th International Conference on Software Engineering (ICSE)},
  pages={712--721},
  year={2013},
  organization={IEEE}
}

@inproceedings{xia2023automated,
  title={Automated program repair in the era of large pre-trained language models},
  author={Xia, Chunqiu Steven and Wei, Yuxiang and Zhang, Lingming},
  booktitle={2023 IEEE/ACM 45th International Conference on Software Engineering (ICSE)},
  pages={1482--1494},
  year={2023},
  organization={IEEE}
}

@article{kumar2025intuition,
  title={Intuition to Evidence: Measuring AI's True Impact on Developer Productivity},
  author={Kumar, Anand and Khare, Vishal and Sharma, Deepak and Kumar, Satyam and Saini, Vijay and Yadav, Anshul and Jain, Sachendra and Rana, Ankit and Verma, Pratham and Meena, Vaibhav and others},
  journal={arXiv preprint arXiv:2509.19708},
  year={2025}
}

@article{dohmke2023sea,
  title={Sea change in software development: Economic and productivity analysis of the ai-powered developer lifecycle},
  author={Dohmke, Thomas and Iansiti, Marco and Richards, Greg},
  journal={arXiv preprint arXiv:2306.15033},
  year={2023}
}

@article{sapkota2025ai,
  title={Ai agents vs. agentic ai: A conceptual taxonomy, applications and challenges},
  author={Sapkota, Ranjan and Roumeliotis, Konstantinos I and Karkee, Manoj},
  journal={arXiv preprint arXiv:2505.10468},
  year={2025}
}

@article{krishnan2025ai,
  title={Ai agents: Evolution, architecture, and real-world applications},
  author={Krishnan, Naveen},
  journal={arXiv preprint arXiv:2503.12687},
  year={2025}
}

@article{hassan2025agentic,
  title={Agentic Software Engineering: Foundational Pillars and a Research Roadmap},
  author={Hassan, Ahmed E and Li, Hao and Lin, Dayi and Adams, Bram and Chen, Tse-Hsun and Kashiwa, Yutaro and Qiu, Dong},
  journal={arXiv preprint arXiv:2509.06216},
  year={2025}
}

@article{sapkota2025vibe,
  title={Vibe coding vs. agentic coding: Fundamentals and practical implications of agentic ai},
  author={Sapkota, Ranjan and Roumeliotis, Konstantinos I and Karkee, Manoj},
  journal={arXiv preprint arXiv:2505.19443},
  year={2025}
}

@article{bolker2015linear,
  title={Linear and generalized linear mixed models},
  author={Bolker, Benjamin M},
  journal={Ecological statistics: contemporary theory and application},
  volume={2015},
  pages={309--333},
  year={2015},
  publisher={Oxford University Press Oxford}
}

@book{stroup2024generalized,
  title={Generalized linear mixed models: modern concepts, methods and applications},
  author={Stroup, Walter W and Ptukhina, Marina and Garai, Julie},
  year={2024},
  publisher={Chapman and Hall/CRC}
}

@inproceedings{borges2016understanding,
  title={Understanding the factors that impact the popularity of GitHub repositories},
  author={Borges, Hudson and Hora, Andre and Valente, Marco Tulio},
  booktitle={2016 IEEE international conference on software maintenance and evolution (ICSME)},
  pages={334--344},
  year={2016},
  organization={IEEE}
}

@article{borges2018s,
  title={What’s in a github star? understanding repository starring practices in a social coding platform},
  author={Borges, Hudson and Valente, Marco Tulio},
  journal={Journal of Systems and Software},
  volume={146},
  pages={112--129},
  year={2018},
  publisher={Elsevier}
}

@article{kumi2024uncovering,
  title={Uncovering concerns of citizens through machine learning and social network sentiment analysis},
  author={Kumi, Sandra and Snow, Charles and Lomotey, Richard K and Deters, Ralph},
  journal={Ieee Access},
  volume={12},
  pages={94885--94913},
  year={2024},
  publisher={IEEE}
}

@article{vayansky2020review,
  title={A review of topic modeling methods},
  author={Vayansky, Ike and Kumar, Sathish AP},
  journal={Information Systems},
  volume={94},
  pages={101582},
  year={2020},
  publisher={Elsevier}
}

@article{alam2025empirical,
  title={An empirical investigation on the challenges in scientific workflow systems development},
  author={Alam, Khairul and Roy, Banani and Roy, Chanchal K and Mittal, Kartik},
  journal={Empirical Software Engineering},
  volume={30},
  number={5},
  pages={151},
  year={2025},
  publisher={Springer}
}

@article{grootendorst2022bertopic,
  title={BERTopic: Neural topic modeling with a class-based TF-IDF procedure},
  author={Grootendorst, Maarten},
  journal={arXiv preprint arXiv:2203.05794},
  year={2022}
}

@article{alam2026analyzing,
  title={Analyzing GitHub Issues and Pull Requests in nf-core Pipelines: Insights into nf-core Pipeline Repositories},
  author={Alam, Khairul and Roy, Banani},
  journal={arXiv preprint arXiv:2601.09612},
  year={2026}
}

@book{tintle2018introduction,
  title={Introduction to Statistical Investigations, AP Edition Workbook},
  author={Tintle, Nathan},
  year={2018},
  publisher={Wiley}
}

@article{morovati2024bug,
  title={Bug characterization in machine learning-based systems},
  author={Morovati, Mohammad Mehdi and Nikanjam, Amin and Tambon, Florian and Khomh, Foutse and Jiang, Zhen Ming},
  journal={Empirical Software Engineering},
  volume={29},
  number={1},
  pages={14},
  year={2024},
  publisher={Springer}
}

@inproceedings{yu2018nbsl,
  title={Nbsl: A supervised classification model of pull request in github},
  author={Yu, Song and Xu, Li and Zhang, Yan and Wu, Jinsong and Liao, Zhifang and Li, Yanbing},
  booktitle={2018 IEEE International conference on communications (ICC)},
  pages={1--6},
  year={2018},
  organization={IEEE}
}

@article{fridh2022classification,
  title={Classification of pull requests using transformers},
  author={Fridh, Oscar and Stypa, Szymon},
  journal={LU-CS-EX},
  year={2022}
}

@article{healy2024uniform,
  title={Uniform manifold approximation and projection},
  author={Healy, John and McInnes, Leland},
  journal={Nature Reviews Methods Primers},
  volume={4},
  number={1},
  pages={82},
  year={2024},
  publisher={Nature Publishing Group UK London}
}

@article{danyal2024sentiment,
  title={Sentiment analysis of movie reviews based on NB approaches using TF--IDF and count vectorizer},
  author={Danyal, Mian Muhammad and Khan, Sarwar Shah and Khan, Muzammil and Ullah, Subhan and Ghaffar, Muhammad Bilal and Khan, Wahab},
  journal={Social network analysis and mining},
  volume={14},
  number={1},
  pages={87},
  year={2024},
  publisher={Springer}
}

@inproceedings{rahimi2024contextualized,
  title={Contextualized topic coherence metrics},
  author={Rahimi, Hamed and Mimno, David and Vigly, Jacob Hoover and Naacke, Hubert and Constantin, Camelia and Amann, Bernd},
  booktitle={Findings of the Association for Computational Linguistics: EACL 2024},
  pages={1760--1773},
  year={2024}
}

@article{lee2025gemini,
  title={Gemini embedding: Generalizable embeddings from gemini},
  author={Lee, Jinhyuk and Chen, Feiyang and Dua, Sahil and Cer, Daniel and Shanbhogue, Madhuri and Naim, Iftekhar and {\'A}brego, Gustavo Hern{\'a}ndez and Li, Zhe and Chen, Kaifeng and Vera, Henrique Schechter and others},
  journal={arXiv preprint arXiv:2503.07891},
  year={2025}
}

@misc{conventional_commits,
    author = {Angular Commit Guidelines},
    title = {Conventional Commits: A specification for adding human and machine readable meaning to commit messages},
    howpublished = {\url{https://www.conventionalcommits.org/en/v1.0.0/}},
    year={2026},
    note = {Accessed: 2026-04-01}
}

@article{alshara2022pi,
  title={Pi-link: A ground-truth dataset of links between pull-requests and issues in github},
  author={Alshara, Zakarea and Shatnawi, Anas and Eyal-Salman, Hamzeh and Seriai, Abdelhak-Djamel and Shatnawi, Maad},
  journal={IEEE Access},
  volume={11},
  pages={697--710},
  year={2022},
  publisher={IEEE}
}

@article{taraghi2026real,
  title={Real Faults in Model Context Protocol (MCP) Software: a Comprehensive Taxonomy},
  author={Taraghi, Mina and Morovati, Mohammad Mehdi and Khomh, Foutse},
  journal={arXiv preprint arXiv:2603.05637},
  year={2026}
}

@article{weng2022identification,
  title={Identification and visualization of key topics in scientific publications with transformer-based language models and document clustering methods},
  author={Weng, Min-Hsien and Wu, Shaoqun and Dyer, Mark},
  journal={Applied Sciences},
  volume={12},
  number={21},
  pages={11220},
  year={2022},
  publisher={MDPI}
}

@article{wang2019evaluating,
  title={Evaluating word embedding models: Methods and experimental results},
  author={Wang, Bin and Wang, Angela and Chen, Fenxiao and Wang, Yuncheng and Kuo, C-C Jay},
  journal={APSIPA transactions on signal and information processing},
  volume={8},
  pages={e19},
  year={2019},
  publisher={Cambridge University Press}
}

@article{zagatti2025investigating,
  title={Investigating the relationship between text vectorization cosine similarity and classification performance},
  author={Zagatti, Fernando R and Shimizu, Gilson Y and Lucr{\'e}dio, Daniel and Caseli, Helena M},
  journal={IEEE Access},
  year={2025},
  publisher={IEEE}
}

@inproceedings{mimno2011optimizing,
  title={Optimizing semantic coherence in topic models},
  author={Mimno, David and Wallach, Hanna and Talley, Edmund and Leenders, Miriam and McCallum, Andrew},
  booktitle={Proceedings of the 2011 conference on empirical methods in natural language processing},
  pages={262--272},
  year={2011}
}

@inproceedings{bellaouar2021topic,
  title={Topic modeling: Comparison of LSA and LDA on scientific publications},
  author={Bellaouar, Slimane and Bellaouar, Mohammed Mounsif and Ghada, Issam Eddine},
  booktitle={Proceedings of the 2021 4th International Conference on Data Storage and Data Engineering},
  pages={59--64},
  year={2021}
}

@article{alagha2021topic,
  title={Topic modeling and sentiment analysis of Twitter discussions on COVID-19 from spatial and temporal perspectives},
  author={AlAgha, Iyad},
  journal={Journal of Information Science Theory and Practice},
  volume={9},
  number={1},
  pages={35--53},
  year={2021},
  publisher={Korea Institute of Science and Technology Information}
}

@inproceedings{meng2013scalable,
  title={Scalable simple random sampling and stratified sampling},
  author={Meng, Xiangrui},
  booktitle={International conference on machine learning},
  pages={531--539},
  year={2013},
  organization={PMLR}
}

@article{morovati2024common,
  title={Common challenges of deep reinforcement learning applications development: an empirical study},
  author={Morovati, Mohammad Mehdi and Tambon, Florian and Taraghi, Mina and Nikanjam, Amin and Khomh, Foutse},
  journal={Empirical Software Engineering},
  volume={29},
  number={4},
  pages={95},
  year={2024},
  publisher={Springer}
}

@article{abidi2021multi,
  title={Are multi-language design smells fault-prone? An empirical study},
  author={Abidi, Mouna and Rahman, Md Saidur and Openja, Moses and Khomh, Foutse},
  journal={ACM Transactions on Software Engineering and Methodology (TOSEM)},
  volume={30},
  number={3},
  pages={1--56},
  year={2021},
  publisher={ACM New York, NY, USA}
}

@article{morovati2023bugs,
  title={Bugs in machine learning-based systems: a faultload benchmark},
  author={Morovati, Mohammad Mehdi and Nikanjam, Amin and Khomh, Foutse and Jiang, Zhen Ming},
  journal={Empirical Software Engineering},
  volume={28},
  number={3},
  pages={62},
  year={2023},
  publisher={Springer}
}

@article{sliwerski2005changes,
  title={When do changes induce fixes?},
  author={{\'S}liwerski, Jacek and Zimmermann, Thomas and Zeller, Andreas},
  journal={ACM sigsoft software engineering notes},
  volume={30},
  number={4},
  pages={1--5},
  year={2005},
  publisher={ACM New York, NY, USA}
}

@inproceedings{spadini2018pydriller,
  title={PyDriller: Python framework for mining software repositories},
  author={Spadini, Davide and Aniche, Maur{\'\i}cio and Bacchelli, Alberto},
  booktitle={Proceedings of the 2018 26th ACM Joint meeting on european software engineering conference and symposium on the foundations of software engineering},
  pages={908--911},
  year={2018}
}

@article{yu2024studying,
  title={Studying the impact of risk assessment analytics on risk awareness and code review performance},
  author={Yu, Xueyao and Cogo, Filipe R and McIntosh, Shane and Godfrey, Michael W},
  journal={Empirical Software Engineering},
  volume={29},
  number={2},
  pages={46},
  year={2024},
  publisher={Springer}
}

@inproceedings{decan2016github,
  title={When GitHub meets CRAN: An analysis of inter-repository package dependency problems},
  author={Decan, Alexandre and Mens, Tom and Claes, Ma{\"e}lick and Grosjean, Philippe},
  booktitle={2016 IEEE 23rd international conference on software analysis, evolution, and reengineering (SANER)},
  volume={1},
  pages={493--504},
  year={2016},
  organization={IEEE}
}

@inproceedings{alshahwan2024automated,
  title={Automated unit test improvement using large language models at meta},
  author={Alshahwan, Nadia and Chheda, Jubin and Finogenova, Anastasia and Gokkaya, Beliz and Harman, Mark and Harper, Inna and Marginean, Alexandru and Sengupta, Shubho and Wang, Eddy},
  booktitle={Companion Proceedings of the 32nd ACM International Conference on the Foundations of Software Engineering},
  pages={185--196},
  year={2024}
}

@inproceedings{siddiq2024using,
  title={Using large language models to generate junit tests: An empirical study},
  author={Siddiq, Mohammed Latif and Da Silva Santos, Joanna Cecilia and Tanvir, Ridwanul Hasan and Ulfat, Noshin and Al Rifat, Fahmid and Carvalho Lopes, Vin{\'\i}cius},
  booktitle={Proceedings of the 28th international conference on evaluation and assessment in software engineering},
  pages={313--322},
  year={2024}
}

@inproceedings{xue2024llm4fin,
  title={Llm4fin: Fully automating llm-powered test case generation for fintech software acceptance testing},
  author={Xue, Zhiyi and Li, Liangguo and Tian, Senyue and Chen, Xiaohong and Li, Pingping and Chen, Liangyu and Jiang, Tingting and Zhang, Min},
  booktitle={Proceedings of the 33rd ACM SIGSOFT International Symposium on Software Testing and Analysis},
  pages={1643--1655},
  year={2024}
}

@article{artola2025your,
  title={Is Your LLM Overcharging You? Tokenization, Transparency, and Incentives},
  author={Artola Velasco, Ander and Tsirtsis, Efstratios and Okati, Nastaran and Gomez Rodriguez, Manuel},
  journal={arXiv preprint arXiv:2505.21627},
  year={2025}
}

@article{jiang2024d,
  title={D-llm: A token adaptive computing resource allocation strategy for large language models},
  author={Jiang, Yikun and Wang, Huanyu and Xie, Lei and Zhao, Hanbin and Zhang, Chao and Qian, Hui and Lui, John C},
  journal={Advances in Neural Information Processing Systems},
  volume={37},
  pages={1725--1749},
  year={2024}
}

@inproceedings{lavrentiev2026token,
  title={Token Cost Optimization in an LLM Agent for JSON-Based System Modeling from Dialogue History},
  author={Lavrentiev, Saveliy and Dukhanov, Alexey},
  booktitle={2026 International Russian Smart Industry Conference (SmartIndustryCon)},
  pages={854--860},
  year={2026},
  organization={IEEE}
}

@article{hsu2003interrater,
  title={Interrater agreement measures: Comments on Kappan, Cohen's Kappa, Scott's $\pi$, and Aickin's $\alpha$},
  author={Hsu, Louis M and Field, Ronald},
  journal={Understanding Statistics},
  volume={2},
  number={3},
  pages={205--219},
  year={2003},
  publisher={Taylor \& Francis}
}

@article{mchugh2012interrater,
  title={Interrater reliability: the kappa statistic},
  author={McHugh, Mary L},
  journal={Biochemia medica},
  volume={22},
  number={3},
  pages={276--282},
  year={2012},
  publisher={Hrvatsko dru{\v{s}}tvo za medicinsku biokemiju i laboratorijsku medicinu}
}

@article{pedregosa2011scikit,
  title={Scikit-learn: Machine learning in Python},
  author={Pedregosa, Fabian and Varoquaux, Ga{\"e}l and Gramfort, Alexandre and Michel, Vincent and Thirion, Bertrand and Grisel, Olivier and Blondel, Mathieu and Prettenhofer, Peter and Weiss, Ron and Dubourg, Vincent and others},
  journal={the Journal of machine Learning research},
  volume={12},
  pages={2825--2830},
  year={2011},
  publisher={JMLR. org}
}

@article{wolter2023open,
  title={Open source license inconsistencies on github},
  author={Wolter, Thomas and Barcomb, Ann and Riehle, Dirk and Harutyunyan, Nikolay},
  journal={ACM Transactions on Software Engineering and Methodology},
  volume={32},
  number={5},
  pages={1--23},
  year={2023},
  publisher={ACM New York, NY}
}

@article{vendome2017license,
  title={License usage and changes: a large-scale study on github},
  author={Vendome, Christopher and Bavota, Gabriele and Penta, Massimiliano Di and Linares-V{\'a}squez, Mario and German, Daniel and Poshyvanyk, Denys},
  journal={Empirical Software Engineering},
  volume={22},
  number={3},
  pages={1537--1577},
  year={2017},
  publisher={Springer}
}

@article{ZEIDMAN201912,
title = {A guide to group effective connectivity analysis, part 2: Second level analysis with PEB},
journal = {NeuroImage},
volume = {200},
pages = {12-25},
year = {2019},
issn = {1053-8119},
doi = {https://doi.org/10.1016/j.neuroimage.2019.06.032},
url = {https://www.sciencedirect.com/science/article/pii/S1053811919305233},
author = {Peter Zeidman and Amirhossein Jafarian and Mohamed L. Seghier and Vladimir Litvak and Hayriye Cagnan and Cathy J. Price and Karl J. Friston},

}

@article{hardcastle2013empirical,
  title={Empirical Bayesian analysis of paired high-throughput sequencing data with a beta-binomial distribution},
  author={Hardcastle, Thomas J and Kelly, Krystyna A},
  journal={BMC bioinformatics},
  volume={14},
  number={1},
  pages={135},
  year={2013},
  publisher={Springer}
}

@inproceedings{lu2008learning,
  title={Learning from mistakes: a comprehensive study on real world concurrency bug characteristics},
  author={Lu, Shan and Park, Soyeon and Seo, Eunsoo and Zhou, Yuanyuan},
  booktitle={Proceedings of the 13th international conference on Architectural support for programming languages and operating systems},
  pages={329--339},
  year={2008}
}

@article{memon2007event,
  title={An event-flow model of GUI-based applications for testing},
  author={Memon, Atif M},
  journal={Software testing, verification and reliability},
  volume={17},
  number={3},
  pages={137--157},
  year={2007},
  publisher={Wiley Online Library}
}

@inproceedings{cotroneo2019bad,
  title={How bad can a bug get? an empirical analysis of software failures in the openstack cloud computing platform},
  author={Cotroneo, Domenico and De Simone, Luigi and Liguori, Pietro and Natella, Roberto and Bidokhti, Nematollah},
  booktitle={Proceedings of the 2019 27th ACM joint meeting on european software engineering conference and symposium on the foundations of software engineering},
  pages={200--211},
  year={2019}
}

@inproceedings{herzig2013impact,
  title={The impact of tangled code changes},
  author={Herzig, Kim and Zeller, Andreas},
  booktitle={2013 10th Working Conference on Mining Software Repositories (MSR)},
  pages={121--130},
  year={2013},
  organization={IEEE}
}

@article{attouche2024validation,
  title={Validation of modern JSON schema: Formalization and complexity},
  author={Attouche, Lyes and Baazizi, Mohamed-Amine and Colazzo, Dario and Ghelli, Giorgio and Sartiani, Carlo and Scherzinger, Stefanie},
  journal={Proceedings of the ACM on Programming Languages},
  volume={8},
  number={POPL},
  pages={1451--1481},
  year={2024},
  publisher={ACM New York, NY, USA}
}

@article{liu2024rethinking,
  title={Rethinking software misconfigurations in the real world: an empirical study and literature analysis},
  author={Liu, Yuhao and Zhou, Yingnan and Zhang, Hanfeng and Chang, Zhiwei and Xu, Sihan and Jia, Yan and Wang, Wei and Hu, Juncheng and Liu, Zheli},
  journal={arXiv preprint arXiv:2412.11121},
  year={2024}
}

@inproceedings{huang2015confvalley,
  title={Confvalley: A systematic configuration validation framework for cloud services},
  author={Huang, Peng and Bolosky, William J and Singh, Abhishek and Zhou, Yuanyuan},
  booktitle={Proceedings of the Tenth European Conference on Computer Systems},
  pages={1--16},
  year={2015}
}

@article{chen2025empirical,
  title={An empirical study on challenges for llm application developers},
  author={Chen, Xiang and Gao, Chaoyang and Chen, Chunyang and Zhang, Guangbei and Liu, Yong},
  journal={ACM Transactions on Software Engineering and Methodology},
  volume={34},
  number={7},
  pages={1--37},
  year={2025},
  publisher={ACM New York, NY}
}

@inproceedings{jha2017developer,
  title={Developer mistakes in writing android manifests: An empirical study of configuration errors},
  author={Jha, Ajay Kumar and Lee, Sunghee and Lee, Woo Jin},
  booktitle={2017 IEEE/ACM 14th International Conference on Mining Software Repositories (MSR)},
  pages={25--36},
  year={2017},
  organization={IEEE}
}

@inproceedings{gunawi2014bugs,
  title={What bugs live in the cloud? a study of 3000+ issues in cloud systems},
  author={Gunawi, Haryadi S and Hao, Mingzhe and Leesatapornwongsa, Tanakorn and Patana-Anake, Tiratat and Do, Thanh and Adityatama, Jeffry and Eliazar, Kurnia J and Laksono, Agung and Lukman, Jeffrey F and Martin, Vincentius and others},
  booktitle={Proceedings of the ACM symposium on cloud computing},
  pages={1--14},
  year={2014}
}

@article{he2023automating,
  title={Automating dependency updates in practice: An exploratory study on github dependabot},
  author={He, Runzhi and He, Hao and Zhang, Yuxia and Zhou, Minghui},
  journal={IEEE Transactions on Software Engineering},
  volume={49},
  number={8},
  pages={4004--4022},
  year={2023},
  publisher={IEEE}
}

@article{wu2024comprehensive,
  title={A comprehensive analysis of challenges and strategies for software release notes on GitHub},
  author={Wu, Jianyu and He, Hao and Gao, Kai and Xiao, Wenxin and Li, Jingyue and Zhou, Minghui},
  journal={Empirical Software Engineering},
  volume={29},
  number={5},
  pages={104},
  year={2024},
  publisher={Springer}
}

@article{zhang2026enhancing,
  title={Enhancing Automated Unit Test Generation with Large Language Models: A Systematic Literature Review},
  author={Zhang, Junwei and Hu, Xing and Gao, Cuiyun and Xia, Xin and Li, Shanping},
  journal={ACM Transactions on Software Engineering and Methodology},
  year={2026},
  publisher={ACM New York, NY}
}

@article{arcuri2014hitchhiker,
  title={A hitchhiker's guide to statistical tests for assessing randomized algorithms in software engineering},
  author={Arcuri, Andrea and Briand, Lionel},
  journal={Software Testing, Verification and Reliability},
  volume={24},
  number={3},
  pages={219--250},
  year={2014},
  publisher={Wiley Online Library}
}

@article{zheng2026github,
  title={Why do GitHub Actions workflows fail? An empirical study},
  author={Zheng, Lianyu and Li, Shuang and Huang, Xi and Huang, Jiangnan and Lin, Bin and Chen, Jinfu and Xuan, Jifeng},
  journal={ACM Transactions on Software Engineering and Methodology},
  volume={35},
  number={5},
  pages={1--29},
  year={2026},
  publisher={ACM New York, NY}
}

@article{macbeth2011cliff,
  title={Cliff's Delta Calculator: A non-parametric effect size program for two groups of observations},
  author={Macbeth, Guillermo and Razumiejczyk, Eugenia and Ledesma, Rub{\'e}n Daniel},
  journal={Universitas Psychologica},
  volume={10},
  number={2},
  pages={545--555},
  year={2011},
  publisher={Pontificia Universidad Javeriana}
}

@article{da2024chronicles,
  title={Chronicles of ci/cd: A deep dive into its usage over time},
  author={Da Gi{\~a}o, Hugo and Flores, Andr{\'e} and Pereira, Rui and Cunha, J{\'a}come},
  journal={arXiv preprint arXiv:2402.17588},
  year={2024}
}

@article{rzig2024empirical,
  title={Empirical analysis on CI/CD pipeline evolution in machine learning projects},
  author={Rzig, Dhia Elhaq and Houerbi, Alaa and Chavan, Rahul Ghanshyam and Hassan, Foyzul},
  journal={arXiv preprint arXiv:2403.12199},
  year={2024}
}

@incollection{wunschiers2025github,
  title={GitHub Repositories and Jupyter Notebooks},
  author={W{\"u}nschiers, R{\"o}bbe},
  booktitle={Computational Biology: A Practical Introduction to Bio Data Juggling with Worked Examples},
  pages={275--289},
  year={2025},
  publisher={Springer}
}

@inproceedings{alrashedy2024software,
  title={How do software engineering researchers use github? an empirical study of artifacts \& impact},
  author={Alrashedy, Kamel and Binjahlan, Ahmed},
  booktitle={2024 IEEE International Conference on Source Code Analysis and Manipulation (SCAM)},
  pages={118--130},
  year={2024},
  organization={IEEE}
}

@article{zhang2025small,
  title={Small-World Phenomenon of Global Open-Source Software Collaboration on Github: A Social Network Analysis},
  author={Zhang, Guoying and Schuessler, Joseph H and Shao, Chris Y},
  journal={Journal of Global Information Management (JGIM)},
  volume={33},
  number={1},
  pages={1--24},
  year={2025},
  publisher={IGI Global Scientific Publishing}
}

@inproceedings{sillito2006questions,
  title={Questions programmers ask during software evolution tasks},
  author={Sillito, Jonathan and Murphy, Gail C and De Volder, Kris},
  booktitle={Proceedings of the 14th ACM SIGSOFT international symposium on Foundations of software engineering},
  pages={23--34},
  year={2006}
}

@inproceedings{kononenko2018studying,
  title={Studying pull request merges: A case study of shopify's active merchant},
  author={Kononenko, Oleksii and Rose, Tresa and Baysal, Olga and Godfrey, Michael and Theisen, Dennis and De Water, Bart},
  booktitle={Proceedings of the 40th international conference on software engineering: software engineering in practice},
  pages={124--133},
  year={2018}
}

@article{kampenes2007systematic,
  title={A systematic review of effect size in software engineering experiments},
  author={Kampenes, Vigdis By and Dyb{\aa}, Tore and Hannay, Jo E and Sj{\o}berg, Dag IK},
  journal={Information and Software Technology},
  volume={49},
  number={11-12},
  pages={1073--1086},
  year={2007},
  publisher={Elsevier}
}

@article{lenarduzzi2021does,
  title={Does code quality affect pull request acceptance? an empirical study},
  author={Lenarduzzi, Valentina and Nikkola, Vili and Saarim{\"a}ki, Nyyti and Taibi, Davide},
  journal={Journal of Systems and Software},
  volume={171},
  pages={110806},
  year={2021},
  publisher={Elsevier}
}

@article{zhang2022pull,
  title={Pull request decisions explained: An empirical overview},
  author={Zhang, Xunhui and Yu, Yue and Gousios, Georgios and Rastogi, Ayushi},
  journal={IEEE Transactions on Software Engineering},
  volume={49},
  number={2},
  pages={849--871},
  year={2022},
  publisher={IEEE}
}

@article{li2021you,
  title={Are you still working on this? An empirical study on pull request abandonment},
  author={Li, Zhixing and Yu, Yue and Wang, Tao and Yin, Gang and Li, Shanshan and Wang, Huaimin},
  journal={IEEE Transactions on Software Engineering},
  volume={48},
  number={6},
  pages={2173--2188},
  year={2021},
  publisher={IEEE}
}

@inproceedings{vasilescu2015quality,
  title={Quality and productivity outcomes relating to continuous integration in GitHub},
  author={Vasilescu, Bogdan and Yu, Yue and Wang, Huaimin and Devanbu, Premkumar and Filkov, Vladimir},
  booktitle={Proceedings of the 2015 10th joint meeting on foundations of software engineering},
  pages={805--816},
  year={2015}
}

@inproceedings{gousios2014exploratory,
  title={An exploratory study of the pull-based software development model},
  author={Gousios, Georgios and Pinzger, Martin and Deursen, Arie van},
  booktitle={Proceedings of the 36th international conference on software engineering},
  pages={345--355},
  year={2014}
}

@article{horikawa2025agentic,
  title={Agentic Refactoring: An Empirical Study of AI Coding Agents},
  author={Horikawa, Kosei and Li, Hao and Kashiwa, Yutaro and Adams, Bram and Iida, Hajimu and Hassan, Ahmed E},
  journal={arXiv preprint arXiv:2511.04824},
  year={2025}
}

@article{wang2025ai,
  title={Ai agentic programming: A survey of techniques, challenges, and opportunities},
  author={Wang, Huanting and Gong, Jingzhi and Zhang, Huawei and Xu, Jie and Wang, Zheng},
  journal={arXiv preprint arXiv:2508.11126},
  year={2025}
}

@article{he2025speed,
  title={Speed at the Cost of Quality: How Cursor AI Increases Short-Term Velocity and Long-Term Complexity in Open-Source Projects},
  author={He, Hao and Miller, Courtney and Agarwal, Shyam and K{\"a}stner, Christian and Vasilescu, Bogdan},
  journal={arXiv preprint arXiv:2511.04427},
  year={2025}
}

@article{yoshimoto2026testing,
  title={Testing with AI Agents: An Empirical Study of Test Generation Frequency, Quality, and Coverage},
  author={Yoshimoto, Suzuka and Fujita, Shun and Horikawa, Kosei and Feitosa, Daniel and Kashiwa, Yutaro and Iida, Hajimu},
  journal={arXiv preprint arXiv:2603.13724},
  year={2026}
}

@article{ogenrwot2026ai,
  title={How AI Coding Agents Modify Code: A Large-Scale Study of GitHub Pull Requests},
  author={Ogenrwot, Daniel and Businge, John},
  journal={arXiv preprint arXiv:2601.17581},
  year={2026}
}

@article{badampudi2023modern,
  title={Modern code reviews—Survey of literature and practice},
  author={Badampudi, Deepika and Unterkalmsteiner, Michael and Britto, Ricardo},
  journal={ACM Transactions on Software Engineering and Methodology},
  volume={32},
  number={4},
  pages={1--61},
  year={2023},
  publisher={ACM New York, NY, USA}
}

@article{gao2026autopilot,
  title={On Autopilot? An Empirical Study of Human-AI Teaming and Review Practices in Open Source},
  author={Gao, Haoyu and Banyongrakkul, Peerachai and Guan, Hao and Zahedi, Mansooreh and Treude, Christoph},
  journal={arXiv preprint arXiv:2601.13754},
  year={2026}
}

@inproceedings{zampetti2017developers,
  title={How developers document pull requests with external references},
  author={Zampetti, Fiorella and Ponzanelli, Luca and Bavota, Gabriele and Mocci, Andrea and Di Penta, Massimiliano and Lanza, Michele},
  booktitle={2017 IEEE/ACM 25th International Conference on Program Comprehension (ICPC)},
  pages={23--33},
  year={2017},
  organization={IEEE}
}

@article{joshi2025swe,
  title={Swe-bench-cl: Continual learning for coding agents},
  author={Joshi, Thomas and Chowdhury, Shayan and Uysal, Fatih},
  journal={arXiv preprint arXiv:2507.00014},
  year={2025}
}

@inproceedings{zheng2025humanevo,
  title={Humanevo: An evolution-aware benchmark for more realistic evaluation of repository-level code generation},
  author={Zheng, Dewu and Wang, Yanlin and Shi, Ensheng and Zhang, Ruikai and Ma, Yuchi and Zhang, Hongyu and Zheng, Zibin},
  booktitle={2025 IEEE/ACM 47th International Conference on Software Engineering (ICSE)},
  pages={1372--1384},
  year={2025},
  organization={IEEE}
}

@inproceedings{moriconi2025ghalogs,
  title={GHALogs: Large-scale dataset of GitHub Actions runs},
  author={Moriconi, Florent and Durieux, Thomas and Falleri, Jean-R{\'e}my and Troncy, Rapha{\"e}l and Francillon, Aur{\'e}lien},
  booktitle={2025 IEEE/ACM 22nd International Conference on Mining Software Repositories (MSR)},
  pages={669--673},
  year={2025},
  organization={IEEE}
}

@inproceedings{tang2022towards,
  title={Towards understanding third-party library dependency in c/c++ ecosystem},
  author={Tang, Wei and Xu, Zhengzi and Liu, Chengwei and Wu, Jiahui and Yang, Shouguo and Li, Yi and Luo, Ping and Liu, Yang},
  booktitle={Proceedings of the 37th IEEE/ACM International Conference on Automated Software Engineering},
  pages={1--12},
  year={2022}
}

@inproceedings{romeo2025uml,
  title={UML is back. Or is it? Investigating the past, present, and future of UML in open source software},
  author={Romeo, Joseph and Raglianti, Marco and Nagy, Csaba and Lanza, Michele},
  booktitle={2025 IEEE/ACM 47th International Conference on Software Engineering (ICSE)},
  pages={2342--2354},
  year={2025},
  organization={IEEE}
}

@inproceedings{sakib2025understanding,
  title={Understanding the popularity of packages in Maven ecosystem},
  author={Sakib, Sadman Jashim and Asaduzzaman, Muhammad and Bright, Curtis and Morgan, Cole},
  booktitle={2025 IEEE/ACM 22nd International Conference on Mining Software Repositories (MSR)},
  pages={364--368},
  year={2025},
  organization={IEEE}
}

@inproceedings{li2020exploratory,
  title={An exploratory study of bugs in extended reality applications on the web},
  author={Li, Shuqing and Wu, Yechang and Liu, Yi and Wang, Dinghua and Wen, Ming and Tao, Yida and Sui, Yulei and Liu, Yepang},
  booktitle={2020 IEEE 31st International symposium on software reliability engineering (ISSRE)},
  pages={172--183},
  year={2020},
  organization={IEEE}
}

@article{aghili2023studying,
  title={Studying the characteristics of AIOps projects on GitHub},
  author={Aghili, Roozbeh and Li, Heng and Khomh, Foutse},
  journal={Empirical Software Engineering},
  volume={28},
  number={6},
  pages={143},
  year={2023},
  publisher={Springer}
}

@inproceedings{rastogi2018relationship,
  title={Relationship between geographical location and evaluation of developer contributions in github},
  author={Rastogi, Ayushi and Nagappan, Nachiappan and Gousios, Georgios and van der Hoek, Andr{\'e}},
  booktitle={Proceedings of the 12th ACM/IEEE international symposium on empirical software engineering and measurement},
  pages={1--8},
  year={2018}
}

@article{farrag2026productivity,
  title={The Productivity-Reliability Paradox: Specification-Driven Governance for AI-Augmented Software Development},
  author={Farrag, Sabry E},
  journal={arXiv preprint arXiv:2605.01160},
  year={2026}
}

@inproceedings{bennett2000software,
  title={Software maintenance and evolution: a roadmap},
  author={Bennett, Keith H and Rajlich, V{\'a}clav T},
  booktitle={Proceedings of the Conference on the Future of Software Engineering},
  pages={73--87},
  year={2000}
}

@article{tang2024enhancing,
  title={Enhancing bug-inducing commit identification: A fine-grained semantic analysis approach},
  author={Tang, Lingxiao and Ni, Chao and Huang, Qiao and Bao, Lingfeng},
  journal={IEEE Transactions on Software Engineering},
  volume={50},
  number={11},
  pages={3037--3052},
  year={2024},
  publisher={IEEE}
}

@article{robbes2026agentic,
  title={Agentic Much? Adoption of Coding Agents on GitHub},
  author={Robbes, Romain and Matricon, Th{\'e}o and Degueule, Thomas and Hora, Andre and Zacchiroli, Stefano},
  journal={arXiv preprint arXiv:2601.18341},
  year={2026}
}

@article{dakhel2023github,
  title={Github copilot ai pair programmer: Asset or liability?},
  author={Dakhel, Arghavan Moradi and Majdinasab, Vahid and Nikanjam, Amin and Khomh, Foutse and Desmarais, Michel C and Jiang, Zhen Ming Jack},
  journal={Journal of Systems and Software},
  volume={203},
  pages={111734},
  year={2023},
  publisher={Elsevier}
}

@article{tambon2025bugs,
  title={Bugs in large language models generated code: An empirical study},
  author={Tambon, Florian and Moradi-Dakhel, Arghavan and Nikanjam, Amin and Khomh, Foutse and Desmarais, Michel C and Antoniol, Giuliano},
  journal={Empirical Software Engineering},
  volume={30},
  number={3},
  pages={65},
  year={2025},
  publisher={Springer}
}

@article{hora2026coding,
  title={Are Coding Agents Generating Over-Mocked Tests? An Empirical Study},
  author={Hora, Andre and Robbes, Romain},
  journal={arXiv preprint arXiv:2602.00409},
  year={2026}
}

@article{peralta2026agentic,
  title={Why Are Agentic Pull Requests Merged or Rejected? An Empirical Study},
  author={Peralta, Sien Reeve O and Hoshi, Fumika and Washizaki, Hironori and Ubayashi, Naoyasu and Kondo, Inase and Higo, Yoshiki and Mukai, Hiroki and Yoshida, Norihiro and Kusama, Kazuki and Tanaka, Hidetake and others},
  journal={arXiv preprint arXiv:2605.22534},
  year={2026}
}

@article{popescu2026investigating,
  title={Investigating Autonomous Agent Contributions in the Wild: Activity Patterns and Code Change over Time},
  author={Popescu, Razvan Mihai and Gros, David and Botocan, Andrei and Pandita, Rahul and Devanbu, Prem and Izadi, Maliheh},
  journal={arXiv preprint arXiv:2604.00917},
  year={2026}
}

@article{huang2026more,
  title={More Code, Less Reuse: Investigating Code Quality and Reviewer Sentiment towards AI-generated Pull Requests},
  author={Huang, Haoming and Jaisri, Pongchai and Shimizu, Shota and Chen, Lingfeng and Nakashima, Sota and Rodr{\'\i}guez-P{\'e}rez, Gema},
  journal={arXiv preprint arXiv:2601.21276},
  year={2026}
}

@article{agarwal2026ai,
  title={AI IDEs or Autonomous Agents? Measuring the Impact of Coding Agents on Software Development},
  author={Agarwal, Shyam and He, Hao and Vasilescu, Bogdan},
  journal={arXiv preprint arXiv:2601.13597},
  year={2026}
}

@article{mchugh2013chi,
  title={The chi-square test of independence},
  author={McHugh, Mary L},
  journal={Biochemia medica},
  volume={23},
  number={2},
  pages={143--149},
  year={2013},
  publisher={Hrvatsko dru{\v{s}}tvo za medicinsku biokemiju i laboratorijsku medicinu}
}

@book{cramer1999mathematical,
  title={Mathematical methods of statistics},
  author={Cram{\'e}r, Harald},
  volume={9},
  year={1999},
  publisher={Princeton university press}
}

@misc{replication_package,
    author = {Mazloomzadeh, Iren and Morovati, Mohammad Mehdi and Khomh, Foutse},
    title = {Replication package of the paper},
    howpublished = {\url{https://github.com/mazloomzadeh/AI-Agent-Analysis}},
    year={2026},
    note = {Accessed: 2026-06-01}
}

@article{lenth2023emmeans,
  title={emmeans: Estimated Marginal Means, aka Least-Squares Means\_.},
  author={Lenth, Russell},
  journal={R package version 1.8. 5},
  year={2023}
}

@article{searle1980population,
  title={Population marginal means in the linear model: an alternative to least squares means},
  author={Searle, Shayle R and Speed, F Michael and Milliken, George A},
  journal={The American Statistician},
  volume={34},
  number={4},
  pages={216--221},
  year={1980},
  publisher={Taylor \& Francis}
}

@inproceedings{popoola2025empirical,
  title={Empirical Analysis of Pull Requests for Google Summer of Code},
  author={Popoola, Saheed},
  booktitle={Proceedings of the 26th ACM Annual Conference on Cybersecurity \& Information Technology Education},
  pages={247--253},
  year={2025}
}

@article{rosa2023comprehensive,
  title={A comprehensive evaluation of SZZ Variants through a developer-informed oracle},
  author={Rosa, Giovanni and Pascarella, Luca and Scalabrino, Simone and Tufano, Rosalia and Bavota, Gabriele and Lanza, Michele and Oliveto, Rocco},
  journal={Journal of systems and software},
  volume={202},
  pages={111729},
  year={2023},
  publisher={Elsevier}
}

\end{document}